\documentclass[apj,numberedappendix,appendixfloats,iop]{emulateapj}

\usepackage{layouts}
\usepackage{amsmath}
\usepackage{natbib}
\makeatletter
\def\tocname{TABLE OF CONTENT}
\def\lofname{LIST OF FIGURES}
\def\lotname{LIST OF TABLES}
\def\tableofcontents{\section*{\tocname}\@starttoc{toc}}
\def\listoffigures{\section*{\lofname}\@starttoc{lof}}
\def\listoftables{\section*{\lotname}\@starttoc{lot}}
\def\tablehead#1#2{
  \@table@not@headedfalse%
  \kill
  \caption[#2]{\\%
    \@tablecaption}%
    \\
  \hline
  \hline\\[-1.7ex]
  #1\hskip\tabcolsep\\[.7ex]
  \hline\\[-1.5ex]
  \endfirsthead
  \caption[]{--- \emph{Continued}}\\
  \hline
  \hline\\[-1.7ex]
  #1\hskip\tabcolsep\\[.7ex]
  \hline\\[-1.5ex]
  \endhead
  \hline
  \endfoot%
}
\makeatother

\makeatletter
\renewcommand*\l@section{\@dottedtocline{0}{0em}{1em}}
\renewcommand*\l@subsection{\@dottedtocline{0}{1em}{1.7em}}
\renewcommand*\l@subsubsection{\@dottedtocline{0}{2.7em}{2.7em}}
\makeatother

\makeatletter
\def\gtappeq{\compoundrel>\over\sim}
\def\ltappeq{\compoundrel<\over\sim}
\def\compoundrel#1\over#2{\mathpalette\compoundreL{{#1}\over{#2}}}
\def\compoundreL#1#2{\compoundREL#1#2}
\def\compoundREL#1#2\over#3{\mathrel
      {\vcenter{\hbox{$\m@th\buildrel{#1#2}\over{#1#3}$}}}}
\makeatother

\slugcomment{Submitted to ApJ Supplement Series}

\shorttitle{The Evolution of Cataclysmic Variables} 
\shortauthors{Knigge, Baraffe \& Patterson}

\begin{document}


\title{The Evolution of Cataclysmic Variables as Revealed by their
  Donor Stars}


\author{Christian Knigge}
\affil{School of Physics \& Astronomy, University of Southampton,
  Southampton SO17 1BJ, UK}   
%\email{C.Knigge@soton.ac.uk}

\author{Isabelle Baraffe}
\affil{School of Physics, University of Exeter, Exeter EX4 4QL, UK}
%\email{I.Baraffe@exeter.ac.uk}

\and

\author{Joseph Patterson}
\affil{Department of Astronomy, Columbia University, 550 West 120th
  Street, New York, NY 10027, USA} 
%\email{jop@astro.columbia.edu}

\email{C.Knigge@soton.ac.uk,\hspace*{0.2cm}I.Baraffe@exeter.ac.uk,\hspace*{0.2cm}jop@astro.columbia.edu} 



\begin{abstract}

We present an attempt to reconstruct the complete evolutionary path
followed by cataclysmic variables (CVs), based on the observed mass-radius
relationship of their donor stars. Along the way, we update the
semi-empirical CV donor sequence presented in Knigge (2006), 
present a comprehensive review of the connection between CV evolution
and the secondary stars in these system, and reexamine most of the
commonly used magnetic braking (MB) recipes, finding that even conceptually
similar ones can differ greatly in both magnitude and functional form.  

The great advantage of using donor radii to infer mass-transfer and
angular-momentum-loss (AML) rates is that they sample the longest
accessible time scales and are most likely to represent the true secular
(evolutionary average) rates. We show explicitly that if CVs exhibit
long-term mass-transfer-rate fluctuations, as is often assumed, the
expected variability time scales are so long that other tracers of the
mass-transfer rate -- including white dwarf (WD) temperatures --
become unreliable.

We carefully explore how much of the radius difference between CV
donors and models of isolated main-sequence stars may be due to
mechanisms other than mass loss. The tidal and rotational deformation
of Roche-lobe-filling stars produces $\simeq 4.5\%$ radius inflation
below the period gap, and $\simeq 7.9\%$ above. A comparison of
stellar models to mass-radius data for non-interacting stars suggests
a real offset of $\simeq 1.5\%$ for fully convective stars
(i.e. donors below the gap) and $\simeq 4.9\%$ for partially radiative 
ones (donors above the gap). We also show that donor bloating due to 
irradiation is probably smaller than, and at most comparable to, these 
effects. 

After calibrating our models to account for these issues, we fit
self-consistent evolution sequences to our compilation of donor
masses and radii. In the standard model of CV 
evolution, AMLs below the period gap are assumed to be driven solely
by gravitational radiation (GR), while AMLs above the gap are usually 
described by a MB law first suggested by Rappaport, Verbunt \& Joss (1983). We
adopt simple scaled versions of 
these AML recipes and find that these are able to match the data quite
well. The optimal scaling factors turn out to be $f_{GR} = 2.47
\pm 0.22$ below the gap and $f_{MB} = 0.66 \pm 0.05$ above (the errors
here are purely statistical, and the standard model corresponds to $f_{GR} =
f_{MB} = 1$). This revised model describes the mass-radius data
significantly better than the standard model.

Some of the most important implications and applications of
our results are as follows. (1) The revised evolution sequence yields 
correct locations for the minimum period and the upper
edge of the period gap; the standard sequence does not. (2) The
observed spectral types of CV donors are compatible with both standard
and revised models. (3) A direct comparison of predicted and observed
WD temperatures suggests an even higher value for $f_{GR}$, but this
comparison is sensitive to the assumed mean WD mass and the possible
existence of mass-transfer-rate fluctuations. (4) The predicted
absolute magnitudes of donors stars in the
near-infrared form a lower envelope around the observed absolute
magnitudes for systems with parallax distances. This is true for all
of our sequences, so any of them can be used to set firm lower limits on (or
obtain rough estimates of) the distance toward CVs based only on 
$P_{orb}$ and single-epoch near-IR measurements. (5) Both standard
and revised sequences predict that short-period CVs should be
susceptible to dwarf nova (DN) eruptions, consistent with
observations. However, both sequences also predict that the fraction
of DNe among long-period CVs should decline with $P_{orb}$ above
the period gap. Observations suggest the opposite behaviour, and we
discuss the possible explanations for this discrepancy. 
(6) Approximate orbital period distributions
constructed from our evolution sequences suggest that the ratio of
long-period CVs to short-period, pre-bounce CV is about $3\times$
higher for the revised sequence than the standard one. This may
resolve a long-standing problem in CV evolution.

Tables describing our donor and evolution sequences are provided in
electronically readable form.

\end{abstract}

\nopagebreak[4]
\keywords{novae, cataclysmic variables, stars: fundamental parameters}

\clearpage
\pagebreak[4]
\newpage


\footnotesize
\tableofcontents
\vfill
\listoffigures
\listoftables
\vfill
\normalsize

%\newpage
%\pagebreak

%\newpage

%\pagebreak[4]

\clearpage
\newpage
\pagebreak

\section{Introduction}
\label{sec:intro}

Cataclysmic variables (CVs) are interacting binary stars in which a
white dwarf (WD) accretes material from a low-mass, near-main-sequence
companion star. The long-term evolution of CVs is driven by angular
momentum losses (AMLs). In what may be called the ``standard model''
of CV evolution, the dominant AML mechanism in long-period systems 
($P_{orb} \gtappeq 3$~hrs) is ``magnetic braking'' (MB), whereas
short-period CVs ($P_{orb} \ltappeq 2$~hrs) are assumed to be
driven by AML associated with the emission of gravitational 
radiation (GR). 

The original motivation for the development of this standard model was
the need to explain the dearth of active CVs in the period range $2~{\rm hrs}
\ltappeq P_{orb} \ltappeq 3~{\rm hrs}$, the famous CV {\em period
gap}. The standard model accomplishes this by assuming that when the donor star 
in a CV becomes fully convective -- which happens at around $P_{orb}
\simeq 3$~hrs -- MB will abruptly shut off. At this point in its
evolution, the donor star has been driven slightly out of
thermal equilibrium and is therefore somewhat oversized for its
mass. When the sudden cessation of MB reduces the mass-loss rate, the
secondary contracts, causing it to lose contact with the Roche lobe
altogether. The system then evolves towards shorter
periods as a detached binary, driven only by GR.  The Roche lobe
eventually catches up with the donor again at $P_{orb} \simeq
2~{\rm hrs}$. At this point mass transfer restarts, and the system
re-emerges as an active CV at the bottom of the period gap. 

The second key feature of the observed orbital period distribution of 
CVs is a sharp cut-off at $P_{min} \simeq 80$~min. In the
standard model, the existence of this minimum period is again
associated with the secondary's mass-loss-induced loss of thermal
equilibrium. More specifically, the donor's thermal time scale 
below the gap ($\tau_{kh} \sim GM_2^2/L_2R_2$) increases faster than
the time scale on which it loses mass ($\tau_{\dot{M}_2} \sim
M_2/\dot{M}_2$). The secondary is therefore driven further and further 
from thermal equilibrium, to the point where its radius no longer
shrinks at all in response to mass loss (and potentially even grows
because of partial degeneracy). In this limit, the size of the binary
orbit must incease again
in order to accomodate the secondary, so there must be a change in the
direction of the system's period evolution. 
Systems that have passed beyond $P_{min}$ are evolving back towards
longer periods and are often called {\em period 
bouncers}. In principle (barring selection effects), $P_{min}$ should
stand out clearly in the $P_{orb}$ distribution of CVs, not just as an
abrupt drop in numbers, but as a narrow spike with a sharp,
short-period cut-off. 
  
The basic disrupted MB scenario described above was conceived more
than 25 years ago \citep{1982ApJ...254..616R, 1983ApJ...275..713R,
1983A&A...124..267S}.
It has survived as the standard model for CV evolution, primarily
because it accounts for the existence of the  
period minimum and the period gap.
\footnote{Actually, it is interesting to note that these original
studies were careful to point out that the existence of the period gap
suggested a sharp {\em reduction} in the strength of MB for fully convective
stars, not necessarily a complete {\em cessation}.}
%Thus our view of the
%standard model today is rather more ``fundamentalist'' than that
%originally put forward.}
However, it has been recognized for some time now that there may be
problems with this picture. Three issues, in particular, have often
been noted as serious challenges for the standard model. 

First, the theoretically predicted minimum period ($P_{min} \simeq
65$~min)
\footnote{Note, however, that this is the value {\em before} any of
the corrections discussed in Section~\ref{sec:skeletons}.} is
substantially shorter than the observed one. Until 
recently, the latter was usually estimated to be around $P_{min} \simeq
75$~min, based on the observed cut-off in the period distribution of
the then available CV sample 
\citep[e.g.][hereafter K06]{2006MNRAS.373..484K}
However, \citet[][hereafter G09]{2009MNRAS.397.2170G} recently located 
the period spike at $P_{min} \simeq 82$~min in the SDSS CV sample,
which makes this discrepancy even worse. Second, the standard
model predicts that the Galactic CV population should be completely
dominated by short-period systems and period bouncers. More
specifically, the model predicts intrinsic abundances ratios of
roughly 1:30:70 for long-period CVs, short-period pre-bounce CVs
and period bouncers, respectively
\citep[e.g.][]{1993A&A...271..149K}. Such a small ratio 
of long-period CVs to short-period (pre-bounce) CVs does not seem
compatible with observations \citep{1998PASP..110.1132P},
even taking into account selection effects 
\citep{2007MNRAS.374.1495P,2008MNRAS.385.1471P,2008MNRAS.385.1485P}.
Third, as discussed in more detail in
Section~\ref{sec:discuss_plausibility}, there is substantial evidence
from non-interacting low-mass stars that fully convective objects (and
perhaps even brown dwarfs) can sustain significant magnetic fields
\citep[e.g.][]{2009IAUS..259..339R,2009AIPC.1094..206B,2009ARA&A..47..333D}
and experience at least some spin-down due to MB
\citep[e.g.][]{2003ApJ...586..464B,2008ApJ...684.1390R}. Whether the
MB torque experienced by these objects is strong enough to matter in a
CV setting is currently still an open question.

Taken at face value, all three of these issues can be viewed as
pointing to an AML mechanism acting in addition to GR below the period
gap \citep{1998PASP..110.1132P,2007MNRAS.374.1495P}. So long as there
remains at least a significant reduction in the 
AML loss rate at $P_{orb} \simeq 3$~hrs, such a modification would not
destroy the model's ability to account for the existence of the
gap. However, there have also been other, more radical 
suggestions for changes to the standard model. These range from
drastic reductions in the AML rate associated with MB 
\citep{2003ApJ...582..358A,2003ApJ...599..516I}, 
to the hypothesis that most CVs
are too young to have reached the theoretical period minimum (in which
case the observed short-period cut-off is simply an age effect;
\citealt{2002ASPC..261..233K}).

Against this background, the goal of the present paper is to construct
a new, comprehensive, semi-empirical CV evolution track based
exclusively on the properties of their donor stars. This is a 
promising pursuit, since the secular evolution of a CV is 
intricately tied to -- and in some sense controlled by -- the
properties of its secondary star. However, the main 
advantage of our method is this: 
%It also has a 
%Donor-based methods for inferring $\dot{M}_2(P_{orb}}$ has one 
%crucial and unique advantage: 
it is likely to yield a valid estimate of the {\em secular}
mass-transfer rate, i.e. the long-term
average $\dot{M}_2$ that drives CV evolution. This is crucial, 
because the mass-transfer and accretion rates in a CV can vary on a
wide range of shorter-than-evolutionary time scales. Methods based on
observational tracers of $\dot{M}_2$ that are sensitive to such
variations will therefore produce noisy estimates of the secular
$\dot{M}_2(P_{orb})$ at best and misleading ones at worst.
\footnote{Misleading estimates will result, for example, if 
bright CVs in temporary high-$\dot{M}_2$ states have a substantially
higher discovery probability.}
In fact, the two most promising alternative methods -- which are based
on time-averaged accretion light 
\citep{1984ApJS...54..443P,1987MNRAS.227...23W,2009arXiv0903.1006P}
and on the effective temperature of the accreting WD
\citep{2002ApJ...565L..35T,2003ApJ...596L.227T,
  2004ApJ...600..390T,2009ApJ...693.1007T}, respectively -- are
both sensitive to variations in 
$\dot{M}_2$ on time scales that are much shorter than those produced
by the most likely types of long-term mass-transfer-rate fluctuations
(see Section~\ref{sec:fluc}). 

In principle, it is quite simple to construct a donor-based CV
evolution track: at fixed mass, the radius of a donor star in
a CV is set primarily by the rate at which it loses mass (since this
determines its degree of thermal 
disequilibrium). Thus, modulo the effects discussed in
Section~\ref{sec:skeletons}, {\em the amount by which a donor is
  bloated, relative to an isolated star of the same mass, is a measure of
$\dot{M}_2$.} Given a set of empirically determined donor masses and
radii spanning the full range of orbital periods, we can therefore
infer how $\dot{M}_2$ (and hence the systemic AML that drives it) must
vary with $P_{orb}$.  

In practice, there are, of course, several obstacles to overcome
before this program can be carried out. However, a good deal of 
the groundwork was already laid by K06, who used the sample
of donor masses and radii presented by 
\citet[][hereafter P05]{2005PASP..117.1204P} to
construct a semi-empirical donor sequence for CVs. This sequence is
built around a broken-power-law approximation to the donor 
mass-radius relation and provides a useful take-off and comparison
point for the present study.

Apart from the obvious hope that our donor-based evolution track will
turn out to be a reasonable description of reality, our main
motivation for constructing it is to provide a new benchmark for work
on CV evolution. As an empirically-based alternative to the standard
model, our revised evolution sequence should provide a useful reference
point for everything from theoretical population synthesis studies to
detailed work on individual systems. In order to enable such uses, we
will make the complete track (including physical and photometric
parameters) available to the community, both as a simple ``recipe''
and as a set of tables in convenient electronic form.

The overall plan of this paper is as follows. We will begin in
Section~\ref{sec:evo} by reviewing the links between the secular
evolution of a CV and the properties of its donor star. This will
include a comparison of some of the most popular MB prescriptions in
the literature. In Section~\ref{sec:update}, we will then revisit and
update the K06 semi-empirical donor sequence to take into 
account new results that have appeared since its construction. In
Section~\ref{sec:fluc} we will discuss how long-term
mass-transfer-rate 
fluctuations can affect observational tracers of CV evolution,
including accretion light, WD temperatures and donor radii. In
Section~\ref{sec:assembly}, we will describe the construction of a
complete, self-consistent CV evolution track that optimally fits the
observed donor mass-radius relation. Along the way, we will explain
how we deal with several calibration 
issues that slightly complicate the interpretation of donor inflation
as a measure of $\dot{M}_2$. In Section~\ref{sec:newtrack}, we will
present the resulting revised CV evolution sequence. More specifically, 
we will show the full range of binary and donor  
properties along this optimal evolution track and check its 
consistency with observed CV properties other than donor radii. In
Section~\ref{sec:implications}, we will consider some of the most
important implications and applications of our track. Finally, in
Sections~\ref{sec:discuss} and ~\ref{sec:discuss}, respectively, we
will discuss our results and summarize our main conclusions. 

\section{CV Evolution And Donor Stars}
\label{sec:evo}

Since the ultimate goal of our study is to infer a global picture of
CV evolution from the properties of the donor stars in these systems,
it is worth reviewing the underlying physics. In the following
sections, we will therefore briefly describe the link between AML
and mass transfer in CVs, the response of a low-mass star to mass 
loss, and the close connection between donor properties and secular CV
evolution. 

\subsection{Angular Momentum Loss and Mass Transfer in CVs}

The evolution of CVs is driven entirely by AMLs. In particular, as
long as the donor stays 
in contact with its Roche lobe, any systemic angular momentum loss at
a rate 
$\dot{J}_{sys} < 0$ will drive a ML rate from the
secondary that is given by \citep[e.g.][]{1995ApJ...439..330K}
\begin{equation}
\frac{\dot{M}_{2}}{M_2} = \frac{\dot{J}_{sys}}{JD} < 0. 
\label{eq:mdot}
\end{equation}
In this evolution equation, $J$ is the angular momentum of the
binary system 
\begin{equation}
J = M_1 M_2 \left(\frac{G a}{M}\right)^{1/2}
\label{eq:j}
\end{equation}
and $D$ is given by
\footnote{The form of $D$ in Equation~\ref{eq:D} assumes
Paczy{\'n}ski's (\citeyear{1971ARA&A...9..183P},
Equation~\ref{eq:pac}) approximation for the 
Roche-lobe radius. For most other purposes in this paper, we will use
the more precise approximation given by Equation~\ref{eq:siro1}. 
However, Paczy{\'n}ski's approximation greatly simplifies the evolution
equation and causes only a minor loss of accuracy (at the level
of a few percent) in the calculation of $\dot{M}_2$. This is entirely
acceptable for our purposes.} 
\begin{equation}
D = \left(\frac{5}{6} + \frac{\zeta}{2}\right) - \frac{M_2}{M_1} +
\alpha\left(\frac{M_2}{M_1} - \frac{1}{3}\frac{M_2}{M}\right) - \nu.
\label{eq:D}
\end{equation}
Here and throughout, $M_1$ denotes the mass of the WD primary, $M =
M_1 + M_2$ the combined mass of the system, $a$ the binary
separation, and $\zeta = d\ln{R_2}/d\ln{M_2}$ is the mass-radius index of
the donor evaluated along its evolution track. 

The parameter $\alpha$ that appears in the evolution
equation is given by 
\begin{equation}
\alpha = \frac{\dot{M}}{\dot{M_2}}
\end{equation}
and measures the fraction of the mass lost from the secondary that is 
ultimately also lost from the system. In the case of conservative mass
transfer, where all of the mass lost from the secondary is permanently
accreted by the primary, $\alpha=0$. Similarly, the parameter 
\begin{equation}
\nu = \frac{\dot{J}_{CAML}/J}{\dot{M}_2 / M_2}
\end{equation}
measures the amount of {\em consequential} AML (CAML) that is associated with
the mass-transfer process  \citep{1995ApJ...439..330K}. Note that this is in
addition to the systemic $\dot{J}_{sys}$, so the total AML rate from
the system is
\begin{equation}
\dot{J} = \dot{J}_{sys} + \dot{J}_{CAML}.
\end{equation}
Throughout this study, we will make the usual
assumption that all of the material accreted by the primary is ejected
again during nova eruptions, taking with it the specific angular
momentum of the primary. This is consistent with the lack
of a clear orbital period dependence in the measured WD masses for CVs 
(K06; also see Section~\ref{sec:update}) and implies $\alpha = 1$ and
$\nu = M_2^2/(M_1 M)$. 

\subsection{The Response of a Low-Mass Star to Mass Loss}

\subsubsection{Basic Physics}
\label{sec:basics}

As briefly noted in the Section~\ref{sec:intro}, the reaction of a
low-mass star to ML depends on two time scales. The first
is the time scale on which the donor is losing mass ($\tau_{\dot{M}_2}
\sim M_2/\dot{M}_2$); the second is the thermal (or Kelvin-Helmholtz)
time scale on which thermal equilibrium is established ($\tau_{kh}
\sim GM_2^2/L_2R_2$). 

If $\tau_{\dot{M}_2} >> \tau_{kh}$, ML is slow, and the donor is
always able to maintain thermal equilibrium. In this limit, the
secondary is indistinguishable from 
an isolated main-sequence (MS) star of the same mass. The 
mass-radius index, $\zeta$, along its evolution track is then set by 
the equilibrium mass-radius relation of lower-MS stars ($R \simeq
M^{0.8}$), i.e. $\zeta = \zeta_{eq} \simeq 0.8$. By contrast, if 
$\tau_{\dot{M}_2} << \tau_{kh}$, ML is fast (adiabatic), and
the donor is completely unable to maintain thermal equilibrium. In
this paper, we are mainly concerned with essentially unevolved
CV donors with $M_2 \ltappeq 0.6~M_{\odot}$, i.e. low-mass stars with
a significant convective envelope. The response of such
stars to adiabatic ML is to grow in size. The effective
mass-radius index in this limit is $\zeta = \zeta_{ad} \simeq -1/3$
\citep[e.g.][]{1982ApJ...254..616R}.

As it turns out, almost all models for CV evolution suggest that
the donor stars in these systems find themselves in the intermediate
regime, $\tau_{\dot{M}_2} \sim
\tau_{kh}$ (see, for example, Figure~23 in
\citet{1984ApJS...54..443P}). 
Thus (pre-bounce) CV secondaries are almost, {\em but not 
quite}, able to maintain thermal equilibrium. They should
therefore be 
somewhat oversized relative to isolated MS stars, with an effective
mass-radius index close to, but slightly below, the equilibrium
value. This theoretical expectation has been confirmed empirically by 
P05 and K06, who found that CV donors 
are indeed up to 20\%-30\% larger than MS stars of the same
mass. Moreover, the observed mass-radius relation for the secondaries
in pre-bounce systems suggests $\zeta \simeq 0.65$ both above and below
the period gap, which is indeed comparable to, but smaller than, the
equilibrium value.

This radius inflation is the main observable effect of the ongoing ML
on the donor star in a CV. It is this effect we will exploit in 
constructing our semi-empirical CV evolution track. As already
explained in Section~\ref{sec:intro}, the basic idea is to use the degree
of donor bloating as a measure of the mass-loss rate from the donor. 

\subsubsection{The Radius Adjustment Time Scale: Sensitivity to Mass-Loss History}
\label{sec:adjust}

If we are going to use the radius of the secondary to reconstruct CV
evolution, it is clearly important to ask 
how quickly the donor adjusts its radius to the prevailing mass-loss
rate. In order for the radius to be a good tracer of the secular
$\dot{M}_2$, we would ideally like this time scale, $\tau_{adj}$, to
be comparable to 
the evolutionary time scale, i.e. the time scale on which the key system
parameters evolve:
\begin{equation}
\tau_{ev} \sim \frac{J}{\dot{J}_{sys}} \sim \frac{M_2}{\dot{M}_2} \sim \frac{P_{orb}}{\dot{P}_{orb}}.
\end{equation}
However, in order for CVs to even {\em have} a unique evolution track,
it is actually required that $\tau_{adj} << \tau_{ev}$. If this were
not the case, donors characterized by the same present-day mass and
mass-loss rate, but starting from different initial
conditions, could have very different radii. This would destroy 
the sharp edges of the period gap and the well-defined cut-off in the CV
distribution at $P_{min}$. Thus, in practice, the best we can hope for
is that $\tau_{adj}$ should be long compared to all other relevant
time scales.

An analytical estimate of $\tau_{adj}$ for stars with a substantial
convective envelope has been derived by \citet{1996MNRAS.279..581S} as
\footnote{For reference, we note that \citet{1996MNRAS.279..581S} refer to
  $\tau_{adj}$ as $\tau_{per}$ and that they actually give $\tau_{per}
  \ltappeq   \tau_{kh,eq} / 18$. However, it is easy 
to show from their work that $\tau_{per} / \tau_{{kh},eq}$ is always
close to 20, justifying the estimate given in
Equation~\ref{eq:tau_adj}.}
\begin{equation}
\tau_{adj} \simeq 0.05 \tau_{{kh},eq}. 
\label{eq:tau_adj}
\end{equation}
Here, $\tau_{{kh},eq}$ is the Kelvin-Helmholtz time scale the star would
have if it were in thermal equilibrium. To the extent that $\tau_{ev}
\sim \tau_{{kh},eq}$ in CVs, $\tau_{adj}$ is generally indeed faster than
$\tau_{ev}$. As noted by \citet{1996MNRAS.279..581S}, this also nicely
explains the rapid convergence of CV evolution tracks characterized by
different initial conditions, which is always observed in numerical
studies \citep[e.g.][]{1983ApJ...268..825P,1992A&A...254..213K,
1993A&A...271..149K}.

The time scale $\tau_{adj}$ can be thought of as the averaging 
time scale for donor radii: the observed radius is a measure of
the mass-transfer rate averaged over the preceeding few
$\tau_{adj}$. Does this mean that donor radii have {\em no}
sensitivity to the mass-loss history on evolutionary time scales?
In other words, is a donor radius measurement at given $M_2$ and
$P_{orb}$ a {\em unique} measure of $\dot{M}_2$, irrespective of
whether, for example, the star experienced faster or slower mass loss
in the past?

\begin{figure*}
\centering 
\includegraphics[height=15cm,angle=-90]{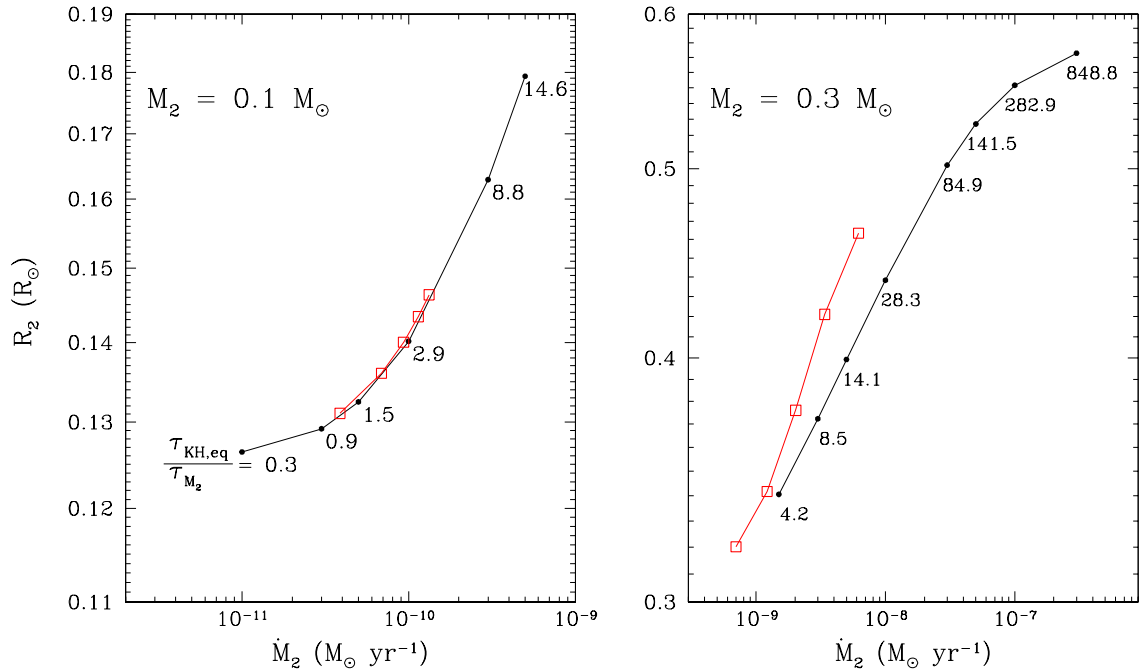}
\caption[The dependence of stellar radius on mass-loss rate for two
representative stellar masses.]{The dependence of stellar radius on
  mass-loss rate for two representative stellar masses: $M_2 = 0.1
  M_{\odot}$ (left panel; typical of short-period CV donors) and $M_2 = 0.3 M_{\odot}$
(right panel; typical of long-period CV donors). The black dots connected by solid lines in
each panel are model radii obtained from self-consistent donor
evolution sequences of the type described in 
Section~\ref{sec:assembly}, under the assumption of {\em constant} mass-loss
rate as a function of time. The red squares connected by dashed lines in each panel correspond
to model radii obtained for sequences in which the mass-loss rates
decrease with time. More specifically, the mass-loss rates in these
sequences are driven by GR-like (Equation~\ref{eq:jdot_gr}) or MB-like
(Equation~\ref{eq:jdot_rvj} with $\gamma = 3$) AML recipes, but adopting a
range of different normalizations for each recipe. 
The numbers below the black
points show the ratio of thermal to mass-loss time scales at these
locations. The radii shown here do not include any of the corrections
discussed in Section~\ref{sec:skeletons}. If mass-loss history had no
effect on stellar radii, the red (dashed) and black (solid) sequences
should form lie on a single, unique curve.} 
\label{fig:convergence}
\end{figure*}

If the condition $\tau_{adj} << \tau_{ev}$ were strictly satisfied,
the answer would be yes. But is this actually true in practice? We try
to answer this question in Figure~\ref{fig:convergence}, where we plot $R_2$ vs
$\dot{M}_2$ for $M_2 = 0.1 M_{\odot}$ (left panel) and $M_2 = 0.3
M_{\odot}$ (right panel). The black dots in each panels are model radii
obtained from donor evolution sequences of the type described in
Section~\ref{sec:assembly}, under the assumption of {\em constant} mass-loss
rate as a function of time. By contrast, the red squares correspond
to model radii obtained 
for sequences in which the evolution is driven by standard GR-like
(left panel) or MB-like (right panel) AML laws, but with variable 
normalizations (so as to produce a range of predicted 
$\dot{M}_2$ at given $M_2$). Both of these AML laws yield a decrease
in $\dot{M}_2$ with time, i.e. mass loss was faster in the
past than at present. The adopted range of mass-loss rates and AML
prescriptions were chosen to be roughly appropriate for 
short-period, pre-bounce CVs (left panel) and long-period CVs
approaching the period gap (right panel). 
If mass-loss history has no effect, all donors with the same
present-day mass and mass-loss rate should have the same radius. 
Thus the red and black dots should fall on a single line.
%even
%though donors evolving along GR-like and MB-like sequences
%experienced higher mass-loss rates in the past.

This is, in fact, what we observe for our low-mass, short-period
donors in the left panel, where the
differences in donor radii at fixed $\dot{M}_2$ 
are always $<$ 1\%. However, the long-period donors in the right 
panel exhibit larger differences, up to 13\% at  
the highest $\dot{M}_2$. There are two reasons for this contrasting
behaviour. First, mass loss is faster compared to the
donor's thermal time scale above the gap. This is shown explicitly by
the numbers below the 
constant-$\dot{M}_2$ sequences, which provide estimates of the ratio
$\tau_{{kh},eq} / \tau_{\dot{M}_2}$ at the specified points. As
expected, we see that the radius difference -- i.e. the impact of mass
loss history -- grows with increasing $\tau_{{kh},eq} /
\tau_{\dot{M}_2}$. In the right panel, we actually reach $\tau_{\dot{M}_2}
\ltappeq \tau_{{kh}}/20 \simeq \tau_{adj}$ for mass-loss rates
$\dot{M}_2 \gtappeq 10^{-8}$, so significant sensitivity
to mass-loss history is no surprise in this regime. Second, even
though $\dot{M}_2$ decreases with time in both GR-like and MB-like 
sequences, the {\em rate} of decrease is faster in the MB-like
sequences (see, for example, Figures~\ref{fig:evo1} and~\ref{fig:evo2}
in Section~\ref{sec:props}). Thus MB-like
sequences will differ more from constant-$\dot{M}_2$ sequences than
GR-like sequences with the same present-day ratio of $\tau_{kh,eq} / \tau_{\dot{M}_2}$. 

It is obvious from Figure~\ref{fig:convergence} that the effect of mass-loss
history cannot be safely ignored for long-period systems. In fact,
despite the good agreement seen in the left panel, it cannot even be
ignored for all short-period systems. The problem is 
that, almost by definition, the ratio $\tau_{{kh}} / \tau_{\dot{M}_2}$
increases throughout the evolution below the gap. In fact, it is this drop
that produces the increasing degree of thermal disequilibrium that
ultimately leads to period bounce (see Section~\ref{sec:bounce}). Thus while
the impact of mass-loss history may be insignificant at $M_2 \simeq
0.1 M_{\odot}$ below the gap, it is expected to grow as we approach period
bounce. We have verified that this is indeed the case. We therefore
conclude that donor radii {\em can} be sensitive to the
{\em shape} of the assumed AML law, both above the gap and below.
\footnote{Note that this does not contradict the observed sharpness of the
period gap and period minimum. After all, presumably all CVs actually
evolve according to the {\em same} AML law. What the sharp edges in
the period distribution require is insensitivity of donor radii to
{\em initial conditions for a given AML law}. This is a 
weaker requirement than insensitivity to {\em different forms of AML}.}

These considerations have practical implications. For example, perhaps
the most obvious way to obtain donor-based estimates of $\dot{M}_2$ for
a set of CVs is to compute several constant-$\dot{M}_2$ sequences and,
for each object with given $M_2$, find the $\dot{M}_2$ that produces
the best-matching radius. However, the sensitivity to mass-loss
history means that the resulting $\dot{M}_2(P_{orb})$ recipe may not
be self-consistent: an evolution sequence constructed with the
inferred recipe may not reproduce the same donor radii (and may
no longer be a good fit to the data). In practice, we therefore use a
simplified method like this only to settle on a basic form for our AML 
recipe (Section~\ref{sec:form}), but then actually fit the data in detail
with self-consistent evolution sequences
(Section~\ref{sec:final}).\footnote{Throughout this paper, the term
``self-consistent'' is used only in this restricted sense. It
should not be confused with ``physically self-consistent'', which is a
much stronger condition. For example, the ``scaled-GR'' models we use
to model enhanced AML below the period gap in Sections~\ref{sec:assembly}
and~\ref{sec:newtrack} are internally self-consistent, but not
physically self-consistent.}

\subsubsection{$M_2$-$L_2$ and $M_2$-$T_{eff,2}$ Relations}
\label{sec:m_vs_t}

%The final aspects of the donor response to ML we need to briefly
%discuss concern the relationships between stellar mass, luminosity and
%effective temperature along the evolutionary track. 
Since the
secondary star is not in thermal equilibrium, it  
does not obey the mass-luminosity relation of isolated MS stars. 
Instead, its effective temperature is, to a good approximation, set
entirely by $M_2$ and equal to the equilibrium temperature of a MS
star of identical mass, $T_{eff,2} \simeq T_{eff,eq}$
\citep{2000MNRAS.318..354B,2001MNRAS.321..544K}.
Since the donor is oversized for its mass, but
maintains its equilibrium temperature, its luminosity, $L_2 =
4 \pi R_2^2 \sigma T_{eff,2}^4$, exceeds the nuclear luminosity
generated in the core. 
%The donor thus runs a luminosity deficit.

The approximation $T_{eff,2} \simeq T_{eff,eq}$ only breaks down as $P_{orb}
\rightarrow P_{min}$ and the donor mass approaches the 
hydrogen-burning limit, $M_2 \rightarrow M_{H}$. In the brown dwarf
regime, there is no thermal equilibrium configuration, so even the
structure and temperature of isolated objects depends strongly on
age \citep{2000ARA&A..38..337C}. In the CV setting, this means that
$T_{eff,2}$ will depend on the mass-loss and thermal history of the
donor in this regime.

\subsection{The Connection between CV Donors and Evolution}

\subsubsection{Magnetic Braking}
\label{sec:mb}

Equation~\ref{eq:mdot} shows that the evolution of CVs as
semi-detached systems is driven entirely by systemic AMLs: without
$\dot{J}_{sys}$, there would be no mass transfer. Now all close
binaries experience AML due to GR, at a rate given by 
\citep{1967AcA....17..287P, 1976ApJ...209..829W}
\begin{equation}
\dot{J}_{GR} =
\frac{-32}{5}\frac{G^{7/2}}{c^5}\frac{M_1^2M_2^2M^{1/2}}{a^{7/2}}.
\label{eq:jdot_gr}
\end{equation}

This acts all the time in every binary system and thus sets a minimum 
baseline AML rate for every CV. However, it is generally agreed that,
at least above the period gap, CVs experience AML rates far in excess
of GR. Thus at least one additional AML mechanism is required, which
is most usually taken to be magnetic braking.  

In general, ``magnetic braking'' describes any AML associated with a
magnetized stellar wind. In the case of CVs, this wind is thought to
be driven from the donor star. This leads to the first important
connection between donor properties and secular CV 
evolution: the dominant systemic AML mechanism invoked in virtually
all evolution scenarios for long-period CVs is entirely associated
with the secondary star in these systems. Moreover,  it is the assumed
cessation of MB when the donor becomes fully convective that produces
the period gap in the standard model of CV evolution.

The basic physics of MB in CVs are easy to understand. Essentially all
low-mass stars drive a relatively weak stellar wind. As long as this
wind is highly ionized, it is effectively forced to co-rotate with the
stellar magnetic field out to the Alfv\'{e}n radius and thus exerts a
significant spin-down torque on the star. In a CV setting, tidal
forces maintain essentially perfect synchronization between the spin
of the Roche-lobe-filling donor star and the binary orbit, so the net
effect of the wind from the donor star is the extraction of angular
momentum from the binary system at a rate of $\dot{J}_{sys} =
\dot{J}_{MB} < 0$.

\begin{figure*}
\includegraphics[height=19cm,angle=-90]{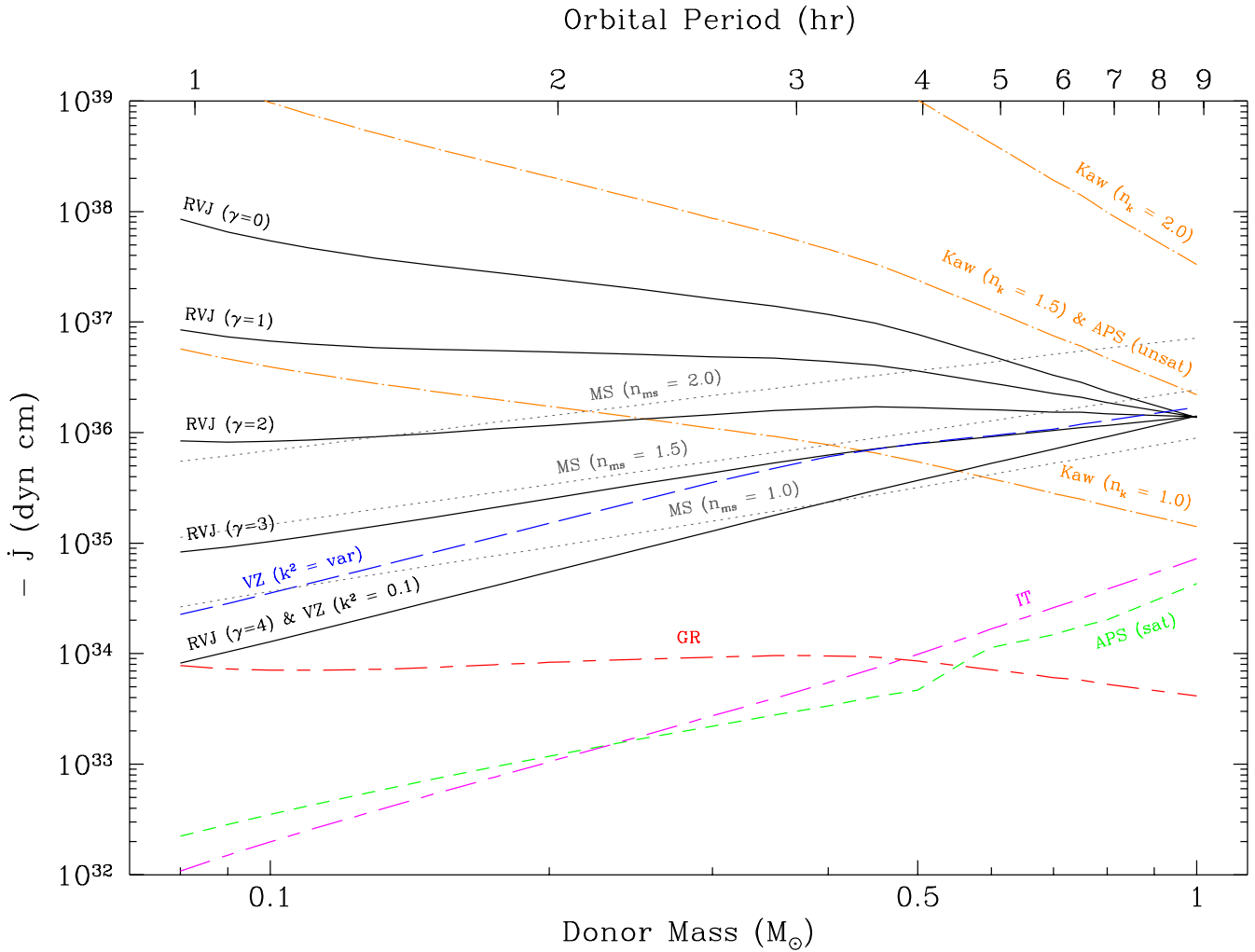}
\caption[A comparison of all angular-momentum-loss
recipes discussed in Appendix~\ref{app:mb}.]
{A comparison of all angular-momentum-loss
recipes discussed in Appendix~\ref{app:mb}. All AML rates are calculated 
assuming a 0.75 $M_{\odot}$ primary and an unevolved
Roche-lobe-filling MS secondary in marginal contact. Thus the
secondary is assumed to follow the standard 5-Gyr BCAH98 mass-radius
relationship, and the orbital period is calculated from the
approximate period-density relation given by Equation~\ref{eq:dense}. None
of the corrections discussed in Section~\ref{sec:skeletons} are included. 
The red long-dash-short-dash line corresponds to GR-driven AML
(Equation~\ref{eq:jdot_gr}). The blue long-dashed line shows the
classic \citet[][Appendix~\ref{app:vz},
  Equation~\ref{eq:jdot_vz}]{1981A&A...100L...7V} prescription
with moment of inertia, $k$,
estimated from stellar models. The solid black lines mark the 
\citet[][Appendix~\ref{app:rvj}, Equation~\ref{eq:jdot_rvj}]{1983ApJ...275..713R}
prescription for different power-law indeces,
$\gamma$, as marked. Note that the $\gamma = 4$ recipe is
identical to a \citet{1981A&A...100L...7V} law with constant $k^2 =
0.1$ by construction. The orange 
dash-dotted line marks the \citet[][Appendix~\ref{app:k88},
Equation~\ref{eq:jdot_k88}]{1988ApJ...333..236K} recipe for
different power-law indices, $n_k$, as marked. Grey dotted lines mark
the \citet[][Appendix~\ref{app:ms87},
Equation~\ref{eq:jdot_ms87}]{1987MNRAS.226...57M} law for different
power-law indices, $n_{ms}$, as marked. The green short-dashed line shows the saturated
model suggested by \citet[][Appendix~\ref{app:aps},
  Equation~\ref{eq:jdot_aps}]{2003ApJ...582..358A}. The unsaturated limit of this model 
is identical to the \citet{1988ApJ...333..236K} recipe with $n_k =
1.5$ by construction. Finally, the saturated AML law suggested by 
\citet[][Appendix~\ref{app:it},
Equation~\ref{eq:jdot_it}]{2003ApJ...599..516I} is shown as a
long-dashed magenta line. The main thing to take away from this figure is the
enormous range in shapes and strengths, even for conceptually similar
AML prescriptions.} 
\label{fig:mb}
\end{figure*}

Unfortunately, the conceptual simplicity of MB as an AML mechanism
does not mean that it is easy to model. In fact, there have been so
many attempts to describe MB in the literature that it can be hard to
keep track of the differences between them. We therefore provide brief
descriptions of some of the most widely-used MB prescriptions in
Appendix~\ref{app:mb}.

In Figure~\ref{fig:mb}, we present a direct comparison of the AML
rates predicted by GR and all of the MB recipes described in
Appendix~\ref{app:mb}. For the purpose
of this simple comparison, we took the stellar parameters from the
standard 5~Gyr MS models of \citet[][hereafter BCAH98]{1998A&A...337..403B}
and then used the period-density relation (Equation~\ref{eq:dense}) to
calculate the corresponding orbital period. Thus we assume that the
donor can be approximated as an ordinary MS star, despite the fact
that it fills the Roche lobe and 
undergoes mass loss. There is then a simple one-to-one relationship
between donor mass and orbital period (and no period gap). This would 
not be an acceptable approximation for self-consistent evolutionary
calculations, which should take into account the ML-driven inflation
of the donor. However, it is well suited to our present purpose, since
it provides for a particularly clean and simple comparison of the
different AML laws. 

The main point to take away from Figure~\ref{fig:mb} is that there are
{\em huge} differences between different MB prescription at fixed
$M_2$/$P_{orb}$. For example, the widely used
\citet[][Appendix~\ref{app:k88},
Equation~\ref{eq:jdot_k88}]{1988ApJ...333..236K} prescription  
with $n_k = 1.5$ produces approximately
100,000 times stronger braking at 
$M_2 \simeq 0.3~M_{\odot}$ than the ``saturated''
prescriptions suggested by \citet[][Appendix~\ref{app:aps},
  Equation~\ref{eq:jdot_aps}]{2003ApJ...582..358A} and  
\citet[][Appendix~\ref{app:it},
Equation~\ref{eq:jdot_it}]{2003ApJ...599..516I}.\footnote{We are using
the term ``saturated'' loosely here, to simply refer to MB
prescriptions in which AML is suppressed at high stellar rotation
rates. In the \citet[][]{2003ApJ...599..516I} recipe, this suppression is
accomplished without any actual saturation of the stellar magnetic field.}
In fact,  the saturated prescriptions are not even
competitive with GR for $M_2 \ltappeq 0.4~M_{\odot}$. 

These are not new results, of course: after all, the saturated
prescriptions were {\em designed} to be weak at high rotation rates. However,
Figure~\ref{fig:mb} also shows that, even among the unsaturated
prescriptions, AML rates can differ by at least 3 orders
of magnitude. For example, compare the classic
\citet[][Appendix~\ref{app:vz}, Equation~\ref{eq:jdot_vz}]{1981A&A...100L...7V}
prescription with $k^2 = 0.1$ to the \citet{1988ApJ...333..236K} one
with $n_k = 1.5$ at $M_2 \simeq 0.3~M_{\odot}$). 
Moreover, the {\em shapes} of the AML laws
(i.e. their dependence of $M_2$/$P_{orb}$) vary widely between
different MB formulations. Some predict a weakening of MB with
decreasing $M_2$/$P_{orb}$ (e.g. \citealt{1981A&A...100L...7V} and
\citealt{1987MNRAS.226...57M}, Appendix~\ref{app:ms87},
Equation~\ref{eq:jdot_ms87}), while others actually
predict a strengthening \citep[e.g.][]{1988ApJ...333..236K}. Clearly,
we cannot rely on the 
existing MB prescriptions to guide us in defining the functional form
and strength of AML in CVs. We will therefore adopt the 
\citet[][Appendix~\ref{app:rvj}, Equation~\ref{eq:jdot_rvj}]{1983ApJ...275..713R}
parameterization when constructing self-consistent evolution tracks
in Section~\ref{sec:final}. This is simply because the strength and
shape can be controlled very easily by varying the normalization and
the power-law index $\gamma$, respectively.
 
In studies of CV evolution, MB is usually assumed to stop abruptly
when the secondary becomes fully convective. In isolated 
stars, this happens at around $M_2 \simeq 0.35~M_{\odot}$
\citep{1997A&A...327.1039C}, 
but in mass-losing stars, the transition will generally occur at
lower masses ($M_2 \simeq 0.2 - 0.3 M_{\odot}$; see
Figure~\ref{fig:extremep} in Section~\ref{sec:pcrit}). The
conventional argument for the assumed cessation of 
MB is that the magnetic field in 
solar-type, low-mass stars is thought to be anchored at the interface
between the radiative core and the convective envelope 
\citep[the tachocline; e.g.][]{1997ApJ...486..484M,
  1997ApJ...486..502C}. Since fully convective stars do not possess
such an interface, the magnetic field -- and with it MB -- is assumed to
vanish at this point.
\footnote{As already noted in Section~\ref{sec:intro},
observations of non-interacting stars do not support this
assumption. Indeed, it is quite clear today that fully convective
objects are capable of sustaining significant magnetic fields. These 
are probably generated via a different {\em type} of dynamo 
than that responsible for the fields of partly radiative stars. See
Section~\ref{sec:discuss_plausibility} for further discussion of this
key point.}
In the standard model, the only remaining AML 
mechanism is then GR, which operates at a much slower rate $|\dot{J}_{GR}| <<
|\dot{J}_{MB}|$ and thus produces a lower mass-transfer rate
$|\dot{M}_{2,GR}| << |\dot{M}_{2,MB}|$.

\subsubsection{The Period Gap}
\label{sec:gap}

The precise origin of the period gap in the standard model is now easy
to understand. The shutdown of strong MB near $P_{orb} \simeq 3~{\rm hrs}$
leads to a sudden reduction in $\dot{J}_{sys}$. This in turn causes a
drop in $\dot{M}_2$ (Equation~\ref{eq:mdot}) and an 
increase in the ratio $\tau_{\dot{M}_2} / \tau_{{kh}}$. This slower
ML cannot sustain the existing level of donor of inflation, so the
secondary will shrink and lose contact with the Roche lobe altogether. This 
loss of contact is a fast process, because the Roche-lobe and donor 
radii are normally equal to within roughly a pressure scale height, $H$,
where $(H/R_2) \sim 10^{-4}$ \citep[e.g.][]{1988A&A...202...93R}. 
Thus the secondary only has
to contract by a few scale heights in order to break contact.
% which
%will happen on a time scale of no more than $(H/R_2) \tau_{{kh}} \sim
%10^5$~yrs. --> should be worked from the equation below...

At this point, ML ceases entirely, and the CV enters the
period gap as a newly detached system. Meanwhile, the donor 
continues to contract towards its thermal equilibrium state, but the 
system nevertheless continues to evolve towards shorter orbital
periods, since it still loses angular momentum due to GR. Since $M_1$,
$M_2$ and $M$ are all constant in a detached system,
Equation~\ref{eq:j} shows that this 
AML will cause the binary orbit to decay. At $P_{orb} \simeq 2~{\rm hrs}$, the
shrinking Roche lobe therefore catches up with the donor star again.
At this point, mass transfer restarts, and the system re-emerges as an
active CV.

Provided that the time it takes a CV to evolve through the period 
gap is sufficiently long, the secondary will be able to
relax completely and emerge from the bottom of the gap with the 
equilibrium radius appropriate for its mass. Is this condition met?
\citet{1995ApJ...439..330K} show that, following 
the cessation of contact, the donor's radius will relax back to its
equilibrium value on a time scale 
\begin{equation}
\tau_{relax} = \frac{R_2 - R_{eq}}{(\zeta - \zeta_{ad})R_2} \tau_{\dot{M}_2},
\end{equation}
where $R_2$, $\zeta$ and $\dot{M}_2$ are evaluated just prior to loss of contact
at the upper edge of the gap, and $\zeta_{ad} = -1/3$ for the low-mass
stars of interest. We find that $\tau_{relax}$ is indeed shorter than
the gap-crossing time for all of the models we consider. We therefore
expect the secondary to emerge from the bottom of the gap
indistinguishable from a MS star of the same mass.

The width of the period gap in this picture is directly determined by
the degree of donor inflation above the gap. Combining the
\citet{1971ARA&A...9..183P} 
approximation for the volume-averaged Roche-lobe radius  
\begin{equation}
\frac{R_2}{a} = 0.462\left(\frac{q}{1+q}\right)^{1/3}
\label{eq:pac}
\end{equation}
(which is valid for mass ratios $q = \frac{M_2}{M_1} \ltappeq 0.8$)
with Kepler's third law yields the well-known period-density
relationship for Roche-lobe-filling stars
\begin{equation}
<\rho_2> = \frac{3 M_2}{4 \pi R_2^3} \simeq 107 \; P^{-2}_{orb,h} ~ \rm{g\;cm^3}.
\label{eq:dense}
\end{equation}
where $P_{orb,h}$ is the orbital period in units of hours. Now,
according to the disrupted AML scenario, $M_2$ is the same at the 
upper and lower edges of the period gap ($P_{gap,\pm}$), so
Equation~\ref{eq:dense} implies that the ratio of the donor radii at the gap
edges must satisfy 
\begin{equation}
\frac{R_{2,+}}{R_{2,-}} =
\left(\frac{P_{gap,+}}{P_{gap,-}}\right)^{2/3}.
\end{equation}
If the donor emerges from the gap in thermal equilibrium, $R_{2,-} =
R_{2,eq}$, the gap width is set entirely by the degree of donor
inflation above the gap. K06 estimated $P_{gap,-} = 2.15 \pm 0.03$~hrs
and $P_{gap,+} = 3.18 \pm 0.04$~hrs, which implies that the radius of
a CV donor entering the gap is about 30\% larger than that of an
equal-mass MS star.

\subsubsection{The Period Minimum and the Period Spike}
\label{sec:bounce}

The existence and location of the period minimum for CVs is also
directly related to the properties of their donor stars. The easiest
way to see this is to logarithmically differentiate the period-density
relation (Equation~\ref{eq:dense}) and use the mass-radius index to
substitute for $(\dot{M}_2/M_2) = \zeta^{-1} (\dot{R_2}/R_2)$. This
yields the evolution equation for the orbital period of the system
\begin{equation}
\frac{\dot{P}_{orb}}{P_{orb}} = \frac{3 \zeta - 1}{2} \frac{\dot{M}_2}{M_2}.
\label{eq:pdot}
\end{equation}

Equation~\ref{eq:pdot} shows that the orbital period decreases for $\zeta >
1/3$, but must increase for $\zeta < 1/3$. As noted above, a CV donor
emerges from the bottom of the period gap in (or close to) thermal
equilibrium, with $\zeta \simeq \zeta_{eq} \simeq 0.8$, but once mass
transfer resumes below the gap, the secondary will again struggle to
maintain this. 

In fact, as the secondary is being whittled down to lower and lower
masses, $\tau_{{kh}}$ increases faster than 
$\tau_{\dot{M}_2}$, even if the mass transfer is driven solely by
GR. Thus $\tau_{\dot{M}_2} / \tau_{{kh}}$ continuously decreases with
time, and the cumulative loss of thermal equilibrium accelerates. The
mass-radius index $\zeta$ must therefore drop away from $\zeta_{eq}$ towards
$\zeta_{ad} \simeq -1/3$. Note that even if mass loss was very slow, the donor 
must eventually pass the Hydrogen-burning limit ($M_{H} \simeq
0.07~M_{\odot}$\ for isolated, solar metallicity stars). Beyond this,
thermal equilibrium is, by definition, impossible, since there is no
internal energy source. In this limit, $\zeta$ will approach the
index of the mass-radius relationship for brown dwarfs, which (at
fixed age) is also $\zeta \simeq -1/3$ \citep{2009AIPC.1094..102C}.
\footnote{Actually, as we shall see later, the mass-radius index
of CV donors below the Hydrogen-burning limit tends to approach
$\zeta \simeq 0$. There are two reasons for this. First, the brown
dwarf mass-radius index is not constant, but goes to $\simeq 0$ with
decreasing mass. Second, in a CV setting,
different donor masses correspond to different ages. More specifically 
higher-mass sub-stellar donors are younger (and hence relatively
larger) than lower-mass donors.} 

These considerations show that $\zeta$ must inevitably pass
through the critical value $\zeta = 1/3$, at which point
$\dot{P}_{orb} = 0$ and hence $P_{orb} = P_{min}$. Since $\zeta$ depends 
on $\tau_{\dot{M}_2} / \tau_{{kh}}$, the exact location of the minimum
period depends on $\dot{M}_2$ and hence $\dot{J}_{sys}$. For fast
ML/AML, $P_{min}$ is generally reached sooner (and is
therefore larger) than for slow ML/AML. Note, however, that
faster ML and AML does not necessarily imply larger donor masses at
$P_{min}$; after all, fast ML also decreases $M_2$ faster than slow
ML. For GR-like AML prescriptions (i.e. with the same functional form,
but different normalizations; see Section~\ref{sec:final}), faster ML/AML 
actually corresponds to {\em lower} $M_2$ at
$P_{min}$. Thus $M_2$ will drop below $M_H$ {\em
before} $P_{min}$ is reached. At least in the context of slow, GR-like
AML prescriptions, $M_2 < M_{H}$ is therefore a necessary, but not
sufficient condition for identifying period bouncers.

The period minimum should be a prominent feature in the orbital period
distribution of the Galactic CV population. In particular, it should
not just produce a sharp cut-off in the distribution, but a strong and
fairly narrow spike. This is easy to understand: the amount of time a
given CV spends in a particular orbital period range ($P_{orb}
\rightarrow P_{orb}+\Delta P_{orb}$) is inversely proportional to the
average value of $\dot{P}_{orb}$ in that range. Now $\dot{P}_{orb}=0$
at $P_{min}$, so systems spend rather a long time there. In a
population close to steady-state, this means that many systems will be
found near $P_{min}$ at any given time. This produces the period spike
that is seen in essentially all CV population synthesis studies 
\citep[e.g.][]{1993A&A...271..149K,
1999MNRAS.309.1034K,2001ApJ...550..897H, 2003MNRAS.340..623B}
and which has now been clearly detected observationally (G09). However, as
already noted in Section~\ref{sec:intro}, the minimum period predicted 
by the standard model of CV evolution ($P_{min} \simeq 65$~min
\footnote{This ``classic'' estimate ignores corrections such as those
described in Section~\ref{sec:skeletons}.})
is substantially shorter than the observed one ($P_{min} \simeq 
82$~min). One possible solution to this problem is to allow for AML
rates in excess of pure GR below the period gap
\citep[e.g.][]{1998PASP..110.1132P}, another is to assume that CV
donors are heavily spotted  \citep[][hereafter L08]{2008MNRAS.388.1582L}.

\section{The Semi-Empirical Donor Sequence Revisited}
\label{sec:update}

In K06 we constructed a complete, semi-empirical donor sequence for
CVs, which provided all physical and photometric properties of a
typical CV secondary as a function of $P_{orb}$. The two key
ingredients for this donor sequence are (i) a broken-power-law fit to 
the empirical $M_2-R_2$ relation and (ii) the theoretically expected
MS-like $M_2-T_{eff}$ relation. These two constraints are sufficient
to fully specify the physical parameters of CV donors as a function of
$P_{orb}$; these can then be used as input to stellar atmosphere
models to obtain the predicted photometric properties of the
secondaries as well. We will refer to donor sequences constructed in
this way generically as ``broken-power-law sequences'' from here on. 

\begin{figure}
\includegraphics[height=9cm,angle=0]{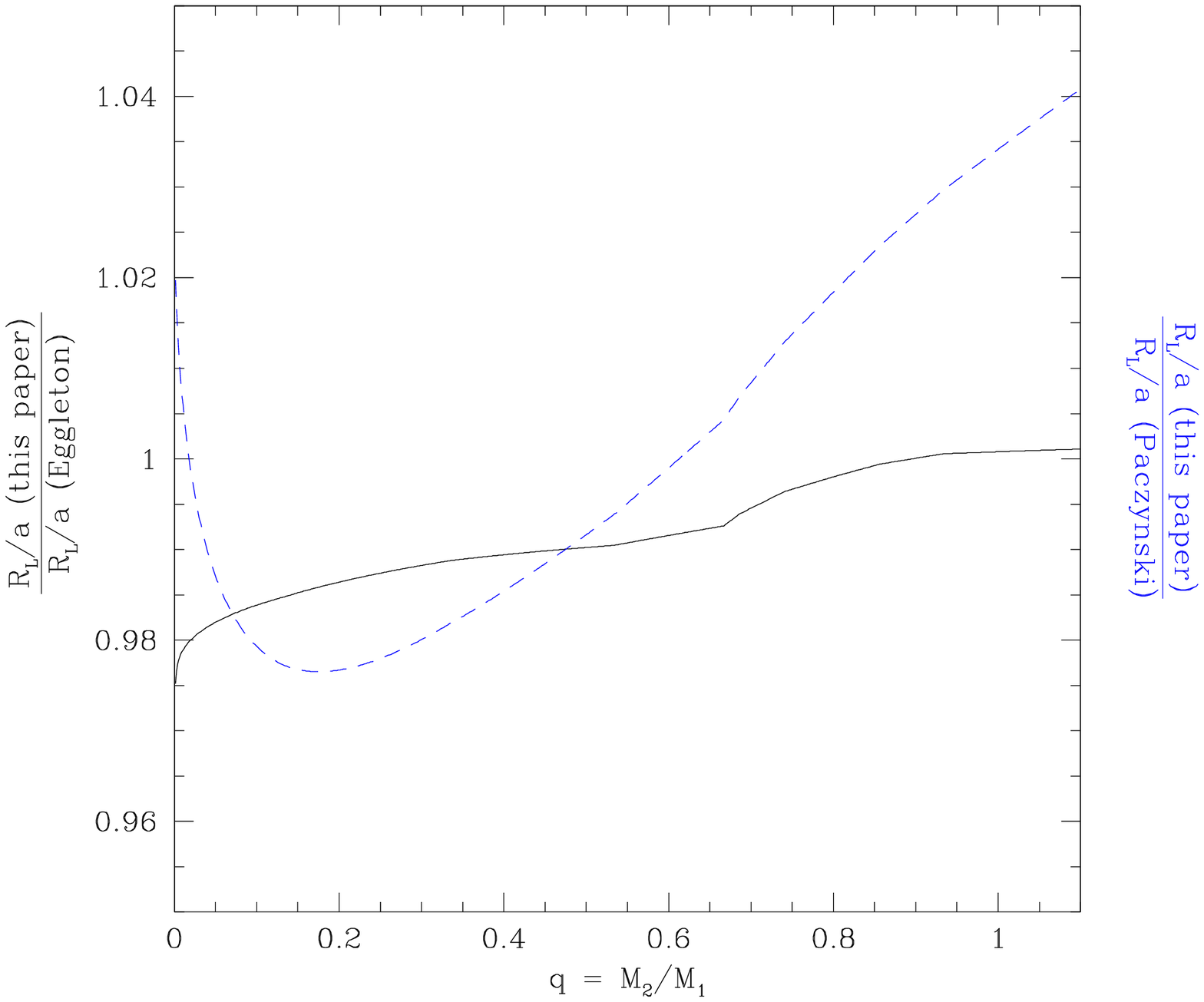}
\caption[The ratios of the approximation to the Roche-lobe radius used
in this paper to those suggested by \citet{1971ARA&A...9..183P} and \citet{1983ApJ...268..368E}.]
{The ratios of the $R_L/a$ approximation used in this paper to
those suggested by \citet[][blue dashed line]{1971ARA&A...9..183P} and
\citet[][black solid line]{1983ApJ...268..368E}.
The latter two approximations treat the
secondary star as a point source in estimating the critical potential,
whereas our approximation, which is based on the work on
\citet{2009ApJ...698..715S}, 
treats the donor as a polytrope. The effective polytropic
index is itself a function of donor mass (and hence mass ratio) in
this prescription, as listed in Table~\ref{tab:siro}.}
\label{fig:siro1}
\end{figure}

The K06 broken-power-law donor sequence has found fairly widespread 
use, both for distance estimation and as a benchmark for what
constitutes a ``normal'' CV secondary star at a given $P_{orb}$. It
also provides an obvious and convenient point of comparison 
for the self-consistent CV evolution tracks presented 
in Section~\ref{sec:newtrack}. However, some updates are in order to
bring the donor sequence in line with recent observational
developments and improve its accuracy. Since the construction of the 
broken-power-law $M_2-R_2$ relation and associated donor sequence
was explained in some detail in K06, we will limit our discussion here
only to {\em differences} between the data and methods used in this
update compared to those used in constructing the original sequence.

Let us begin with the $M_2-R_2$ relation. First, we have added
the system parameters of six of the new, short-period, eclipsing SDSS
CVs analysed by L08. We exclude only SDSS J1507, since 
this is probably a halo CV \citep{2008PASP..120..510P} or
possibly a recently born system \citep{2007MNRAS.381..827L}.
However, we do use L08's revised system parameters for OY Car, which 
are slightly different from those given in \citet{1990MNRAS.242..606W}. 

Second, we now adopt the location of the period spike, $P_{min} = 82.4
\pm 0.7$~min (G09), as the best estimate of the minimum period. This
effectively sets the location of one of the two breaks in the
broken-power-law  parameterization (the other is set by the donor mass
just above and below the period gap).

\begin{deluxetable*}{lllllll}
%\begin{table}
%\caption[Coefficients for the Roche-lobe approximation
%and the tidal/rotation deformation factor as functions of the
%effective polytropic index.]{}
%\end{table}
\tablecolumns{7} 
%\tablewidth{20}
\tablecaption
%[Coefficients for the Roche-lobe approximation
%and the tidal/rotation deformation factor as functions of the
%effective polytropic index.]
{Coefficients for the Roche-lobe approximation
(Equation~\ref{eq:siro1}) and the tidal/rotation deformation factor
(Equations~\ref{eq:siro2} and Equations~\ref{eq:siro2a}) as functions
of the effective polytropic index, $n$. Data from Sirotkin \& Kim
(2009) and Sirotkin (private communication).} 
\tablehead{
\colhead{Coefficient} &
\colhead{Eggleton (1983)} &
\multicolumn{5}{c}{Effective Polytropic Index}\\
\colhead{} &
\colhead{approximation} &
\colhead{$n=1.50$} &
\colhead{$n=1.75$} &
\colhead{$n=2.00$} &
\colhead{$n=2.50$}\\ \hline \\
\colhead{} &
\colhead{} &
\multicolumn{5}{c}{Corresponding Stellar Mass}\\
\colhead{} &
\colhead{} &
\colhead{$\leq 0.1 M_{\odot}$} &
\colhead{$0.51 M_{\odot}$} &
\colhead{$0.56 M_\odot$} &
\colhead{$0.64 M_\odot$} \\
\multicolumn{7}{c}{Roche lobe coefficients (Equation~\ref{eq:siro1})}
}{Coefficients for the Roche-lobe approximation
(Equation~\ref{eq:siro1}) and the tidal/rotation deformation factor
(Equations~\ref{eq:siro2} and Equations~\ref{eq:siro2a}) as functions
of the effective polytropic index.}
\startdata
$c_1$     & 0.49         & 0.5126 & 0.5296 & 0.5303 & 0.5310 \\
$c_2$     & $\frac{2}{3}$  & 0.7388 & 0.7661 & 0.7640 & 0.7616 \\
$c_3$     & 0.6          & 0.6710 & 0.7112 & 0.7100 & 0.7083 \\
$c_4$     & $\frac{2}{3}$  & 0.7349 & 0.7577 & 0.7554 & 0.7526 \\
$c_5$     & $\frac{1}{3}$  & 0.3983 & 0.4232 & 0.4213 & 0.4191 \\
\cutinhead{Tidal and rotational deformation coefficients (Equation~\ref{eq:siro2a})}
$N_{iso}$  & -            & 0.42422 & 0.38589 & 0.36475 & 0.35150  \\ 
$d_1$     & -            & 0.0191 & 0.0144 & 0.01084   & 0.00588  \\
$d_2$     & -            & 0.9561 & 0.9850 & 0.9980    & 1.025    \\
$d_3$     & -            & 0.3557 & 0.3880 & 0.3945    & 0.4086   \\
$d_4$     & -            & 0.9130 & 0.9333 & 0.9470    & 0.9767   \\
$d_5$     & -            & 1.1635 & 1.205  & 1.202     & 1.198   
\enddata
\label{tab:siro}
\end{deluxetable*}

Third, we have replaced Paczy{\'n}ski's
\citeyearpar{1971ARA&A...9..183P} approximation for the 
volume-averaged Roche-lobe size (Equation~\ref{eq:pac}) with a more 
precise approximation based on the results of
\citet[][hereafter SK09]{2009ApJ...698..715S}. 
This new approximation actually goes beyond the pure Roche
model, in that it does not represent the donor as a point source in
the potential, but as a polytrope with polytropic index $n$. The
functional form adopted for  $R_L/a$ is similar to Eggleton's
\citeyearpar{1983ApJ...268..368E} improved approximation,  
\begin{equation}
\frac{R_{L}}{a} = \frac{c_{1} q^{c_{2}}}{c_{3}q^{c_{4}} + \ln{(1 + q^{c_{5}})}},
\label{eq:siro1}
\end{equation}
but with fit parameters, $c_{i}(n)$, that depend on the
polytropic index of the Roche-lobe-filling star. When using
Equation~\ref{eq:siro1} in a CV setting, it is therefore 
necessary to specify the polytropic index that best
describes the donor under consideration. Fully convective stars
(i.e. donors below  
the period gap) are well-described by $n=3/2$ polytropes. For more
massive donors, we interpolate on the relationship between stellar
mass and effective polytropic index given in Table~3 of
\citet{1994ApJ...423..344L}. This was derived by matching the properties of
polytropes to those of MS stellar models. In Table~\ref{tab:siro}, we
provide the stellar masses, $M_2$, and coefficients, $c_{i}$, for a
small grid of polytropic indices that span the parameter space
relevant for CVs. The coefficients for $n=3/2$ were taken from SK09;
coefficients for other $n$ were kindly provided to us by Fedir
Sirotkin (private communication).

Implementing this more complex form of $R_L/a$ is just about worth 
it. Figure~\ref{fig:siro1} shows how $R_L/a$ as calculated via
Equation~\ref{eq:siro1} (with polytropic index $n$ varying
with mass ratio as given in Table~\ref{tab:siro}) compares to both the
\citet{1971ARA&A...9..183P} and the \citet{1983ApJ...268..368E} 
approximations. The difference between
the $n$-dependent formulation and the standard ones (in which the
donor is represented as a point source in the potential) can amount to a
few percent and varies systematically with mass ratio. Note that the
point-source Roche model is least accurate for fully convective
low-mass star, i.e. donors below the gap. Physically, this is because
such stars are not very centrally concentrated.

Fourth, we have slightly modified the $\chi^2$-fitting method we use
to infer the optimal broken-power-law 
parameters. More specifically, we found that explicitly including
systematic uncertainties in the fit (see K06 for details) could
sometimes produce unphysical results in the period bouncer
regime. This is probably due to bias affecting the method 
in particular limiting cases. We therefore now use a standard
$\chi^2$ fit (without a systematic error term), but still allowing for
intrinsic dispersion at the level needed to achieve $\chi^2_{\nu} =
1$. 

\begin{figure*}
\includegraphics[height=19cm,angle=-90]{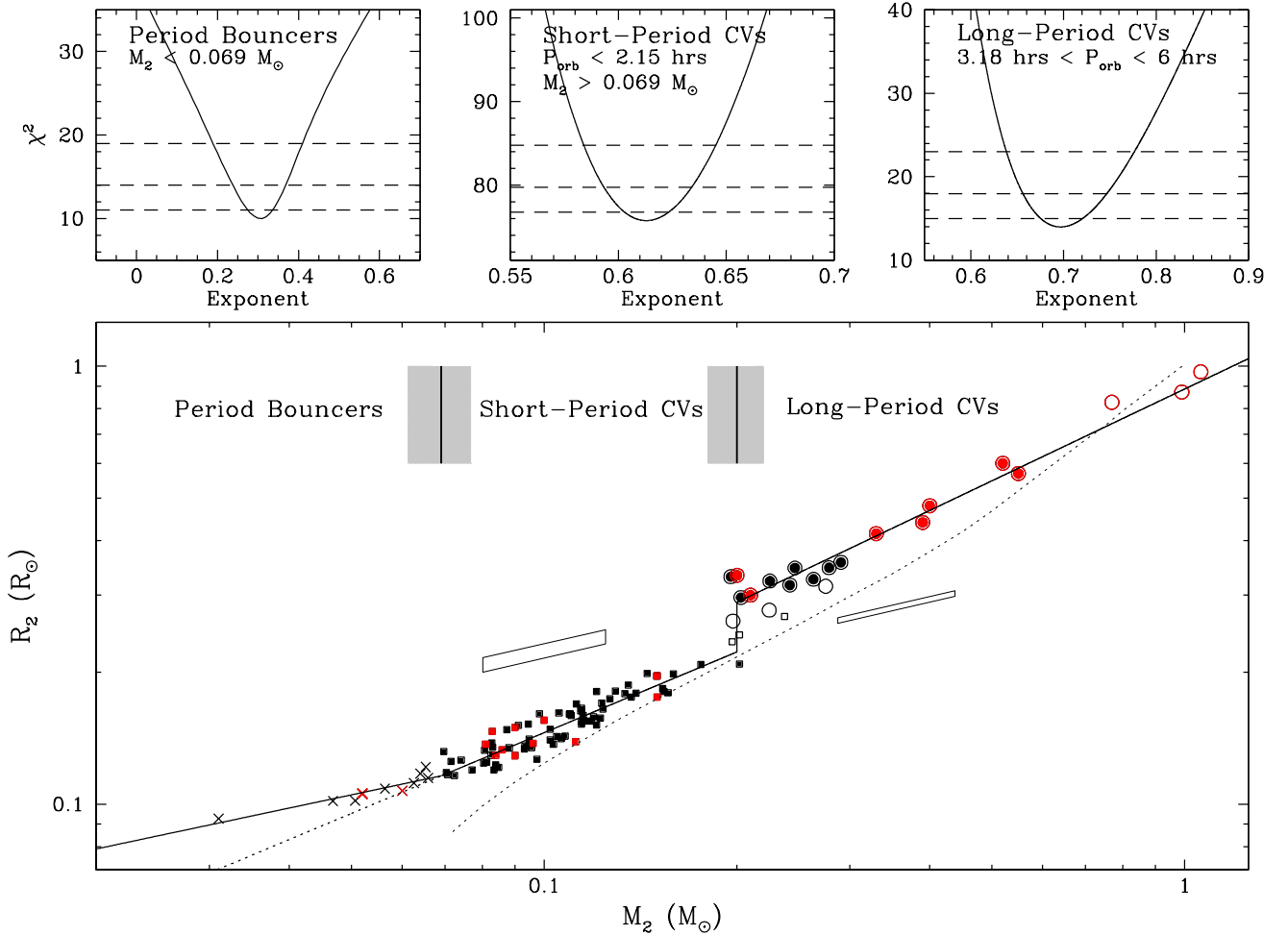}
\caption[The mass-radius relation of CV donor stars.]
{The mass-radius relation of CV donor stars. {\em Bottom
panel:} Points shown are empirical mass and radius estimates for CV
donors. Superhumpers are shown in black, eclipsers in red. Filled
squares correspond to short-period CVs, filled circles to long-period
systems, and crosses to candidate period bouncers. The parallelograms
illustrate the typical error on a single short-period or long-period
CV. Open symbols were ignored in fits to the data since they
correspond to systems in the period gap or long-period 
\citep[probably evolved,][]{2003MNRAS.340.1214P} 
systems. The solid lines show the optimal broken-power-law
fit to the data in the period bouncer, short-period and long-period
regions. The dotted line is the mass-radius relation for main sequence stars taken from
the 5~Gyr BCAH98 isochrone.
{\em Top panels:} Constraints on the power law exponents of the
$M_2-R_2$ relations in the three period/mass regimes. For each regime,
we plot $\chi^2$ vs exponent and indicate the $\chi^2$
corresponding to 1$\sigma$, 2$\sigma$ and 3$\sigma$ around the
minimum with horizontal dashed lines.}
\label{fig:brokenpower}
\end{figure*}

Figure~\ref{fig:brokenpower} shows the new broken-power-law fit to the updated 
$M_2 - R_2$ data set. The optimal mass-radius relation is
%I'm ignoring the slight asymmetry in the errors on the bouncer power law
%index for simplicity.
\begin{equation}
\frac{R_2}{R_{\odot}} = 
\left\{ \begin{array}{ll}
\scriptstyle 0.118 \pm 0.004\left(\frac{M_2}{M_{bounce}}\right)^{0.30\pm 0.03} & 
\scriptstyle {{\rm ~for ~ } M_2 < M_{bounce}} \\   & \\ 
\scriptstyle 0.225 \pm 0.008\left(\frac{M_2}{M_{conv}}\right)^{0.61\pm0.01} & 
\scriptstyle {\begin{array}{l}
\scriptstyle {\rm for ~ } M_{bounce} < M_2 < M_{conv} \\
\scriptstyle P_{min} < P_{orb} < P_{gap,-} 
\end{array}} \\ ~ \\ 
\scriptstyle 0.293 \pm 0.010\left(\frac{M_2}{M_{conv}}\right)^{0.69\pm0.03} & 
\scriptstyle {\begin{array}{l}
\scriptstyle {\rm  for ~ } M_{conv} < M_2 < M_{evol}  \\ 
\scriptstyle P_{gap,+} < P_{orb} < P_{evol}, 
\end{array}} 
\end{array} \right.
\label{eq:m2r2}
\end{equation}
where the following quantities have been assumed
\begin{equation}
\begin{array}{lll}
M_{bounce} &=& 0.069 \pm 0.009 ~ M_{\odot}\\
M_{conv} &=& 0.20 \pm 0.02 ~ M_{\odot}\\
M_{evol} &\simeq& 0.6-0.7 ~ M_{\odot} \\
P_{min} &=& 82.4 \pm 0.7 ~ {\rm min} \\
P_{gap,-} &=& 2.15 \pm 0.03 ~ {\rm hr} \\
P_{gap,+} &=& 3.18 \pm 0.04 ~ {\rm hr} \\
P_{evol} &\simeq& 5-6 ~ {\rm hr}.
\end{array}
\label{eq:defs}
\end{equation}
The notation here is the same as in K06, i.e. $P_{min}$, $P_{gap,-}$,
$P_{gap,+}$ and $P_{evol}$, respectively, are the orbital periods at
period minimum, at the bottom and top of the period gap, and at the
point where systems with evolved secondaries are expected to
start dominating the Galactic CV population. Similarly, $M_{bounce}$,
$M_{conv}$ and $M_{evol}$ are the corresponding donor masses at these
points (with $M_2 = M_{conv}$ at both $P_{gap,-}$ and $P_{gap,+}$,
since the donor mass does not change during evolution through the gap).

\begin{deluxetable*}{cccccclcccccccc}
\tabletypesize{\footnotesize}
%\rotate
%\tablewidth{620pt}
%\tablecaption[The updated semi-empirical broken-power-law donor sequence for CVs.]
\tablecaption{The updated semi-empirical broken-power-law donor sequence for CVs.}
\tablehead{
\colhead{$P_{orb} (hr)$} &
\colhead{$M_2 (M_{\odot})$} &
\colhead{$R_2 (M_{\odot})$} &
\colhead{$T_{eff} (K)$} &
\colhead{$\log{g}$} &
\colhead{$\log{L/L_{\odot}}$} &
\colhead{$M_U$} &
\colhead{$M_B$} &
\colhead{$M_V$} &
\colhead{$M_R$} &
\colhead{$M_I$} &
\colhead{$M_J$} &
\colhead{$M_H$} &
\colhead{$M_K$} &
\colhead{$SpT$} }{The updated semi-empirical broken-power-law donor sequence for CVs.}
\startdata
 1.390  & 0.035  & 0.094  &  1115  & 5.036  & 28.67  & 35.88  & 30.01  & 25.97  & 21.51  & 18.55  & 14.60  & 14.71  & 14.63  &      T   \\
 1.387  & 0.040  & 0.098  &  1320  & 5.058  & 29.00  & 33.07  & 27.71  & 24.82  & 20.59  & 17.64  & 13.77  & 13.74  & 13.60  &      T   \\
 1.384  & 0.045  & 0.102  &  1537  & 5.077  & 29.30  & 31.24  & 25.99  & 23.49  & 19.42  & 16.51  & 13.00  & 12.88  & 12.81  &      T   \\
 1.382  & 0.050  & 0.105  &  1738  & 5.094  & 29.54  & 30.28  & 25.47  & 22.51  & 18.97  & 16.38  & 12.82  & 12.25  & 11.90  &      T   \\
 1.379  & 0.055  & 0.108  &  1934  & 5.110  & 29.75  & 28.80  & 24.46  & 21.28  & 18.19  & 15.77  & 12.28  & 11.56  & 11.07  &   L0.0   \\
 1.377  & 0.060  & 0.111  &  2110  & 5.125  & 29.93  & 27.45  & 23.25  & 20.52  & 17.82  & 15.34  & 11.82  & 11.13  & 10.71  &   M9.6   \\
 1.375  & 0.065  & 0.114  &  2265  & 5.138  & 30.07  & 25.76  & 21.64  & 19.51  & 17.42  & 14.87  & 11.31  & 10.64  & 10.32  &   M9.2   \\
 1.373  & 0.069  & 0.116  &  2377  & 5.148  & 30.17  & 24.41  & 20.64  & 18.71  & 16.89  & 14.40  & 11.04  & 10.39  & 10.08  &   M8.3   \\
 1.382  & 0.070  & 0.117  &  2402  & 5.146  & 30.20  & 24.08  & 20.43  & 18.52  & 16.74  & 14.28  & 10.98  & 10.33  & 10.02  &   M8.1   \\
 1.423  & 0.075  & 0.122  &  2520  & 5.140  & 30.32  & 22.62  & 19.44  & 17.63  & 16.01  & 13.71  & 10.72  & 10.08  &  9.76  &   M7.3   \\
 1.464  & 0.080  & 0.127  &  2618  & 5.133  & 30.42  & 21.49  & 18.65  & 16.92  & 15.41  & 13.23  & 10.50  &  9.87  &  9.55  &   M6.9   \\
 1.503  & 0.085  & 0.132  &  2699  & 5.127  & 30.50  & 20.58  & 18.01  & 16.34  & 14.91  & 12.84  & 10.31  &  9.69  &  9.38  &   M6.6   \\
 1.540  & 0.090  & 0.137  &  2768  & 5.122  & 30.58  & 19.89  & 17.50  & 15.87  & 14.50  & 12.52  & 10.15  &  9.54  &  9.23  &   M6.4   \\
 1.576  & 0.095  & 0.141  &  2827  & 5.116  & 30.64  & 19.31  & 17.07  & 15.47  & 14.15  & 12.25  & 10.01  &  9.41  &  9.09  &   M6.2   \\
 1.611  & 0.100  & 0.146  &  2877  & 5.111  & 30.70  & 18.83  & 16.71  & 15.14  & 13.86  & 12.02  &  9.89  &  9.29  &  8.97  &   M6.0   \\
 1.678  & 0.110  & 0.154  &  2957  & 5.102  & 30.80  & 18.11  & 16.15  & 14.62  & 13.39  & 11.65  &  9.67  &  9.08  &  8.77  &   M5.7   \\
 1.741  & 0.120  & 0.163  &  3020  & 5.094  & 30.88  & 17.56  & 15.71  & 14.21  & 13.03  & 11.36  &  9.49  &  8.90  &  8.59  &   M5.4   \\
 1.801  & 0.130  & 0.171  &  3069  & 5.086  & 30.95  & 17.14  & 15.37  & 13.89  & 12.74  & 11.13  &  9.33  &  8.74  &  8.44  &   M5.2   \\
 1.858  & 0.140  & 0.179  &  3115  & 5.078  & 31.02  & 16.76  & 15.06  & 13.59  & 12.47  & 10.92  &  9.18  &  8.60  &  8.30  &   M5.0   \\
 1.912  & 0.150  & 0.187  &  3160  & 5.072  & 31.08  & 16.39  & 14.76  & 13.31  & 12.22  & 10.71  &  9.04  &  8.46  &  8.17  &   M4.7   \\
 1.964  & 0.160  & 0.194  &  3200  & 5.065  & 31.14  & 16.07  & 14.50  & 13.06  & 11.99  & 10.53  &  8.92  &  8.34  &  8.05  &   M4.5   \\
 2.014  & 0.170  & 0.202  &  3227  & 5.059  & 31.18  & 15.86  & 14.32  & 12.89  & 11.84  & 10.40  &  8.81  &  8.23  &  7.95  &   M4.4   \\
 2.062  & 0.180  & 0.209  &  3250  & 5.054  & 31.23  & 15.67  & 14.15  & 12.73  & 11.69  & 10.27  &  8.71  &  8.13  &  7.85  &   M4.2   \\
 2.108  & 0.190  & 0.216  &  3272  & 5.048  & 31.27  & 15.49  & 13.99  & 12.58  & 11.55  & 10.16  &  8.61  &  8.04  &  7.76  &   M4.1   \\
 2.152  & 0.200  & 0.223  &  3290  & 5.043  & 31.30  & 15.33  & 13.85  & 12.45  & 11.43  & 10.05  &  8.53  &  7.96  &  7.67  &   M4.0   \\
 3.183  & 0.200  & 0.289  &  3300  & 4.817  & 31.53  & 14.64  & 13.23  & 11.89  & 10.87  &  9.49  &  7.95  &  7.37  &  7.09  &   M4.1   \\
 3.578  & 0.250  & 0.338  &  3372  & 4.779  & 31.71  & 14.03  & 12.66  & 11.34  & 10.35  &  9.02  &  7.54  &  6.96  &  6.69  &   M3.8   \\
 3.931  & 0.300  & 0.383  &  3428  & 4.747  & 31.85  & 13.55  & 12.20  & 10.90  &  9.94  &  8.64  &  7.20  &  6.63  &  6.37  &   M3.5   \\
 4.249  & 0.350  & 0.427  &  3489  & 4.721  & 31.97  & 13.12  & 11.79  & 10.50  &  9.57  &  8.30  &  6.91  &  6.33  &  6.08  &   M3.2   \\
 4.540  & 0.400  & 0.469  &  3550  & 4.698  & 32.08  & 12.75  & 11.42  & 10.14  &  9.23  &  8.00  &  6.65  &  6.07  &  5.83  &   M2.9   \\
 4.808  & 0.450  & 0.509  &  3614  & 4.678  & 32.18  & 12.41  & 11.07  &  9.80  &  8.91  &  7.72  &  6.41  &  5.82  &  5.60  &   M2.6   \\
 5.058  & 0.500  & 0.547  &  3690  & 4.660  & 32.28  & 12.08  & 10.72  &  9.46  &  8.58  &  7.44  &  6.18  &  5.58  &  5.38  &   M2.3   \\
 5.291  & 0.550  & 0.585  &  3790  & 4.644  & 32.39  & 11.74  & 10.34  &  9.08  &  8.23  &  7.16  &  5.95  &  5.33  &  5.15  &   M1.9   \\
 5.511  & 0.600  & 0.622  &  3911  & 4.629  & 32.49  & 11.39  &  9.93  &  8.67  &  7.85  &  6.86  &  5.72  &  5.07  &  4.93  &   M1.3   \\
 5.718  & 0.650  & 0.657  &  4055  & 4.615  & 32.61  & 10.99  &  9.48  &  8.22  &  7.42  &  6.55  &  5.49  &  4.84  &  4.72  &   M0.5   \\
 5.914  & 0.700  & 0.692  &  4204  & 4.603  & 32.71  & 10.55  &  9.01  &  7.77  &  7.01  &  6.24  &  5.26  &  4.63  &  4.53  &   K7.3   
\enddata
\tablecomments{UBVRI
magnitudes are given on the Johnson-Cousins system
\citep{1990PASP..102.1181B}, JHK are given on the CIT system
\citep{1982AJ.....87.1029E,1982AJ.....87.1893E}. The sequence provided
here is abbreviated. A more complete sequence, using steps of
$0.001M_{\odot}$ and including the far-infrared L (CIT), L$^\prime$
and M$^\prime$ (Mauna Kea) bands is available in electronic form.}
\label{tab:seq_bpl}
\end{deluxetable*}

The fit parameters and errors are quite similar to those derived in 
K06, but there are some differences in the bouncer and
short-period regimes. These differences arise primarily 
because the larger value of $P_{min}$ has caused some data to be
shifted from the short-period regime into the bouncer regime, and also
because of the addition of high-precision data points from
L08. However, all of the main conclusions of K06 still hold. In
particular, the intrinsic dispersion in radius is still only 0.01-0.02
dex in all three regimes, in line with the theoretical expectation that 
all unevolved CVs should quickly converge onto a unique evolution
track (with some WD-mass-dependent scatter). Also, the power-law slope
derived for period bouncers is (just) consistent with an increasing
period in this regime (i.e. $\zeta < 1/3$, c.f. Equation~\ref{eq:pdot}). 

We can now use this new $M_2-R_2$ relation to create our updated donor
sequence. Again, we will only discuss how our implementation here
differs from that described in K06. First, in K06 we adopted the 
$M_2-T_{eff}$ relation suggested by the BCAH98 5~Gyr MS models down to
$T_{eff} \simeq 2000$~K, switching to the 1~Gyr ``DUSTY''
\citep{2002A&A...382..563B} 
and, ultimately, 1~Gyr ``COND'' \citep{2003A&A...402..701B} 
isochrones at lower temparatures. We adopted younger isochrones in the
brown dwarf regime, because CV secondaries are essentially born as brown
dwarfs when their masses drop beyond the Hydrogen burning limit. This 
need to adopt a characteristic ``age'' for donors in the brown dwarf
regime was an obvious weakness of the original donor
sequence.\footnote{Note that this weakness {\em only} affects donors
at the lowest masses, in or close to the brown dwarf regime. As
explained in Section~\ref{sec:m_vs_t}, at higher
masses ($M_2 \gtappeq 0.1~M_{\odot}$), the $M_2 - T_{eff}$ relation of CV
donors is virtually identical to that of isolated MS stars (Kolb, King 
\& Baraffe 2001), with no significant dependence on age.}
Here, we instead adopt the $M_2 - T_{eff}$ 
relation suggested by the revised (best-fit) evolution track presented
in Section~\ref{sec:props}. Clearly, this still does not amount to a
self-consistent treatment, since the donor properties along this track 
are not identical to those on the broken-power-law donor
sequence. Moreover, the stellar models on which this track is based do
not account for dust formation and settling (see discussion in
Section~\ref{sec:final}). Nevertheless, we expect this to be at
least a mild improvement over the ad-hoc adoption of a particular
age for the lowest-mass stellar models. Second, we use a
slightly updated set of stellar 
atmosphere grids to compute the photometric donor properties. More
specifically, we use the standard NextGen atmosphere models
\citep{1999ApJ...512..377H} down to
$T_{eff} = 2300$~K, the AMES-DUSTY models 
\citep{2000ApJ...542..464C}
between $1800~{\rm K} \leq T_{eff} \leq 2000~{\rm K}$, and the
AMES-COND models 
\citep{2003A&A...402..701B} 
for temperatures
$T_{eff} \leq 1700~{\rm K}$. The transitions between these regimes are
bridged by interpolation. See Section~\ref{sec:final} for a brief
discussion of the differences between the different types of
atmosphere models.

A short version of the revised donor sequence is provided in
Table~\ref{tab:seq_bpl}; a more finely spaced version is available in
electronic form. We defer a closer look at the physical and
photometric properties along the sequence to Section~\ref{sec:props}. There, 
we will directly compare the old and new versions of the 
broken-power-law donor sequences to the donor properties suggested by our
full evolution track. However, we note here already that the new
version of the donor sequence should not be used for masses lower than
$M_2 \simeq 0.05~M_{\odot}$, since the power-law approximation to the
mass-radius relation probably breaks down in this regime
(c.f. Figure~\ref{fig:phys}). 

Finally, here and throughout this paper, we have followed K06 in adopting a
constant value of $M_{1} =  0.75 M_{\odot}$ for the mass of a 
WD in a typical CV. In K06, we showed that the mean WD mass among the 
eclipsing CV sample available at the time was $<M_{1}> = 0.75 \pm
0.05~M_{\odot}$, with no evidence for evolution in WD mass as a
function of $P_{orb}$, but with a roughly 20\% intrinsic dispersion
($\sigma_{int} = 0.16~M_{\odot}$). If we add the new data points from
L08, there is still no evidence for evolution, but the mean WD mass
increases slightly to  $<M_{1}> = 0.79 \pm 0.03~M_{\odot}$. This is
not significantly different from the earlier estimate, and, since we
had already begun to assemble our grid of donor sequences and
evolution tracks at the time of L08's publication, we chose to retain 
$M_{1} =  0.75~M_{\odot}$ as a representive WD mass.

\section{Reconstructing Secular Evolution in the Presence of
$\dot{M}_2$ Fluctuations}   
\label{sec:fluc}

We noted in Section~\ref{sec:intro} that our donor-based method for
reconstructing $\dot{M}_2(P_{orb})$ is likely to yield {\em secular}
mass-transfer rates. This is a key advantage, since the mass-transfer
and accretion rate in a CV can vary on a wide range of time scales. As
long as the variability time scale, $\tau_{var}$, is shorter than the
evolutionary time scale, $\tau_{ev} \simeq \dot{J}/J \simeq
\dot{M}_2/M_2$, these fluctuations will not significantly affect the
system's overall evolution. However, they can confound observational
tracers of $\dot{M}_2$ that track the mean mass-transfer or accretion
rate on time scales $\ltappeq \tau_{var}$. 

The possible existence of mass-transfer-rate fluctuations on
unobservably long time scales has been widely discussed 
in the CV community. In this section, we will explain the origin of
this idea, briefly discuss the main mechanisms that could produce such
fluctuations, compare the variability time scales associated with
these mechanisms to the averaging time scales of several observational 
tracers of $\dot{M}_2$, and finally assess the evidence for such
fluctuations in the light of the latest data.  

\subsection{The Historical Case for Long-Term $\dot{M}_2$ Fluctuations}

The first attempts to estimate CV mass-transfer rates found
substantial scatter in the inferred average $\dot{M}_2$ values at
given $P_{orb}$, especially above the period gap
\citep{1984ApJS...54..443P,1987MNRAS.227...23W}.  These studies used
time-averaged accretion light as a tracer of $\dot{M}_2$ (see
Section~\ref{sec:accretion_light}), but additional evidence
for significant scatter in the present-day mass-transfer rates at
fixed $P_{orb}$ appeared to come from the co-existence of  
dwarf novae and nova-likes at orbital periods above the gap. According
to the disk instability model for dwarf nova eruptions
\citep[e.g.][]{1996PASP..108...39O,2001NewAR..45..449L}, these CV
sub-classes are differentiated primarily by the rate at which mass is
supplied to the accretion disk. If all CVs follow a unique evolution
track, one might therefore expect CV sub-types to populate distinct
period ranges. Taken at face value, scatter in sub-types must therefore
reflect scatter in $\dot{M}_2(P_{orb})$ (but see
Section~\ref{sec:stability} for a more detailed look at this issue).

Such observationally inferred dispersion in $\dot{M}_2(P_{orb})$ is
unlikely to reflect a real spread in the {\em secular}
mass-transfer rates within the CV population. Theoretically, all CVs with   
initially unevolved donors are expected to quickly join onto a {\em
unique} evolution track, whose properties are determined solely by the
mechanism for AML from the system 
\citep{1983ApJ...268..825P, 1992A&A...259..159R, 1993A&A...271..149K,
1996MNRAS.279..581S}. Empirically, a 
unique track is also necessary in order to explain the existence of a
period gap with sharp edges and a well-defined minimum period. The
observed scatter in $\dot{M}_2$ was therefore quickly interpreted as
evidence for mass-transfer-rate fluctuations on unobservably
long time scales, but still satisfying  $\tau_{var} << \tau_{ev}$
\citep{1984MNRAS.209..227V,1989MNRAS.237...39H}.

\subsection{Physical Causes of Long-Term $\dot{M}_2$ Fluctuations} 

How could such fluctuations be produced? The mass-loss rate from the
donor depends exponentially on the distance between the stellar radius
and the Roche lobe \citep{1988A&A...202...93R}
\begin{equation} 
\dot{M}_2 = \dot{M}_0 \exp^{-\frac{\Delta R}{H}},
\label{eq:ritter}
\end{equation} 
where $\Delta R = R_L - R_2$, and $H \simeq 10^{-4} R_2$ is the scale
height near $R_2$. Quite generally, fluctuations in $\dot{M}_2$ may
therefore be associated with variations in $H$ and/or changes in
$\Delta R$.

Fluctuations in $\dot{M}_2$ driven by spatial and/or temporal
variations in $H$ are, in fact, likely to occur. However, their time
scales are much  
shorter than we are interested in here. For example,
the strongly magnetic polars, which lack an accretion disk entirely,
exhibit high- and low-state behaviour with characteristic time scales
of weeks to years \citep[e.g.][]{2000A&A...361..952H,
2005AJ....130..742K}. 
These low states may well be linked to the $L_1$ point becoming
covered by star spots. This can quench $\dot{M}_2$, because star spots are
characterized by lower $H$ than the surrounding photospheric
regions (\citealt{1994ApJ...427..956L,1998ApJ...499..348K}, but also
see \citealt{2000ApJ...530..904H}).
Similarly, there is some
observational evidence that the mass-transfer rate from the secondary
may be enhanced during dwarf nova eruptions 
\citep{2002PASP..114..721P, 2004AN....325..185S, 2004AcA....54..221S}.
This may be due to irradiation-driven
heating of the upper atmospheric layers in the donor star, causing an
increase in $H$ and hence $\dot{M}_2$
(\citealt{2004AcA....54..181S,2004AcA....54..429S}; but also see 
\citealt{2003A&A...401..325O,2004A&A...428L..17O}).
However, neither of these types of variations
occur on the long time scales relevant to us here. In fact, the
longest plausible time scale on which $H$ variations may be driven is
probably that of magnetic activity cycles. Even this is ``only'' on
the order of years and therefore still accessible to observations.

The only promising way to produce unobservably slow fluctuations in
$\dot{M}_2$ is therefore to invoke variations in $\Delta R$. Such
variations might be driven by changes in either $R_2$ or $R_L$. The
only two mechanisms that have been considered in any detail in this
context -- irradiation-driven $\dot{M}_2$ cycles and nova-induced
hibernation -- can, in fact, be distinguished along these lines.

\subsubsection{Varying $R_{2}$: Irradiation-Driven Mass-Transfer Rate
Fluctuations}
\label{sec:irrad_cycle}

It can be shown on fairly general grounds that a long-term limit cycle
on the basis of stellar radius variations is possible, provided the
donor's thermal relaxation time scale depends explicitly on $\dot{M}_2$
\citep{1995ApJ...444L..37K,1996ApJ...467..761K}. The only obvious
physical mechanism for producing such a coupling is irradiation.

A full discussion of the finer details of irradiation-driven
mass-transfer cycles is far beyond the scope of the present
study. Such discussions are provided by 
\citet{1995ApJ...444L..37K,1996ApJ...467..761K,
1995ASSL..205..479R,1996IAUS..165...65R,
1995ASSL..205..315W,1995PASA...12...60W,
1998ApJ...500..923M,
2000A&A...360..969R,
2000NewAR..44..167K} and most recently \citet[][hereafter
BR04]{2004A&A...423..281B}. 
For our present purposes, the two questions that really matter are:
(i) on what time scale are such 
cycles expected to be driven?; (ii) what is the amplitude of the $R_2$
variations expected during these cycles?

Let us start with the first question. In order to produce significant
fluctuations in $\dot{M}_2$, $\Delta R$ must fluctuate by at least
$H$. Moreover, the low states of these cycles correspond to an
(almost) detached state, which is terminated when the Roche lobe
catches up with the stellar radius again. Thus a firm lower limit on
the time scale of such cycles is given by the time it takes the
Roche lobe to move through $H$ in the absence of any mass loss. This
is given by \citep[e.g.][]{1995ApJ...439..330K}
\begin{equation}
\tau_{h} \simeq \frac{H}{R_2} \frac{J}{2\dot{J}}.
\label{eq:tau_h}
\end{equation}

In order to obtain a more realistic estimate of the cycle time scale,
we need to answer the second question. As we will show in
Section~\ref{sec:irrad}, irradiation in CVs is only capable of inflating
donor radii by $\simeq 1\%$. For mass-transfer cycles driven by
irradiation, this is then also a rough estimate of the amplitude of
the radius variations the donor experiences during these cycles,
i.e. $\delta R / R_2 \simeq 0.01$.
\footnote{Note that the donor does not relax all the way back to its thermal
equilibrium radius during the low states, because the lobe catches up
with the stellar radius well before this can happen. This is also why
a significant ($\gtappeq 5\times$) reduction in the AML rate is needed
in the disrupted MB braking 
model in order to produce a period gap of the observed width. Without
such a reduction, the Roche lobe would catch up with the
donor before it has relaxed all the way back to its thermal
equilibrium radius. The result would be a premature emergence from the
gap, i.e. a gap of insufficient width.}
A more realistic estimate of the duration of mass transfer cycles 
is therefore
\begin{equation}
\tau_{\delta R} \simeq \frac{\delta R}{R} \frac{J}{2\dot{J}}.
\label{eq:tau_deltar}
\end{equation}
Finally, an independent estimate of the time scale associated with 
irradiation-driven mass-transfer cycles is provided by BR04. They show
that {\em low-amplitude} cycles are expected to have characteristic
time scales of 
\begin{equation}
\tau_{bh} \simeq 2\pi \left(\frac{H}{R_2} \frac{J}{\dot{J}} \frac{\tau_{CE}}{10}\right)^{0.5},
\label{eq:tau_bh}
\end{equation}
where $\tau_{CE}$ is the thermal time scale of the convective envelope
of the donor. This depends primarily on the mass of the convective
envelope, $M_{CE}$, and the thermal time scale of the undisturbed
star, $\tau_{{kh,eq}}$, and is given explicitly by 
\begin{equation}
\tau_{CE} \simeq  \frac{3}{7} \frac{M_{CE}}{M_2} \tau_{{kh},eq}.
\label{eq:tau_ce}
\end{equation}
For reference, in our standard model sequence, $M_{CE}$ can be
approximated roughly by 
\begin{equation}
M_{CE} \simeq M_2 - \exp^{-(3M_2+0.75)}
\end{equation}
across most of the mass range above the period gap. Note that 
$\tau_{bh}$ is really a lower limit, since it only refers to
low-amplitude cycles. The numerical examples in BR04 show that
large-amplitude cycles are characterized by longer time scales.

As noted above, the donor radius fluctuates by only $\simeq 1\%$
across the full mass-transfer cycle. This is a small fluctuation on top
of the larger radius increase that is due to the secular
$\dot{M}_2$. After all, the donor must be $\simeq 30\%$ inflated at
the upper edge of the gap (Section~\ref{sec:gap}). Thus, as advertised 
above, mass-transfer rates inferred from donor radii should remain
largely unaffected by such cycles. 

This last statement comes with some fine print. A key problem with
irradiation-driven cycles is that CVs with donor masses $M_2 \ltappeq 0.65  
M_{\odot}$ -- i.e. essentially all the CVs we are 
interested in -- are not susceptible to the instability that produces
these cycles (e.g. \citealt{1996ApJ...467..761K,2000A&A...360..969R};
BR04). However, it has been shown that the combination of
irradiation-driving with CAML \citep{1995ApJ...439..330K}
can destabilize CVs with lower donor masses, even possibly including
systems below the period gap (e.g. \citealt{1996ApJ...467..761K,
  1998ApJ...500..923M, 2000A&A...360..969R}; BR04). It is not obvious
to us whether CAML-assisted 
irradiation-driven cycles would necessarily also show only small
radius excursions. In the low state of such a cycle, the AML is
reduced (because the ``consequential'' part of it depends, by
definition, on $\dot{M}_2$). But the AML time scale also sets the time
scale for Roche-lobe shrinkage in the detached state. It is therefore 
conceivable that some models of CAML-assisted cycles could predict 
considerably longer low states -- perhaps long enough to allow significant
shrinkage of the donor. If so, then radius measurements of active CVs,
would trace the average high-state $\dot{M}_2$, rather than the
secular value. Note, however, that even the standard interpretation of
the period gap itself would have to be modifed in this case. The
gap width, for example, would reflect only the high-state
properties of donors above and below the gap in such a scenario. If
the cycle time scale were comparable to the evolutionary time scale,
it would also become difficult to explain the sharpness of the period
gap and the period minimum. 

\subsubsection{Varying $R_{L}$: Nova-Induced Cycles and Hibernation}
\label{sec:hibernation}

The second way to drive long-term $\dot{M}_2$ fluctuations in CVs is
via cyclic variations in the Roche-lobe radius, $R_L$ (which must be
superposed on the slow evolutionary changes in $R_L$). There is really
just one obvious recurring event in the life of a CV that might
produce such variations: a nova eruption. 

The idea that nova eruptions might explain the large inferred scatter
in $\dot{M}_2$ at fixed $P_{orb}$ was first put forward by
\citet{1986ApJ...305..251M}. He argued that the dominant effect of
these eruptions would be to {\em reduce} the size of the binary
orbit. More specifically, he suggested that during a nova eruption,
the binary would effectively orbit inside a common envelope composed
of material lifted off the WD surface. This material would therefore
be frictionally heated, spun up, and ejected, with the energy and AML 
associated with this process being drained from the binary orbit. The
dominant effect on binary evolution in this picture is {\em
  frictional angular momentum loss} (FAML). This shrinks the binary
orbit and, with it, the Roche lobe. MacDonald's theory therefore
predicts that novae should exhibit enhanced mass-transfer rates in the
aftermath of nova eruptions. 

At almost the same time, \citet{1986ApJ...311..163S} presented a
completely different take on the same basic idea. 
%They also noted 
%that the ejection of mass and angular momentum associated with
%nova eruptions could have a significant effect on CV
%evolution. However, the direction they inferred for this effect was
%opposite to that suggested by \citet{1986ApJ...305..251M}. 
In effect, they argued that the dominant effect of the
nova eruption is the removal of {\em mass} from the binary system (not
the removal of angular momentum). 
In this case, the orbit (and the Roche lobe) must
{\em expand} in the aftermath of an eruption, leaving the secondary
temporarily in detached (or almost detached) low state. Thus
\citet{1986ApJ...311..163S} suggested that most CV may ``hibernate''
for most of the time between nova eruptions.

As with irradiation-induced mass-transfer cycles, a full discussion
of nova-induced mass-transfer cycles and hibernation is beyond the
scope of the present study. The effect of  
nova eruptions on binary parameters is examined in 
\citet{1987ApJ...319..819L} and \citet{1991A&A...246...84L}, while
\citet{1983ApJ...268..710S} and \citet{2010arXiv1003.4207M} discuss
attempts to measure this effect directly. The observational case {\em
for} hibernation is presented in \citet{1990ApJ...356..609V,
1992ASPC...29..379D} and \citet{2005A&A...432..199S}, while the observational
case {\em against} is made by
\citet{1992MNRAS.258..449N,1994MNRAS.266..761W,
1996MNRAS.278..845S} and \citet{2008A&A...483..547T}. Finally, the impact of
nova-induced $\dot{M}_2$ fluctuations on long-term CV evolution has
been studied by \citet{1998MNRAS.297..633S} and \citet{2001ApJ...563..958K}. 
A comprehensive review of nova theory and observations,
including hibernation, is given by \citet{1989PASP..101....5S}, while
a very readable introduction and take-off point for further study is
provided by \citet{2006A&G....47a..29W}. 

Here, we simply wish to know on what time scale nova eruptions might
cause $\dot{M}_2$ to fluctuate. This is fairly easy to answer: the
relevant time scale is simply the nova recurrence time, which can be 
calculated from nova models as 
\begin{equation}
\tau_{nova} \simeq \frac{M_{ign}}{\dot{M}_{acc}}.
\end{equation}
Here, $\dot{M}_{acc}$ is the accretion rate onto the WD, averaged over
the nova cycle. If nova 
eruptions are the only cause of long-term variability, we can take 
$\dot{M}_{acc} = \dot{M}_2$, where, as usual, $\dot{M}_2$ is the
secular mass-transfer rate. The nova ignition mass, $M_{ign}$, is
the critical mass of the accreted envelope on the WD above which a 
thermonuclear runaway is triggered. $M_{ign}$ depends primarily on three
parameters: (i) the accretion rate, $\dot{M}_{acc}$; (ii) the WD mass,
$M_{1}$; and (iii) the WD core temperature, $T_{wd,c}$. In a CV, the
equilibrium WD core temperature is set by the compressional release of
energy associated with accretion and is therefore itself a function of
$M_{1}$ and $\dot{M}_{acc}$ \citep{2004ApJ...600..390T}.
\footnote{The WDs in long-period CVs actually may not have
time to fully equilibrate, but \citet{2005ApJ...628..395T} argue that
$M_{ign}$ becomes insensitive to $\dot{M}_{acc}$ in this parameter
regime anyway.}

Strictly speaking, $\tau_{nova}$ is clearly an upper limit on
the time scale associated with nova-induced mass-transfer rate
cycles. However, at least for the most popular incarnation of such
cycles -- hibernation -- it is easy 
to see that the actual time scale cannot be much less than
this. Suppose, for example, that hibernation causes complete
detachment of the donor from the Roche lobe, but that this detached
state lasts for only 1\% of the nova recurrence time. In this case,
the mass-transfer rate in the ``high state'' would, by definition, be
just 1\% above the secular mean. Similarly, the space density of 
active CVs would amount to a full 99\% of the intrinsic CV space
density. Relaxing the assumption of complete detachment makes little
difference: there is then a wider range of {\em possible} $\dot{M}_2$
values at fixed $P_{orb}$, but 99\% of the active CV population would
have $\dot{M}_2$ within better than 1\% of the secular
mean. This shows that in order to have any significant effect on our
understanding of CV evolution at all, the hibernating phase must take
up a large fraction of the nova cycle.

A rough {\em lower} limit on the time scale associated with hibernation
cycles is once again provided by $\tau_{h}$
(Equation~\ref{eq:tau_h}). After all, if the system is detached in the
hibernating state, the Roche lobe has to have time to catch up to the
stellar radius again before the next nova eruption.

\subsection{Averaging Time Scales Associated with Observational Tracers of $\dot{M}_2$}

Let us now take a look at the different observables that can be used
to estimate $\dot{M}_2$, focusing particularly on the time scale over
which they effectively average the mass-transfer rate.

\subsubsection{Accretion Light}
\label{sec:accretion_light}

The most straightforward tracer of a CV's mass-transfer rate is its
brightness. To first order, this is primarily a measure   
of the accretion luminosity, $L_{acc} \simeq GM_{1}\dot{M}_{acc}/R$,
and thus of the {\em accretion rate} through the disk and onto the WD,
$\dot{M}_{acc}$. This can vary dramatically in a CV. For example, in a
dwarf nova, $\dot{M}_{acc}$ changes by orders of magnitude over the 
outburst cycle. If the full time scale of such variations is covered
and resolved observationally -- as is true for many, but not all dwarf  
nova cycles -- one can attempt to average over this variability. The
relevant averaging time-scale associated with the use of accretion
light as a tracer of $\dot{M}_2$ is therefore simply the length of the
observational record. A typical number for this might be $\tau_{obs}
\sim {\rm 10~yrs}$.

\subsubsection{White Dwarf Temperatures}

Another promising method to estimate $\dot{M}_2$ is based on the quiescent
effective temperature of the accreting WD in a CV 
\citep{2002ApJ...565L..35T,2003ApJ...596L.227T, 2004ApJ...600..390T,
2009ApJ...693.1007T}. This is set primarily by the mean accretion
rate over the last thermal time scale of the non-degenerate layer on
the WD surface 
\citep{2003ApJ...596L.227T} 
\begin{equation}
\tau_{wd} \simeq 10^4 
\left(\frac{\dot{M}_{acc}}
{10^{-10} M_{\odot} yr^{-1}}\right)^{-3/4} 
~{\rm  yrs },
\label{eq:tau_wd}
\end{equation}
for a $M = 0.8 M_{\odot}$ WD. 

\subsubsection{Donor Radii}

The method adopted in the present study is to infer $\dot{M}_2$ from
the mass-loss-induced bloating of the donor radii in CVs. As already
discussed in Section~\ref{sec:adjust}, the time scale on which a CV
secondary is able to adjust its radius to the prevailing mass-loss
rate from its surface is given by $\tau_{adj} \simeq 0.05
\tau_{{kh},eq}$ (Equation~\ref{eq:tau_adj};
\citealt{1996MNRAS.279..581S}).

\begin{figure*}
\centering
\includegraphics[height=13cm,angle=-90]{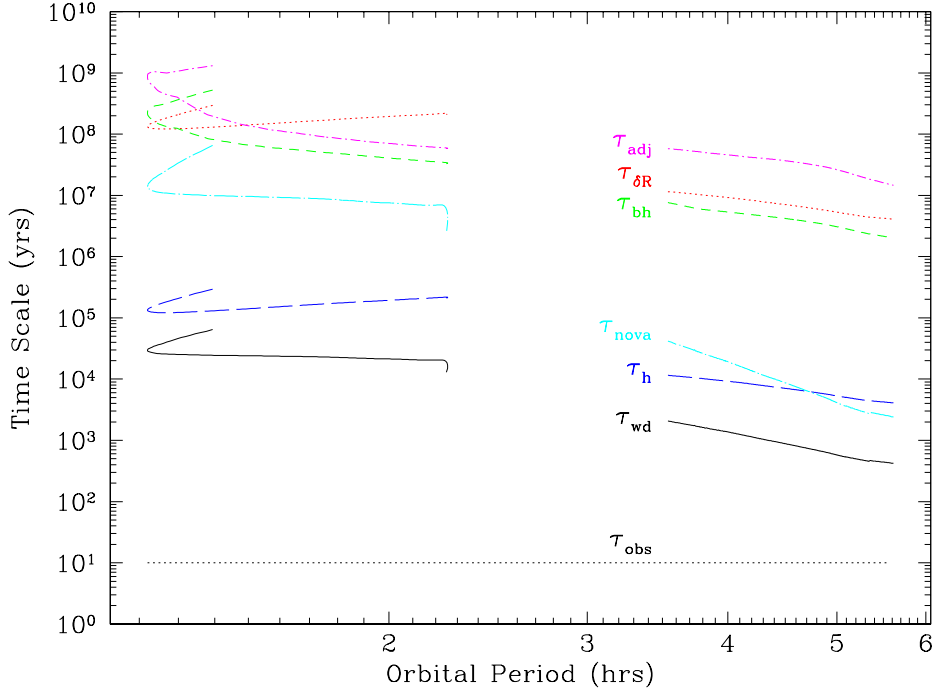}
\caption[Several key time scales as calculated along a standard-model
CV evolution track.]
{Several key time scales as calculated along a standard-model
CV evolution track. The thick solid black horizontal line at the
bottom of the plot marks $\tau_{obs} = 10$~yrs, which is a rough
estimate of the typical averaging time scale associated with
mass-transfer-rate estimates based on accretion liminosity (see
Section~\ref{sec:accretion_light}). The thin solid black line shows
the thermal time scale
of the non-degenerate surface layer on the accreting WD, $\tau_{wd}$
(Equation~\ref{eq:tau_wd}). This is roughly the averaging time scale for
mass-transfer-rate estimates based on (quiescent) WD effective
temperatures. The long-dashed blue line corresponds to the time it takes
the Roche lobe to move through one scale height in the donor's
atmosphere, $\tau_{H}$ (Equation~\ref{eq:tau_h}). This is the shortest
possible time scale for mass-transfer-rate fluctuations driven by
donor radius changes. The remaining three curves show several more
realistic estimates of the time scale for long-term mass-transfer rate
fluctuations: $\tau_{bh}$ (Equation~\ref{eq:tau_bh}; green short-dashed
line), $\tau_{\delta R}$ (Equation~\ref{eq:tau_deltar}; red dotted line) and
$\tau_{adj}$ (Equation~\ref{eq:tau_adj}; magenta dash-dotted line).} 
\label{fig:tau}
\end{figure*}

\subsection{A Direct Comparison of Variability and Observational Averaging Time Scales}
\label{sec:tau_comp}

We now have estimates for the time scales on which long-term $\dot{M}_2$
fluctuations may be driven, and also for the effective averaging time
scales associated with different observational tracers of
$\dot{M}_2$. How do these time scales compare? 

Figure~\ref{fig:tau} shows all the time-scale estimates we arrived at in the
previous sections as a function of $P_{orb}$. All of these curves have
been calculated for a CV evolving according to the standard model. In
the case of $\tau_{nova}$, we estimated appropriate ignition masses
for $M_{1} = 0.75 M_{\odot}$ by interpolating on the calculations
shown in Figure~8 of \citet{2004ApJ...600..390T}.
%\footnote{The equilibrium core temperatures predicted by
%\citet{2004ApJ...600..390T} for CVs are always $\ltappeq
%10^7$~K. We have checked that the corresponding nova recurrence
%times agree fairly well with those calculated by
%\citet{2005ApJ...623..398Y} for $T_{wd,c} = 10^{7}$~K and the same
%$M_{1}$ and $\dot{M}_{acc}$.}

As one might expect, $\tau_{obs}$ is by far the shortest of all of
these time scales. Thus accretion light will only yield reliable
estimates of secular mass-transfer rates if long-term cycles do not
exist at all. However, $\tau_{wd}$ is also about an order of
magnitude faster than even $\tau_{h}$, which is quite a firm
lower limit on {\em any} likely cycle time scale. 

Strictly speaking, if $\tau_{var} > \tau_{wd}$, 
$\tau_{wd}$ itself will vary with the varying $\dot{M}_{acc}$. 
Thus $\tau_{wd}$ will be much longer in the low state of a cycle than
in the high state, an effect that is not taken into account in 
Figure~\ref{fig:tau}. Perhaps as a result
of this, $\tau_{wd}$ may perform slightly better than suggested by
Figure~\ref{fig:tau}. The detailed calculations presented in 
\citet[][hereafter TG09]{2009ApJ...693.1007T} show, for example, that if
$\dot{M}_{acc}$ were to vary by a factor of 9 with a period of $8\times
10^4$~yrs around a secular mean of $\dot{M}_2 = 5 \times
10^{-11}~M_{\odot}~yr^{-1}$ (i.e. a variation between $\dot{M}_2 = 1
\times 10^{-11}~M_{\odot}~yr^{-1}$ and $\dot{M}_2 = 9 \times
10^{-11}~M_{\odot}~yr^{-1}$), the quiescent WD luminosity -- and hence
the inferred $\dot{M}_2$ -- would fluctuate by a factor 2-3 over the
course of this cycle. 

In any case, Figure~\ref{fig:tau} shows that if mass-transfer rate
fluctuations exist at all in CVs, their time scale could be very long
indeed: $10^6 - 10^9$~yrs for irradiation-driven cycles, and $10^4 -
10^8$~yrs for nova-induced variations.
\footnote{The fact that $\tau_{nova}$ and $\tau_{h}$ cross at the
longest periods suggests that it is not quite self-consistent to
assume that nova-induced hibernation would leave the secular mean
$\dot{M}_2$ unaffected in this case.}
Neither accretion light nor WD
temperatures can be expected to yield reliable estimates of secular
$\dot{M}_2$ if either mechanism drives large-amplitude
accretion-rate variations over a large fraction of a CV's life. 

Finally, what about donor radii? Figure~\ref{fig:tau} shows that
$\tau_{adj}$ is comparable to our estimates for the time scale
associated with 
irradiation-driven cycles. This is as expected, since $\tau_{adj}$ can
actually itself be thought of as an estimate of (or limit on) the time
scale of such cycles . After all, $\tau_{adj}$ is a
measure of the time it takes the stellar radius to adjust to the
changing $\dot{M}_2$ across the cycle. However, as explained in
Section~\ref{sec:irrad_cycle}, donor radii are nevertheless likely to
trace the secular $\dot{M}_2$ in this case, because the donor radius
variations that drive these cycles are much smaller than the
mass-loss-induced donor bloating. In the case of hibernation, there is
no worry at all: $\tau_{adj}$ is comfortably longer than $\tau_{nova}$ at all
orbital periods.

\subsubsection{The Observational Case for Long-Term $\dot{M}_2$
Fluctuations: An Update}

Having spent so much time discussing the theory and implications of
mass-transfer cycles, let us provide a brief update on the
observational evidence for their existence. The most recent  
compilations of empirical $\dot{M}_2$ tracers are contained in TG09,
which is based on the WD effective temperature, and in 
\citet{2009arXiv0903.1006P}, 
which is based on the time-averaged accretion disk luminosity. Let us first
consider long-period systems. These are not included in Patterson's
study at all, and only a handful are contained in TG09. Those few points do seem to show
quite significant scatter, although this is mainly driven by just
three objects. All of these have exceptionally high inferred accretion
rates, and all are located in the period range $3$~hrs - $4$~hrs (see their 
Figure~5 and also Figure~\ref{fig:WD} in Section~\ref{sec:WD}). This
{\em might} be evidence for mass-transfer-rate fluctuations above the
period gap.  

Below the gap, there is still some scatter in the WD-based
$\dot{M}_2$ estimates, but the error bars are also large. More
importantly, however, it seems possible that much or all of the
scatter could be due to within-sample variations in white dwarf mass and
the mass of 
the non-degenerate layer on the WD surface (see
Section~\ref{sec:fluc}). The 
time-averaged absolute magnitudes, $M_V$, presented in 
\citet{2009arXiv0903.1006P} show considerable variation as well, by about $1~mag$
at fixed $P_{orb}$. However, the systematic uncertainties affecting
these $M_V$ estimates are also quite large, so it is hard to assess
whether this really provides evidence for long-term mass-transfer
cycles below the gap.

Finally, what about the co-existence of dwarf novae and nova-likes at
the same orbital period? The current data on this is presented and
interpreted in more detail in Section~\ref{sec:stability}. It turns out that
this may provide {\em some} evidence for $\dot{M}_2$ fluctuations
above the gap, although it would probably not be compelling in its own
right. Overall, we think the latest data provide marginally convincing
evidence for long-term mass-transfer-rate variations above the period
gap. Below the gap, it is unclear whether such cycles are needed to
explain the observations. 

\section{CV Evolution From Donor Properties}
\label{sec:assembly}

Let us remind ourselves of our main goal. We wish to construct a
complete evolution track for CVs based solely on the observed properties of
their donor stars. This is possible because mass loss on a time scale
comparable to the donor's thermal time scale will lead to a moderate
expansion of the star. Thus the degree by which a donor is bloated
(relative to an equal-mass MS star) can be used as a tracer of the
rate at which it is losing mass.

In this section, we will first deal with several 
effects other than mass loss that might make Roche-lobe-filling CV
secondaries appear larger than models of ordinary MS stars would
predict (Section~\ref{sec:skeletons}). We will then describe how we
actually construct our final evolution track by fitting
self-consistent models of mass-losing stars to the observational data
(Sections~\ref{sec:form} and~\ref{sec:final}).

\subsection{Overview of the Method}

The key challenge in reconstructing CV evolution from donor properties 
is to obtain a reliable calibration of the relationship between
$\dot{M}_2$ and observed donor inflation. In 
principle, this relationship can be provided entirely by stellar
models, and this is indeed the backbone of our method. However, in
practice, two obstacles must be overcome. First, there are several
effects other than mass loss that might make Roche-lobe-filling CV
secondaries appear larger than models of ordinary 
MS stars would predict. We will deal with these effects in
Section~\ref{sec:skeletons}. 

Second, as explained in Section~\ref{sec:adjust}, donor bloating is not
strictly a measure of the instantaneous $\dot{M}_2$, but is also 
somewhat sensitive to the mass-loss {\em history} of the
donor. Unfortunately, it is too computationally expensive to calculate 
full stellar model evolution sequences for 
the ML histories corresponding to all plausible AML recipes. We
therefore use an approximate method to first determine a suitable
functional form and approximate normalization for $\dot{J}_{sys}$; this is  
briefly described in Section~\ref{sec:form}. Once this form is fixed,
self-consistent model sequences can be used to determine its optimal
normalization and evaluate the goodness-of-fit to the data. The
resulting CV evolution track can then again be coupled to stellar
atmosphere models to simultaneously provide a complete set of
photometric donor properties. These final steps are explained in
Section~\ref{sec:final}.

\subsection{Apparent Donor Bloating Unrelated to Mass Loss}
\label{sec:skeletons}

There are at least three questions that need to be addressed before
the radii predicted by stellar structure models can be applied to CV
donor stars. First, to what extent are these models capable of
accounting for the observed radii of {\em isolated} low-mass stars?
Second, what is the net effect of tidal and rotational deformation on
the radii of semi-detached donor stars? Third, could irradiation cause
significant radius inflation in CV secondaries? 

\subsubsection{Larger-Than-Expected Radii in Non-Interacting Low-Mass Stars} 
\label{sec:bloat}

Over the last few years, it has become increasingly clear that the 
empirically-inferred radii of at least some lower-MS stars are
significantly larger than predicted by stellar structure models (by up
to $30\%$ in extreme cases; see \citealt[][]{2007ApJ...660..732L} for a
fairly recent overview). If we are going to use such models to infer
mass-loss rates from observed donor radii, this discrepancy clearly
needs to be taken into account. 

\begin{figure*}
\includegraphics[height=19cm,angle=-90]{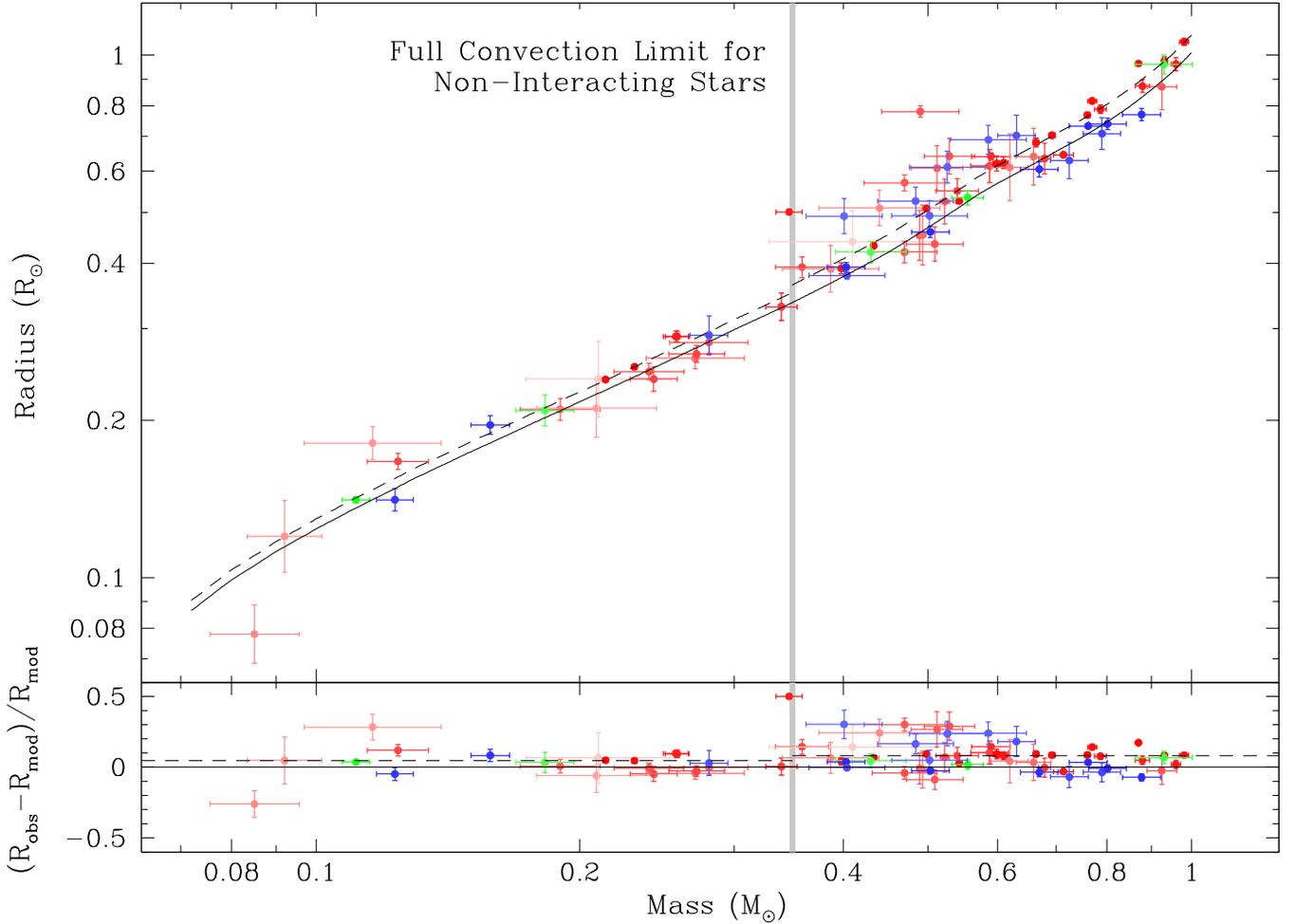}
\caption[The mass-radius relationship of non-interacting low-mass 
stars.]
{{\em Top panel:} The mass-radius relationship of non-interacting low-mass 
stars. Points and errors are based on the observational data listed in
Table~\ref{tab:msmr}. Different colours correspond to different types of
objects: single stars are shown in blue, stars in binaries with WD
companions are plotted in green; stars in other binaries are shown in
red. Observations with larger errors bars are plotted with slightly
lighter/fainter colours. The solid black line shows the predicted
mass-radius relation at 5~Gyrs from BCAH98. The dashed black lines
show the same sequence after re-scaling it slghtly to optimally match
the data. The re-scaling is done independently for fully convective and
partially radiative stars and amounts to 4.5\% and 7.9\% adjustments,
respectively. {\em Bottom panel:} Residuals of the data points with
respect to predicted mass-radius relationship. The dashed lines once
again show the optimal re-scalings. The vertical grey line in both
panels marks the dividing line between fully convective and partially
radiative stars (0.35 $M_{\odot}$; \citealt{1997A&A...327.1039C}).}
\label{fig:mr_iso}
\end{figure*}

In order to test and calibrate the BCAH98 models we are using, we have
compiled our own, fairly comprehensive list of empirically determined
masses and radii for lower MS stars. This is given in
Appendix~\ref{app:msmr}. The resulting empirical mass-radius
relation is shown in Figure~\ref{fig:mr_iso}, along with a standard
5~Gyr BCAH98 MS track.

Figure~\ref{fig:mr_iso} confirms that at least some lower-MS stars have radii 
significantly in excess of the standard BCAH98 model
prediction. However, it also reveals two other important
points. First, there is no evidence that {\em all} 
observational radii are larger than predicted. Instead, it appears
that there is considerable radius {\em scatter} in the $M-R$ relation,
with some stars falling quite close to the model track, but others
lying well above it. This finding mirrors that of
\citet{2006A&A...460..783B}, who dubbed those stars found close to the
theoretical MS ``immaculate dwarfs''. 

Second, it is striking that fully convective stars appear to
be much closer to the model predictions, on average, and exhibit much
less radius scatter, than stars with a radiative core. A hint along
these lines had first been noted by \citet{2006Ap&SS.304...89R} and 
\citet{2007ApJ...660..732L}, but 
our larger data base here allows us to confirm and quantify this
impression. In particular, we can directly estimate the average
fractional radius excess, $<f_R>$ 
and the intrinsic dispersion around this excess, $\sigma_{int}$,
both above and below the fully convective limit ($\simeq
0.35M_{\odot}$). We take the errors on mass and radius to be
uncorrelated and thus estimate $<f_R>$ and $\sigma_{int}$ by
minimizing the $\chi^2$-statistic 
\begin{eqnarray}
\chi^2 & = &
\sum^{N_{data}}_{i=1}
\frac{\log{R_{data,i}}- \log{f_R
    R_{model,i}}}{\sigma_{\log{R_{data,i}}}^2  + \sigma_{int}^2} 
\nonumber
\\ & + &
\frac{\log{M_{data,i}}- \log{M_{model,i}}}{\sigma_{\log{M_{data,i}}}^2}
\label{eq:chi1}
\end{eqnarray}
with respect to $f_R$, while adjusting the intrinsic dispersion so
that $\chi^2_{\nu} = \chi^2 / (N_{data} - 1) = 1$ for the best
estimate. In evaluating Equation~\ref{eq:chi1}, we take the 
model point ($\log{M_{model,i}}$, $f_R R_{model,i}$) associated with
the $i$-th data point to be that closest to it in  $\chi^2$ sense
(i.e. that  which minimizes this datum's contribution to the total
$\chi^2$).

For stars with radiative cores and masses below $0.7~M_{\odot}$ (the
regime relevant to unevolved, long-period CV donors), we obtain $<f_R>
= 1.079 \pm 0.012$ with  $\sigma_{int} = 0.022$ dex (equivalent to a 
5\% dispersion). However, for fully convective stars, we find a
much smaller average radius excess, $<f_R> = 1.045 \pm 0.005$, with no
need for any intrinsic dispersion ($\chi^2_\nu = 0.79$ with
$\sigma_{int} = 0$). Note that the one data point virtually {\em on}
the fully convective boundary has been excluded from both estimates.

These numbers agree fairly well with those found in the recent
study by Morales et al. (2010). Based on their own compilation of 
of mass-radius measurements, \citet{2010ApJ...718..502M}
estimated $f_R = 1.081$
with $\sigma_{int} = 12.2\%$ for partially radiative stars, and $f_R =
1.057$ with $\sigma_{int} = 3.1\%$ for fully convective stars. The
slight differences between their numbers and ours -- especially with
respect to the intrinsic dispersions -- are probably due to somewhat
different samples and statistical methods being adopted in the two
analyses. For example, had we ignored the mass uncertainties, or
assumed them to be strongly positively correlated with the radius
errors, our intrinsic dispersion estimates would also have been
somewhat larger.

The evidence for intrinsic radius scatter among (at least) stars with
a radiative core suggests that there are one or more hidden parameters
that determine if a particular star is 
``immaculate'' (radius close to the expected value) or not. 
The two most promising candidates for such parameters are starspot
coverage and convective efficiency \citep{2007A&A...472L..17C}. 
More specifically, both a very high starspot covering fraction
and a reduced  convective efficiency in the stellar  envelope can
decrease the heat flux from the stellar interior and thus lead to a
swelling of the star. The underlying cause of reduced convective
efficiency could be either fast rotation and/or a strong magnetic
field.

\citet{2007A&A...472L..17C} explored both of these
possibilities, with findings that may be relevant to CV donors
(c.f. L08). In particular, it turns out that reduced convective
efficiency has only a modest effect on fully convective stars, whereas
spot coverage affects stars with and without a radiative core in a
broadly similar way. This suggests that slight radius excesses among fully
convective stars may be due to spot coverage, whereas the 
considerably larger excesses seen in stars with radiative cores
may be primarily due to reduced convective efficiency. It is also easy 
to account for the observed radius scatter in this picture, since the
efficiency of convective energy transport will depend on factors that
vary from star to star (e.g. rotation and magnetic field strength). 

A quantitative analysis of these ideas has recently been carried out
by \citet{2010ApJ...718..502M}. Crucially, their study also considered
whether some of the inferred radius excesses may be artifacts of the
way in which the 
radii were estimated. Most of the precise radius estimates for
low-mass stars to date -- particularly those for fully convective
stars -- are  based on light curve analyses of eclipsing binary
systems. The results of such analyses can be sensitive to asymmetries
in the surface brightness distribution across the stellar surface. Of
particular concern here are high-latitude or polar star spots, which
are common amongst active stars \citep[e.g.][]{2004AN....325..216H}. 
The application
of standard eclipse analyses to light curves affected by polar spots
will tend to overestimate the stellar radii. The main finding of
\citet{2010ApJ...718..502M} is that the observed radius excesses are
probably due to all three factors: (i) observational radius estimates
derived from eclipsing binaries are probably overestimated by 
$\simeq$ 3\% (actually between 0\% and 6\%) due to the effect of polar
spots; (ii) the effect of star spots on the outward heat flux accounts
for $\simeq 3\%$ of the observed differences; (iii) reduced convective
efficiency accounts for any remaining excess ($0\% - 4\%$).

The results of \citet{2007A&A...472L..17C} and \citet{2010ApJ...718..502M}
suggest a simple, physically motivated calibration for our
theoretical donor models. Following \citet{2010ApJ...718..502M}, we assume
that the average radius excesses inferred above need to 
be corrected downward to account for the 3\% bias associated with
polar spots. This leaves us with a physical excess of 1.5\% for fully
convective stars (which is probably driven by star-spot-induced
suppression of outward heat flux) and an excess of 4.9\% for partially
radiative stars (which is probably driven by both star spots and a
reduced convective efficiency due to rotation and/or magnetic
fields). We can therefore apply these correction factors to our
theoretical donor models, i.e. we will adjust the predicted radii
upward by a factor $f_R = 1.015$ below the gap and $f_R = 1.049$
above.

There are, of course, systematic uncertainties associated with this
calibration. For example, CV donors are faster rotators than most of
the stars in our calibration sample. This might matter, 
because fast rotation could be linked to all of the physical
mechanisms that may drive stellar inflation unrelated to mass loss
\citep{2007A&A...472L..17C}: it may directly suppress
convection, it may produce stronger magnetic fields, and it may also
produce higher star spot coverage. If so, then the radii of CV
donors might be more inflated than those of the stars in our
calibration sample.

There is not enough observational data on CV donors to test this idea
directly. For example, there is very little evidence for even the
presence or absence of star spots on CV
secondaries. \citet{2002ApJ...568L..45W}
spectroscopically inferred a spot coverage of 
$\simeq 20\%$ for the donor in SS Cyg, but the likely nuclear-evolved
nature of this star \citep{2007ApJ...662..564B} makes it an unreliable
calibration point for 
our purposes. Similarly, Roche tomography has been used to detect
star spots in the donor stars of AE Aqr \citep{2006MNRAS.368..637W},
BV Cen \citep{2007MNRAS.382.1105W} and V426 Oph
\citep{2007AN....328..813W}, but one of these is a strongly magnetic
system (AE Aqr) and at least one of the others contains an evolved
donor (BV Cen). The only application of Roche tomography to a fully
convective donor is presented by \citet{2008A&A...480..199B} for the
magnetic system EX~Hya. Their analysis reveals evidence of
irradiation, but not of star spots or chromospheric emission.
 
Fortunately, however, our calibration sample does actually include a
few synchronously rotating stars in very short-period binaries. These 
do not support the idea that such stars exhibit massively larger
inflation. For example, the fully convective secondaries in the
pre-CVs NN~Ser ($P_{orb} \simeq 3.1$~hrs; \citealt{2010MNRAS.402.2591P}) and
RR~Cae ($P_{orb} \simeq 7.3$~hrs; \citealt{2007MNRAS.376..919M}) both exhibit
rather modest radius excesses between 3\% and 4\%, quite consistent
with other fully convective stars in the calibration sample. In the
absence of evidence to the contrary, the best we can do is assume that
the calibration factors derived from our detached star sample will apply
to CVs donors as well. The systematic uncertainties associated with
this assumption will need to be explored, however, and we will do so
in Section~\ref{sec:discuss_bloat}.

\subsubsection{Tidal and Rotational Deformation}
\label{sec:deform}

\begin{figure}
\includegraphics[height=9cm,angle=0]{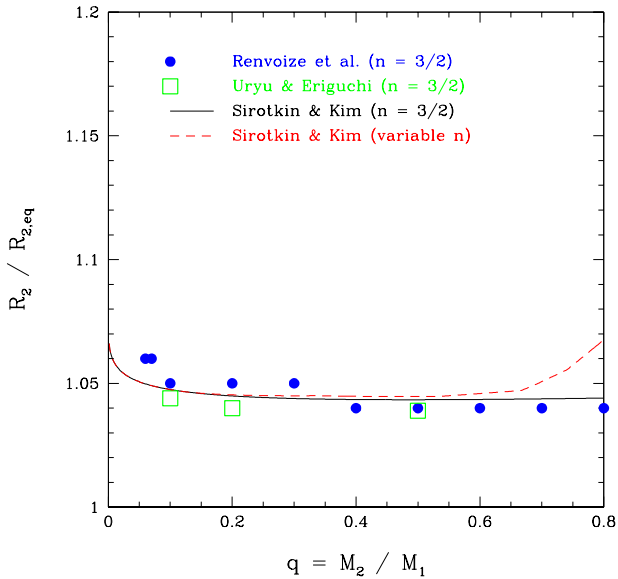}
\caption[The effect of tidal and rotational deformation on the
equilibrium stellar radius of a Roche-lobe-filling star.]
{The effect of tidal and rotational deformation on the
equilibrium stellar radius of a Roche-lobe-filling star. The solid
black line is the analytical approximation derived by SK09 (our
Equation~\ref{eq:siro2}) for polytropes with $n = 3/2$. This is 
appropriate for fully convective donors below the period gap. The blue
points and green open squares show numerical results obtained for the
$n = 3/2$ case in the hydrodynamic simulations by
\citet{2002A&A...389..485R}
and \citet{1999MNRAS.303..329U}, respectively. The red dashed
line shows the correction factor for CV donors if $n$ is allowed to
vary self-consistently with stellar mass and hence mass ratio, from $n
= 3/2$ at the lowest mass ratios to $n = 2.8$ at the highest.}
\label{fig:siro2}
\end{figure}

CV secondaries are not spherical. More specifically, the critical
equipotential surface that defines the outer boundary of a
Roche-lobe-filling star is distorted into the well-known ``tear-drop''
shape by strong tidal and rotational forces. This is a key difference
between CV donors and isolated MS stars. In practice, this difference
is usually ignored, and the donors are treated as spherical
objects with radius equal to the volume-averaged Roche-lobe radius,
$R_L$, i.e. the radius of a sphere with the same volume as the
lobe. Both Equations~\ref{eq:pac} and \ref{eq:siro1} are numerical
approximations for $R_L$. Moreover,  it is usually assumed that the
{\em equilibrium} radius of the star is unaffected by its distorted
shape. 

For many purposes, this treatment is sufficient. However, our goal
here is to use the modest amounts of donor bloating induced by mass
loss to infer $\dot{M}_2$. Moreover, all of the stellar models we will
use to accomplish this are purely 1-dimensional, i.e. spherical and
non-rotating. In this context, it is clearly important to check
whether the distorted shape of the donor may affect its equilibrium 
radius. The precise question we need to answer is this: if we take an
isolated spherical star with mass $M_2$ and equilibrium radius $R_{2,eq}$,
place it in a binary, and then slowly contract the Roche lobe around
it, what will its volume-averaged radius be at the point when it 
just reaches contact? 

There have been several attempts to address this question in recent
years. For example, \citet{2002A&A...389..485R} directly implemented
the thought experiment described above via SPH
simulations. Approximating MS stars as polytropes, they found that for
polytropic index $n = 3/2$ 
(particularly appropriate for fully convective donors), 
$R_L$ at contact was $\simeq 5\%$ larger than the 
equilibrium radius of the isolated star, with only a weak
dependence on mass ratio. Similar results were found by
\citet{1999MNRAS.303..329U}, using a different computational
technique.

Most recently, the question has been addressed again by SK09, who used
a self-consistent field method 
that should be 
both accurate and computationally efficient. They show that the radii
of Roche-lobe-filling polytropes exceed those of isolated ones by a
factor 
\begin{equation}
\frac{R_{RL}}{R_{iso}} =
\left(\frac{N_{RL}}{N_{iso}}\right)^{n/(n-3)},
\label{eq:siro2}
\end{equation}
where $N_{iso}(n)$ and $N_{RL}(n,q)$ are dimensionless coefficients
appearing in the polytropic mass-radius relationships (see Equations
23 and 24 in SK09). Both of
these coefficients 
depend on on the polytropic index, $n$, while $N_{RL}$ additionally 
depends the mass ratio of the system. This latter dependence can be
approximated by (Equation~40 in SK09)
\begin{equation}
N_{RL} = N_{iso} -\frac{d_1 q^{d_2}}{d_3 q^{d_4} + \ln{(1 + q^{d_5})}}.
\label{eq:siro2a}
\end{equation}
Appropriate values for $N_{iso}(n)$ and $d_i(n)$ are provided in
Table~\ref{tab:siro} for the parameter space relevant to CV donors. The
values for $n=3/2$ were taken from SK09, while
the rest were kindly provided to us by Fedir Sirotkin (private
communication).

Figure~\ref{fig:siro2} shows the predicted deformation-induced radius
inflation for $n=3/2$ (appropriate for a fully convective donor) and
for $n$ varying self-consistently with stellar mass (and hence mass ratio) as
given in Table~\ref{tab:siro}. For $n = 3/2$, tidal and rotational
deformation inflates donor radii by $\simeq 4.5\%$ across most of the mass 
ratio range. This is in quite reasonable agreement with the numerical 
results obtained by \citet{2002A&A...389..485R} and
\citet{1999MNRAS.303..329U}, which are also plotted in
Figure~\ref{fig:siro2}. For the more 
realistic case of variable $n$, the deformation effects can be
stronger. For the highest-mass donors we consider ($M_2 = 0.6
M_{\odot}$, for which we adopt $n \simeq 2.3$), the radius inflation
can be as large as 6.8\%. \footnote{For donors of even higher mass,
the predicted level of deformation-induced inflation rises rapidly,
reaching $\simeq$ 14\% for $M_2 \simeq 0.75$ ($n = 2.8$). We do not
consider this regime, partly because we encounter numerical
difficulties in applying such a large and fast-changing correction to
our 1-dimensional stellar models.} In our self-consistent evolution
sequences, we account for deformation-induced donor inflation by
applying the correction factor predicted by Equation~\ref{eq:siro2}
with self-consistently varying $n$.

Before we move on, we note three caveats regarding our implementation
of deformation-induced donor inflation. First, while there is good
agreement between \citet{2002A&A...389..485R} 
and SK09 for $n=3/2$ polytropes, the
same is not true for $n=3$ polytropes. This is probably due to the
fact that the radius of an $n=3$ polytrope does not depend on its mass,
which may lead to computational difficulties for one or both
methods. However, the precise reason for this discrepancy remains to
be established. In practice, we do not think this is a serious
problem, because the inflation predicted by the SK09 
formalism for $n = 2.8$ agrees well with the inflation obtained for $n
= 3$ by \citet{2002A&A...389..485R}. Thus our results would be
virtually unchanged if we had interpolated on the results of
\citet{2002A&A...389..485R} instead of using the SK09 formalism.  

Second, \citet{2009A&A...494..209L} have also recently
modelled the effect of Roche deformation on low-mass MS
stars. Surprisingly, they find that the net effect of this deformation
is to slightly {\em reduce} the stellar radius. The results of
\citet{2009A&A...494..209L} are based on the application of the
\citet{1970stro.coll...20K}
perturbative method to full stellar models. It seems 
likely that their discrepant results are due to the breakdown of this 
method in the limit of strong distortion. However, we
cannot rule out that the discrepancy may be related to the polytropic
approximation, in which case the \citet{2009A&A...494..209L} results
would be preferred. 

Third and finally, \citet{2002A&A...389..485R} 
have shown that, in addition to the mechanical effects of Roche
deformation, there are also what they call ``thermal
effects''. Briefly, the deformation-induced expansion of the donor
changes its characteristic thermal time scale and hence its detailed
response to mass loss. \citet{2002A&A...389..485R} show that these
effects may partially compensate for deformation-induced inflation,
i.e. they will tend to slightly reduce the effect of deformation on CV
evolution sequences. We ignore these thermal effects in our models.

\subsubsection{Irradiation}
\label{sec:irrad}

\begin{figure}
\includegraphics[height=9cm,angle=0]{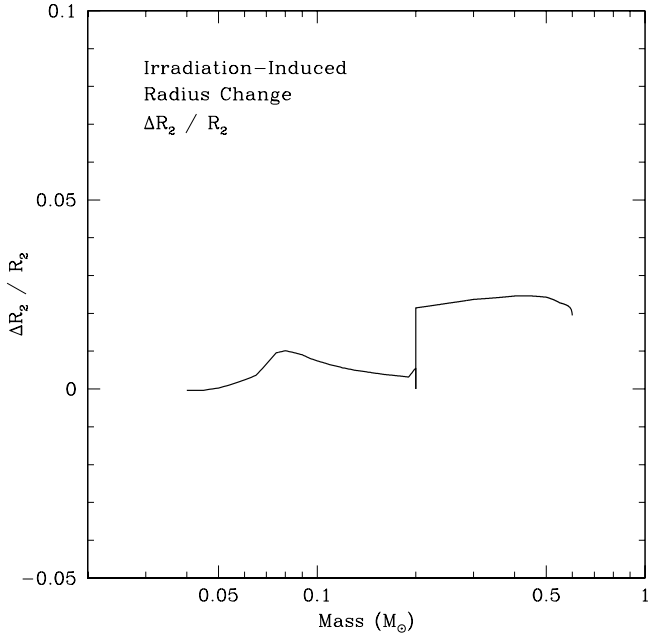}
\caption[The expected amount of radius inflation due to irradiation
along a standard-model CV evolution track.]
{The expected amount of radius inflation due to irradiation
along a standard-model CV evolution track. The net efficiency factor
was assumed to be $\alpha = 0.1$ across the entire track; see text for
details.}
\label{fig:irrad}
\end{figure}

Another key difference between isolated MS stars and CV secondaries is
that the latter may be exposed to irradiation from the vicinity of the
WD primary. This irradiation can block the release of internally
generated energy over some fraction of the donor's surface, causing
the donor to swell. Could this irradiation-induced donor bloating be
large enough to matter in the present context?

Donor irradiation has been studied extensively in the literature,
mainly in the context of long-term mass-transfer cycles 
(see Section~\ref{sec:fluc}; \citealt{1995ApJ...444L..37K,1996ApJ...467..761K, 
1995ASSL..205..479R,1996IAUS..165...65R,
1995ASSL..205..315W,1995PASA...12...60W,
1998ApJ...500..923M,
2000A&A...360..969R,
2000NewAR..44..167K}; BR04).
The key result for our purposes is that if irradation blocks
energy release over a fraction $s$ of the donor's surface, the donor 
radius will be inflated by a factor of about \citep{2000A&A...360..969R}
\begin{equation}
\frac{R_{2}}{R_{2,0}} = (1-s)^{0.1}.
\label{eq:irrad0}
\end{equation}
A firm upper limit on the degree of irradiation-induced
bloating can be obtained by noting that $s_{max} \leq 1/2$, with the
equality obtaining if the entire hemisphere facing the primary is
blocked. In this case, the 
radius of the donor would be inflated by about 7\%. In reality, the
primary does not ``see'' the entire facing hemisphere of the
donor star, and so a better estimate is $s_{max} \simeq 0.3 - 0.4$
(BR04). The maximum possible irradiation-induced
radius inflation is then $4\% - 5\%$.

In order to obtain a more realistic estimate of donor bloating due to
irradiation, we have to estimate the blocked surface fraction 
$s$. This is controlled primarily by the ratio of irradiating flux to
intrinsic flux at the donor's surface; the relevant dimensionless
parameter can be written as \citep[c.f.][hereafter HR97]{1997A&AS..123..273H}
\begin{equation}
x_0 = \alpha_{acc} \alpha_{irr}\alpha_{d}
\frac{(G M_1\dot{M}_1)/(4\pi R_1 a^2)}{\sigma T_{eff,0}}.
\label{eq:irrad1}
\end{equation}
Here, the $\alpha$ terms are efficiency factors which we discuss 
further below, 
$\dot{M}_1$ is the instantaneous accretion rate onto the WD,
$T_{eff,0}$ is the effective temperature of the donor in the absence
of irradiation, and all other symbols have their usual meaning. 

Equation~\ref{eq:irrad1} is identical to Equation~8 in HR97, except
that we have split their single efficiency factor into three separate
terms, following BR04. 
The first, $\alpha_{acc} \leq 1$, allows for the non-isotropic
release of 
accretion luminosity from the vicinity of the WD, and also for the
possibility that the accretion disk may shield substantial portions of
the donor from irradiation. The second, $\alpha_{irr} \ltappeq 1$,
measures the efficiency with which the irradiating flux penetrates to the
sub-photospheric layers. The third, $\alpha_{d} \leq 1$, allows for the
possibility that the release of accretion energy fluctuates on short
time scales, in which case $\alpha_{d}$ is roughly equal to the duty
cycle of the fluctuations (BR04).

HR97 have numerically calculated the function $s(x_0)$ for a large
grid of stellar models spanning a wide range of $T_{eff}$ and
$\log{g}$. They have also provided a convenient numerical 
fitting formula for their results, along with an electronic table
giving the relevant coefficients as a function of $T_{eff}$,
$\log{g}$ and $R_2/a$.
\footnote{Since the original HR97 grid does not
quite cover the full range of stellar properties we are interested in, 
Jean-Marie Hameury kindly provided us with an expanded table of
coefficients.}
%spanning an even wider range of $T_{eff}$ and $\log{g}$. 

We can now assess the impact of irradiation on any particular CV
donor. As a representative example, let us estimate the magnitude of
irradiation-induced donor bloating along a standard-model-like CV
evolution track, 
calculated with $\dot{J}_{sys} = \dot{J}_{GR} + \dot{J}_{MB}(\gamma =
3)$ above the gap and $\dot{J}_{sys} = \dot{J}_{GR}$ below the 
gap. (We have checked that the effect of irradiation on the optimal
track discussed in Section~\ref{sec:props} is qualitatively similar.)
We conservatively take all efficiency parameters to be unity along
the entire sequence, with two exceptions. First, we set
$\alpha_{acc} = 0.1$ above the period gap; this roughly accounts
for the anisotropic radiation and donor shielding produced by an
optically thick accretion disk. Second, we take $\alpha_{d} =
0.1$ below the gap, since most short-period systems are dwarf novae
with low duty cycles. With these choices, the net efficiency -- 
$\alpha_{acc} \alpha_{irr}\alpha_{d} = 0.1$ -- is constant along
the entire evolution track. For each model CV along this track, we
can now use Equation~\ref{eq:irrad1} to estimate $x_0$, calculate
$s(x_0)$ from the HR97 fitting formula (using coefficients appropriate
for the current set of $T_{eff}$, $\log{g}$ and $R_2/a$) and finally
use Equation~\ref{eq:irrad0} to determine the expected degree of donor
inflation. 

The results of this exercise are shown in Figure~\ref{fig:irrad}. We find
that, even with our fairly conservative set of efficiency factors,
irradiation-induced donor inflation is $<3\%$ along the entire
sequence, and $\ltappeq 1\%$ below the period gap. The corresponding
numbers for the optimal track in Section~\ref{sec:props} are $<2.5\%$
above the gap and $<2\%$ below. As discussed in
Section~\ref{sec:bloat}, we already explicitly allow for the possibility
that our model radii systematically underestimate real MS star radii by
$1.5$\% (for fully convective stars) or $4.9$\% (for stars with
radiative cores). Since irradiation-induced donor inflation is usually
smaller, and at most comparable, to this, yet is tricky to implement
self-consistently, we draw the line here and neglect it in our modeling.
\footnote{Implementing a self-consistent correction for 
irradiation-induced donor bloating is non-trivial, because the
strength of irradiation depends itself on the mass-loss rate from the
donor. Thus there is feedback between irradiation- and ML-induced
donor inflation, necessitating the use of an iterative computational
scheme.}
In any case, the
impact on our results of somewhat larger (or smaller) radius 
corrections -- regardless of their cause -- is discussed explicitly in 
Section~\ref{sec:discuss_bloat}. 

\subsection{Settling on the Form of $\dot{J}_{sys}$}
\label{sec:form}

In principle, we are now ready to construct our semi-empiricial CV
evolution track, but there is one last practical obstacle to
overcome. The problem is that the construction of 
self-consistent donor evolution sequences is computationally
expensive. Ideally, we would create large grids of self-consistent
sequences for a wide range of AML 
recipes and normalizations, and then directly fit all of these to the 
empirical mass-radius data. However, this is simply not feasible.

We therefore adopt the following procedure. First, as noted in
Section~\ref{sec:mb}, we consider only the
\citet[][hereafter RVJ83]{1983ApJ...275..713R} formulation for 
MB. With its adjustable normalization and power-law index $\gamma$
(see Equation~\ref{eq:jdot_rvj}), this formulation allows us to
efficiently cover the relevant parameter space in full. Second, we obtain
a first estimate of $\dot{M}_2(P_{orb})$ by interpolating across a set
of constant-$\dot{M}_2$ sequences to roughly match the observed donor
mass-radius relationship. As discussed in Section~\ref{sec:adjust}, this 
ignores the effect of mass-loss history and is therefore not
self-consistent. However, it does provide a good starting point for
detailed modelling. Third, we compare the resulting approximate
$\dot{M}_2(P_{orb})$ relation to those predicted by a range of
RVJ83-like  (Equation~\ref{eq:jdot_rvj}) with $0 \leq \gamma \leq 4$)
and GR-like  (Equation~\ref{eq:jdot_gr}) AML prescriptons. We then
finally identify those prescriptions that appear to best 
match the shape of the approximate $\dot{M}_2(P_{orb})$ relation above
and below the period gap. Based on these steps, we adopted an
RVJ83-based AML recipe with $\gamma = 3$ above the gap, and a GR-based
one (Equation~\ref{eq:jdot_gr}) below the gap in our self-consistent
models. It should be understood, however, that we do not consider the
exact shape of the AML recipes to be particularly well constrained
(see, for example, the discussion in Section~\ref{sec:reconcile}).

\subsection{Constructing the Final Track}
\label{sec:final}

With the AML prescription fixed, the final evolution track can be
constructed via the following steps. First, we calculate two small
grids of self-consistent evolution 
tracks. The set of normalization factors 
covered by the RVJ83-based grid for CVs above the gap was
$f_{MB} = 0.25, 0.5, 1.0, 
2.0, 4.0$; the corresponding set of normalization factors for the
GR-based grid for short-period CVs was $f_{GR} = 1.0, 2.0, 3.0, 4.0, 5.0$.
\footnote{Normalization factors below unity are unphysical, as GR-driven AML
must always be active. Also, Figure~\ref{fig:mb} already shows that AML
rates in excess of GR are likely needed below the gap.} Tracks for
other normalization factors are constructed via linear interpolation
on these grids.

The donor models used in creating these evolution sequences employ
the BCAH98 stellar physics, which still represents the state of the art
in modelling of low-mass stars. In a CV context, BCHA98-based donor
models have previously been used by 
\citet{1999MNRAS.309.1034K,2000NewAR..44...99K,2000MNRAS.318..354B,2001MNRAS.321..544K}, 
and we refer the reader to these papers for 
additional details. However, one point that does deserve explicit
mention is that the BCAH98 models do 
not include a treatment for atmospheric dust formation and
settling. Dust is expected to form in the atmospheres of very cool 
($T_{eff} \ltappeq 2500$~K), low-mass ($M_2 \ltappeq 0.1 M_{\odot}$
donors, and to eventually settle out again at even lower temperatures
($T_{eff} \ltappeq 1700$~K). However, these processes will primarily 
affect the SED of the donor, rather than its radius and response to
mass loss, and this can be taken into account by calculating
photometric properties from model atmosphere grids that do account for
dust.
\footnote{Of course, one hemisphere of the donor is also bathed
in UV and X-ray radiation from the vicinity of the WD primary, which
might inihibit dust formation in its atmosphere.}
We have checked that the radius
difference between the BCAH98 5-Gyr isochrone and an 
equivalent one that accounts for dust (based on the DUSTY00 models
described by \citealt{2000ApJ...542..464C}) is less
than 1\% down to $M_2 = 0.075M_{\odot}$. Thus our BCAH98-based
evolution sequences and associated donor properties should be quite 
reliable for essentially the entire parameter space covering
long-period and pre-bounce short-period CVs. It should nevertheless be
kept in mind that our BCAH98-based models will become less reliable as
we move deeply into the regime of sub-stellar secondaries and
post-period-minimum CVs.

Next, we apply the donor radius correction factors discussed in
Section~\ref{sec:skeletons}. For the secondaries in short-period CVs, this
means a $1.5\%$ upward correction to 
bring the BCAH98 model radii in line with observations of non-interacting 
stars, and a $\simeq 4.5\%$ upward correction to account for
tidal and rotational deformation. For donors in long-period CVs, the
corresponding corrections are $4.9\%$ and $4.5\% - 6.9\%$,
respectively. For comparison, we 
have also carried out calculations with smaller and larger correction
factors; the impact of the adopted correction factors on our
results will be discussed in Section~\ref{sec:discuss_bloat}. Note
also that, as 
discussed in Section~\ref{sec:deform}, the corrections for tidal and
rotation distortion are actually a function of donor mass and mass
ratio.

Since all radius corrections change the mean density of the donor,
we also need to adjust the orbital period of the 
system in order to retain consistency with the period-density
relation. Moreover, since our AML recipes are functions of $P_{orb}$
and $R_{2}$, changing these parameters also produces a slight
inconsistency between the mass- and angular-momentum-loss rates that
were actually used in the models ($\dot{J}_{sys}$ and $\dot{M}_2$) and the
values that these recipes would now produce, given the adjusted donor
radius and orbital period ($\dot{J}_{sys}^{\prime}$ and
$\dot{M}_2^{\prime}$). In order to get back 
to a mutually consistent set of parameters, we use an iterative
scheme. Thus we adjust the radius again, this time by the amount
expected from considering the ratio $\dot{M}_2^{\prime}$/$\dot{M}_2$.  We
then recalculate the orbital period and re-estimate the expected
angular-momentum-loss and mass-loss rates for the new set of system
parameters. This sequence of adjustments is repeated until mass,
radius, period, AML- and ML-rates are all mutually consistent again. 

Third, we fit the adjusted models to the empirical data. We do this 
via a $\chi^2$-minimization in the 
$P_{orb}-M_2$ plane. Since errors on orbital periods are usually much 
smaller than those on donor masses, we neglect them. However, we
do add an intrinsic dispersion term and allow this to affect both
$M_2$ and $P_{orb}$ isotropically. The goodness-of-fit statistic we
minimize is thus 
\begin{eqnarray} 
\chi^2 & = & \sum_{i=1}^N 
\frac{(\log{M_{2,i}} - \log{M_{2,mod}})^2}{\sigma_{\log{M_{2,i}}}^2 +
  \sigma_{int}^2} 
\nonumber
\\ & + &
\frac{(\log{P_{orb,i}} - \log{P_{orb,mod}})^2}{\sigma_{int}^2},
\label{eq:chi}
\end{eqnarray}
where $N$ is the number of data points, and $\sigma_{int}$ is adjusted
so as make $\chi^2 = N - 1$ (the only free parameter we fit is the
normalization of our adopted AML recipes). The $\chi^2$ associated with each
individual datum is evaluated at the point along the model sequence
that minimizes this $\chi^2$. Thus we can calculate meaningful
$\chi^2$ values even for sequences that do not extend to the lowest
observed orbital periods, for example. Errors on the normalization
factors are derived in the usual way, by considering the change in
normalization required to reduce the goodness-of-fit by $\Delta \chi^2 =
1$.

Up to this point, long-period and short-period systems have been
treated independently, since they are characterized by different AML
recipes. The fourth step in constructing the final evolution track is
therefore to merge the two sequences. In doing so, we include
the detached evolution through the period gap, under the assumption
that AML in the gap follows the same AML recipe and normalization as
derived for CVs below the gap. We do not, however, follow the
relaxation of the donor star back to its thermal equilibrium radius in
detail. In reality, this happens on a time scale of $\tau_{adj}$
(Equation~\ref{eq:tau_adj}), but we simply reset $R_2$ to its equilibrium MS
value immediately after the system has entered the gap. We have
checked, however, that $\tau_{adj}$ is significantly shorter than the
time it takes a CV to pass through the gap in our models, 

The fifth and final step is to calculate the photometric donor
properties along the evolution track. We do this by interpolating in
$T_{eff}$ and $\log{g}$ on the same set of stellar atmosphere grids
that were also used for the updated broken-power-law donor sequence
(Section~\ref{sec:update}). Thus we again adopt  the standard NextGen
atmosphere models \citep{1999ApJ...512..377H} 
down to $T_{eff} = 2300$~K, the AMES-DUSTY models 
\citep{2000ApJ...542..464C} between $1800~{\rm K} \leq T_{eff} \leq
2000~{\rm K}$, and the AMES-COND models \citep{2003A&A...402..701B} 
for temperatures $T_{eff} \leq 1700~{\rm K}$. The boundaries
between these regimes are bridged by interpolation. 

Physically, these model grids differ mainly in their treatment of 
dust
\footnote{They also employ different molecular line lists; see
\citet{2000ApJ...542..464C} and \citet{2003A&A...402..701B} for
details.}
:
the NexGen models assume that no dust formation has taken place
in the atmosphere; the DUSTY models allow for the formation of dust
grains and include their opacity; the COND models also include 
dust formation, but do not include its opacity. The assumption behind
the COND models is that rapid gravitational settling efficiently
removes dust from the atmosphere.

Our choices here are based partly on a comparison to the MS star data
of \citet{1999A&A...348..524B}, and partly on 
a desire to keep the photometric donor properties relatively
smooth along our model track. Thus the NextGen grid does
the best job of reproducing the observed location of the MS in the
$(I-K)~vs~M_K$ plane down to $I-K \simeq 4.5$, which corresponds to
about $T_{eff} \simeq 2300$~K. Moreover, there can be fairly
substantial photometric differences between the different grids at the
same temperature and surface gravity, so an abrupt switch at the
recommended temperatures would have introduced awkward discontinuities
into the model sequences. These considerations also imply, of course,
that the photometric donor properties below $T_{eff} \simeq 2300$~K 
(i.e. in systems near or beyond period bounce) should be treated with
considerable caution. The underlying problem is that the physics
governing the gradual formation and settling of atmospheric dust have
still not been comprehensively modelled, even in isolated MS
stars and brown dwarfs. Given that CV secondaries are additionally exposed to
anisotropic UV and X-ray radiation fields, it is quite unclear whether
and how one should account for dust formation and settling in their
atmospheres.

\section{A Complete, Semi-Empirical, Donor-Based CV Evolution Track} 
\label{sec:newtrack} 

\begin{figure*}
\centering
\includegraphics[height=22cm,angle=0]{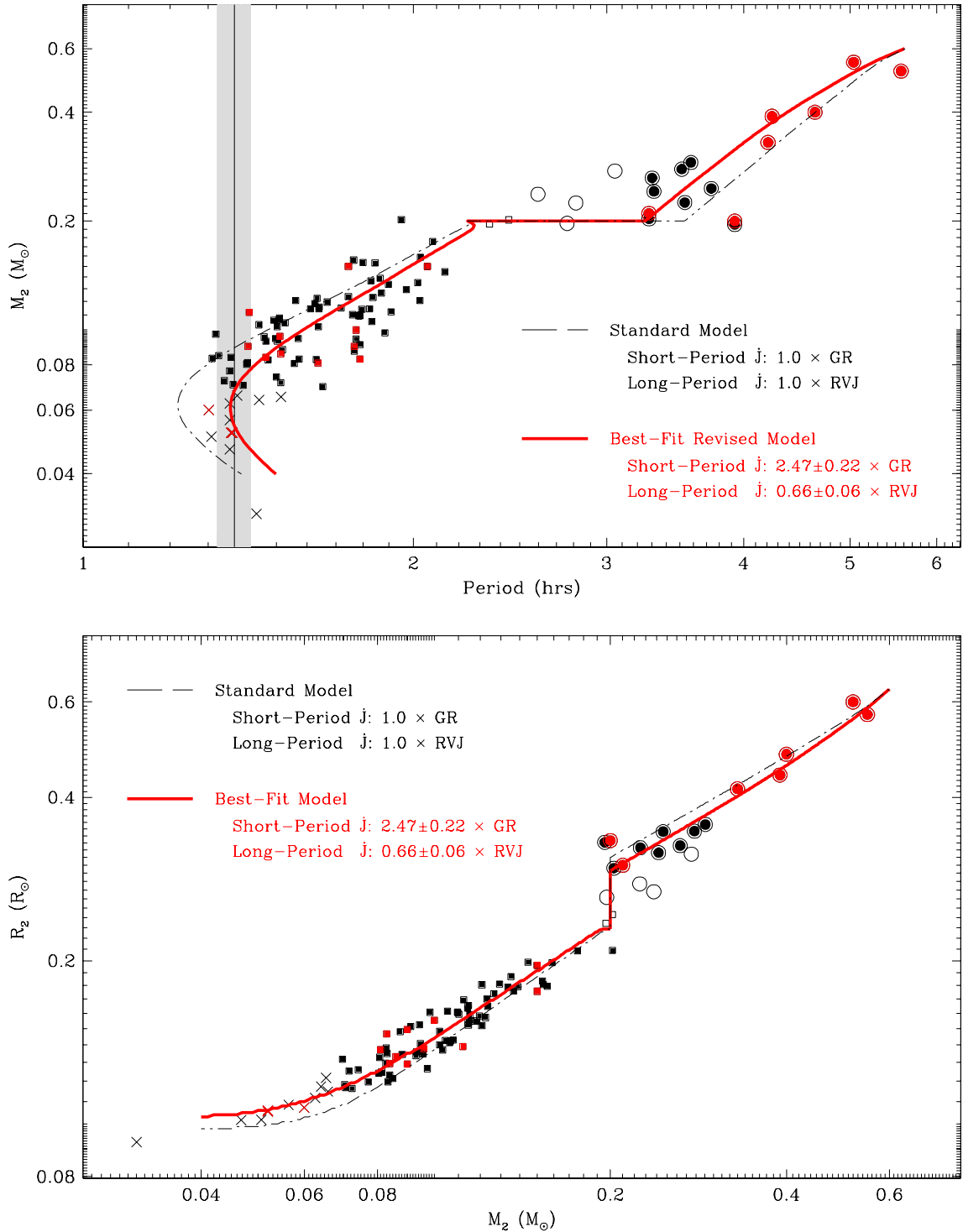}
\caption[Self-consistent model fits to the observed properties of CV
donors.]
{Self-consistent model fits to the observed properties of CV
donors in the 
period-mass (top panel) and mass-radius (bottom panel) planes. The
same symbols as in Figure~\ref{fig:brokenpower} are used for the data
points. Error bars are suppressed for clarity, but were used in the
fits. The black dash-dotted line shows the predicted evolution of
donor properties according to the standard model for CV
evolution, which is characterized by $f_{GR} = f_{MB} = 1$. The red
solid line shows the evolution of donor properties along the best-fit
model track. This revised model is characterized by $f_{GR} = 2.47 \pm
0.22$ (i.e. enhanced AML) below the gap and $f_{MB} = 0.66 \pm 0.05$
(i.e. slightly reduced AML) above. The vertical line in the top panel
marks the observd location of the period spike (which presumably
corresponds to $P_{min}$; the width of the shaded area around this
corresponds to the FWHM of the spike as measured in G09).}
\label{fig:fits}
\end{figure*}

The final result of this model construction and fitting process is shown in
Figure~\ref{fig:fits}. This represents the main result of our
study. The top panel shows the data and the best-fitting
model sequence in the $P_{orb}-M_2$ plane; the bottom panel shows the
same information in the $M_2-R_2$ plane. For comparison, we also show 
a classic ``standard model'' sequence, in which we adopt the same AML recipes as
for our best-fit model (GR below the gap, RVJ83 with $\gamma = 3$
above the gap), but fix the normalization of these recipes to
unity ($f_{GR} = f_{MB} = 1$). Note that we do apply the radius
correction factors discussed in Section~\ref{sec:skeletons} to the standard
sequence, so it is still an improvement on similar sequences published
previously.  

The normalization factors derived for our best-fit sequence are $2.47
\pm 0.22$ below the gap and $0.66 \pm 0.05$ above the gap.
\footnote{The errors quoted here are purely statistical and, in
particular, do not include the systematic uncertainties associated
with the corrections discussed in Section~\ref{sec:skeletons}.}
Thus our results imply that AML is somewhat stronger than pure GR in 
short-period CVs, but somewhat weaker than a typical MB formulation
in long-period CVs. No intrinsic dispersion at all was needed for
long-period systems (we obtained $\chi^2_\nu = 1.0$ with $\sigma_{int}
= 0$); for short-period CVs, a small intrinsic dispersion of
$\sigma_{int} = 0.02$~dex was sufficient to achieve $\chi^2_\nu =
1.0$. 

Figure~\ref{fig:fits} shows that our best-fit donor evolution sequence is a
substantial improvement on the standard model in terms 
of its match to the data, especially below the period gap. Here, the
standard model clearly underpredicts the donor radii at fixed mass
(or, equivalently, 
overpredicts their masses at fixed $P_{orb}$). It also predicts a
minimum period that is inconsistent with the period spike
(G09), whose location is marked by the vertical
line in the top panel, and also with the data shown in
Figure~\ref{fig:fits}. By contrast, the best-fit sequence describes
the data quite 
well. In particular, it predicts a minimum period that matches the
location of the period spike (see also Section~\ref{sec:pcrit} below). This
was not guaranteed, since the value of $P_{min}$ is not imposed as a
boundary condition in the construction of the best-fit evolution
sequence.

In order to make our calculations as useful as possible to the
community, we provide comprehensive listings of all relevant physical
and photometric binary parameters along both the standard model track
and along our new best-fit evolution sequence in
Tables~\ref{tab:seq_phys_stan} - \ref{tab:seq_donor_rev}

As a supplement to our donor-based evolution tracks, we have also
calculated the corresponding physical properties of the WD
primary. For this, we first estimated the effective temperature of the
$0.75$~$M_{\odot}$ WD assumed in our models by interpolating on the
theoretical $\dot{M}_{acc}$ vs $T_{eff,1}$ calibration for $0.6$~$M_{\odot}$
and $1.0$~$M_{\odot}$ WDs shown in Figure~10 of \citet{2004ApJ...600..390T}.
\footnote{Note, however, that the \citet{2004ApJ...600..390T} 
calculations do not account for the dependence of the WD radius,
$R_{1}$, on
$T_{eff,1}$.} 
As discussed in their paper (and shown in their results), even
for fixed mass and accretion rate, the predicted $T_{eff,1}$ depends
somewhat on the mass of the accreted layer on the surface of the
WD, $\Delta M_{acc}$. Following \citet{2004ApJ...600..390T}, 
we have therefore calculated three sets of WD temperatures for
each model track, corresponding to $\Delta M_{acc} \simeq (0.05, 0.5,
0.95) M_{ign}$, where $M_{ign}(M_{1}, \dot{M}_{acc})$ is the ignition mass
required to trigger a nova eruption. Our default set with $\Delta
M_{acc} \simeq 0.5 M_{ign}$ corresponds to systems about halfway
between nova eruptions, while the other two sets represent the upper
and lower limits corresponding to system just before or after a nova
outburst.
\footnote{The heating of the WD by the thermonuclear burning it
undergoes during a nova outburst is not taken into account,
however. See TG09 for a discussion of this effect.}
Given $M_{1}$ and $T_{eff,1}$, we then estimate $R_{1}$ and
all photometric properties by interpolating on the
\citet{2006AJ....132.1221H} grid of WD models.
\footnote{http://www.astro.umontreal.ca/~bergeron/CoolingModels/}
The predicted properties of the accreting WD are listed as a function
of $P_{orb}$ in Tables~\ref{tab:seq_wd_stan} and~\ref{tab:seq_wd_rev}.

In the following sections, we will take a closer look at the detailed
properties of our revised, donor-based CV evolution track. Where
appropriate, we will also compare it to both the broken-power-law
donor sequence derived in Section~\ref{sec:update} and a conventional,
``standard model'' evolution track (i.e. $f_{GR} = f_{MB} = 1$).

\subsection{Physical and Photometric Properties}
\label{sec:props}

\begin{figure*}
\includegraphics[height=17cm,angle=0]{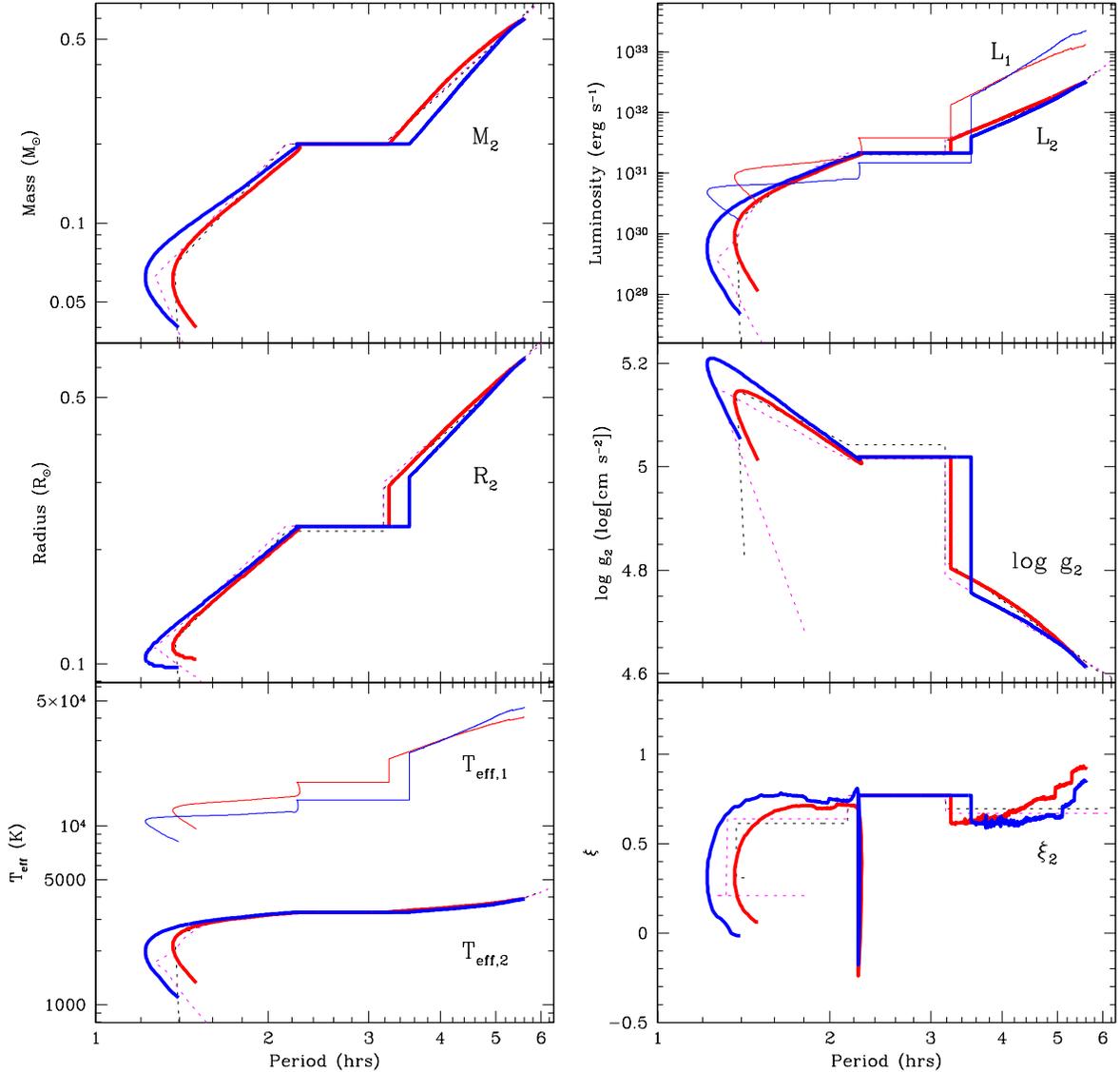}
\caption[The run of physical donor properties as a function of orbital
period along standard and revised evolution tracks.]
{The run of physical donor properties as a function of orbital
period along standard and revised evolution tracks. The left-hand
column of panels shows, from top to bottom, the donor mass, radius and
effective temperature. The right-hand column of panels shows, from top
to bottom, the donor luminosity, surface gravity and effective
mass-radius index. In each panel, the thick blue solid line is the track
corresponding to the standard model ($f_{GR} = f_{MB} = 1$) while the
thick red solid line is the evolution sequence predicted by the revised
model ($f_{GR} = 2.47$; $f_{MB} = 0.66$). In each panel, we also show
the original (thin magenta dashed line) and  updated (thin black
dashed line) broken-power-law donor sequences (see
Section~\ref{sec:update}). For comparison, the (long-term average) temperature
and luminosity predicted for the accreting WD in our evolution
sequences are also plotted in the relevant panels (thin solid blue
line: standard model; thin solid red line: revised model).} 
\label{fig:phys}
\end{figure*}

Figure~\ref{fig:phys} shows the physical properties of the donor star as a
function of orbital period along the standard model track (thick blue
line) and along our new best-fit model track (thick red line). For
comparison, we also show the physical donor properties along both the
original (thin black dashed line) and updated (thin magenta dashed
line) broken-power-law donor sequences. The donor properties along
both broken-power-law sequences are generally very similar to those
along our new best-fit model, except in the poorly constrained
period-bounce regime. This is reassuring, since all of these sequences
are based on almost the same set of donor mass-radius data.  

The physical differences between the standard and best-fit models are
also easily understood. Above the gap, AML is stronger in the standard
model, resulting in a larger donor radius, lower surface gravity,
lower luminosity and larger mass-radius index. Moreover, due to the 
period-density relation, the mass at fixed period is higher along the
standard track, and the effective temperature slightly lower (this is
hard to see on the scale shown in the Figure~\ref{fig:phys}). Below the gap, the
situation is exactly reversed. In particular, the higher mass-transfer
rate in the best-fit model causes the mass-radius index to drop to 1/3
at a considerably longer orbital period than in the standard model (82
min vs 73 min), although at nearly the same donor mass ($M_{bounce} = 0.061
M_{\odot}$ vs 0.062 $M_{\odot}$, respectively). It is worth noting
that, even in our version of the standard model, the minimum period is
considerably longer than that found in most previous
investigations. This is because of 
our corrections for tidal/rotational deformation and for the radius
offset between stellar models and non-interacting stars, which is probably 
related to magnetic activity (see Section~\ref{sec:skeletons}). Below
the gap, these upward radius corrections amount to about 4.5\% and
1.5\%. A a result of the period-density relation, the combined 6\%
upward radius correction produces a $\simeq$9\% upward shift in the
location of period bounce.

Figure~\ref{fig:phys} also shows the temperature and luminosity of the
accreting WD as a function of $P_{orb}$ along the model tracks. Since 
$M_{1}$ is constant along our tracks (by assumption), other WD
parameters do not evolve appreciably and are therefore not
plotted. It is interesting to note that the bolometric luminosity of
the accreting WD dominates over that of the donor star in essentially
all model systems. The only exception occurs in the standard model
sequence, where there is a small region just below the period gap
where the secondary's bolometric output exceeeds that of the WD.

\begin{figure*}
\centering
\includegraphics[height=17cm,angle=0]{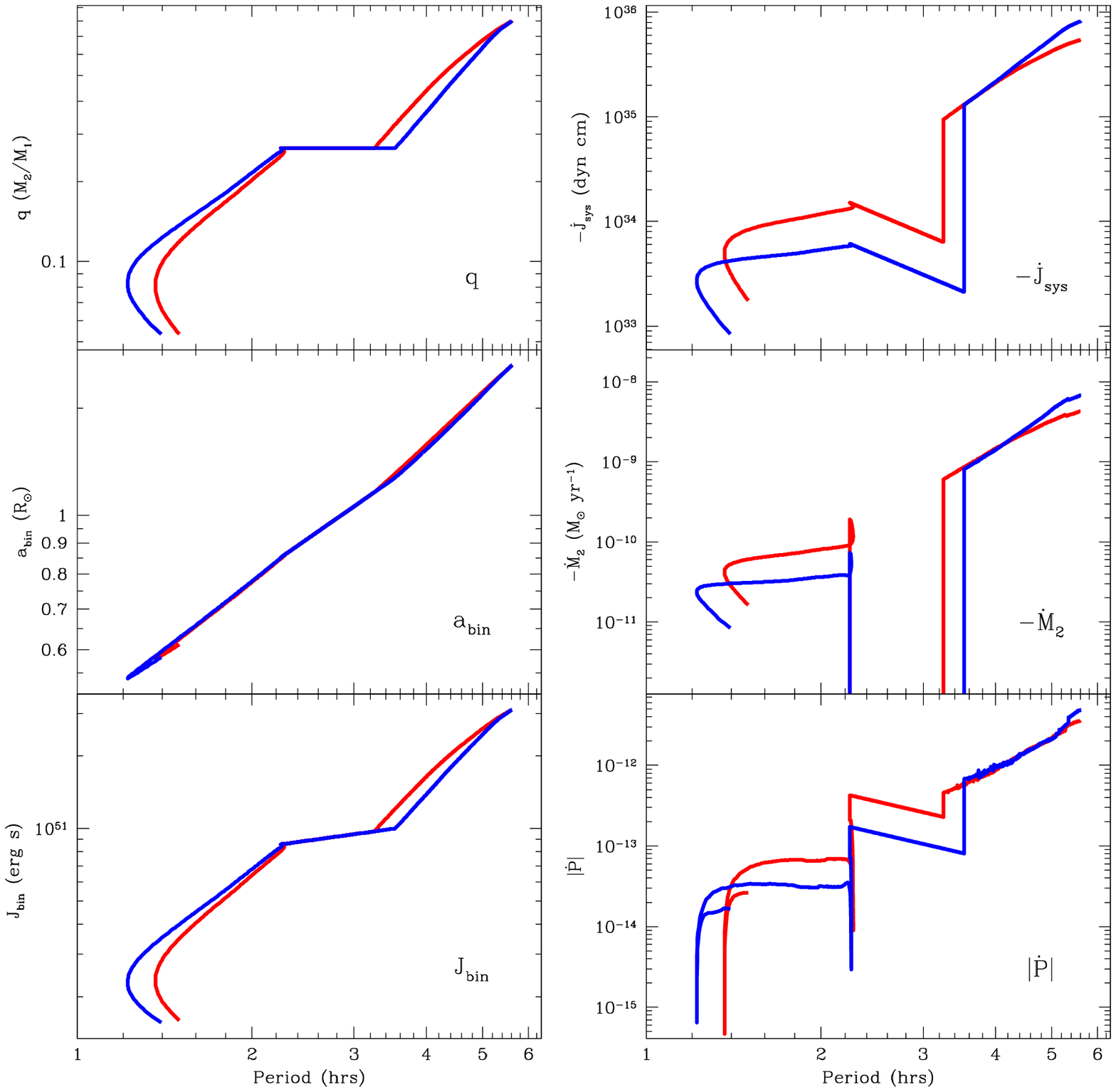}
\caption[The evolution of key binary parameters as a function of
orbital period along our self-consistent model tracks.]
{The evolution of key binary parameters as a function of
orbital period along our self-consistent model tracks. In all panels,
the thick blue line is the standard model ($f_{GR} = f_{MB} = 1$),
while the  thick red line is the revised best-fit model ($f_{GR} =
2.47$; $f_{MB} = 0.66$). The left-hand column of panels shows, from top
to bottom, the evolution of the mass ratio, binary separation and
orbital angular momentum. The right-hand column of panels shows, from
top to bottom, the evolution of the systemic angular-momentum-loss
rate, the mass-transfer rate and the orbital period derivative. }
\label{fig:evo1}
\end{figure*}

\begin{figure*}
\centering
\includegraphics[height=17cm,angle=0]{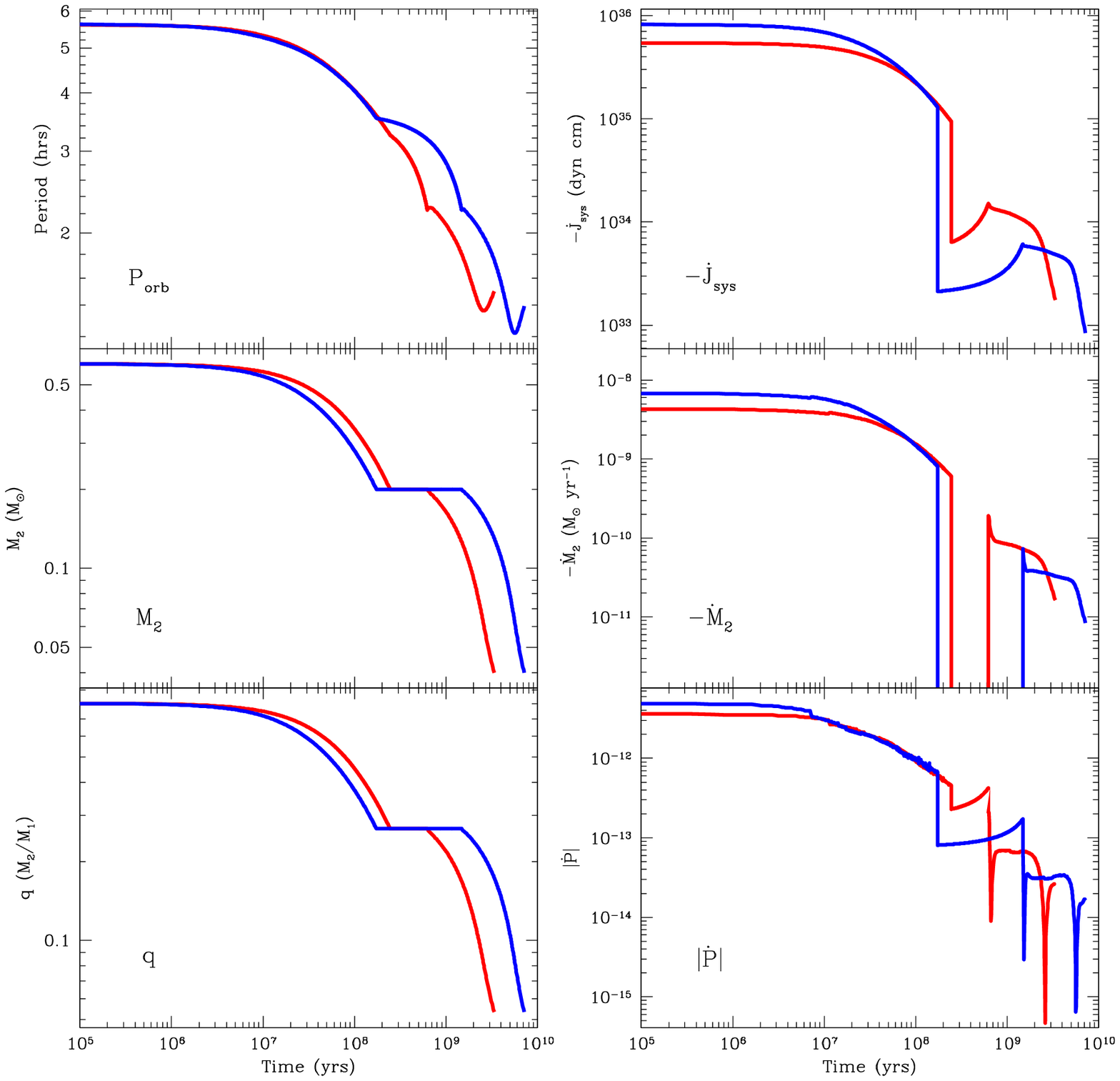}
\caption[The evolution of key binary parameters as a function of
time along our self-consistent model tracks.]
{The evolution of key binary parameters as a function of
time along our self-consistent model tracks. In all panels,
the thick blue line is the standard model ($f_{GR} = f_{MB} = 1$),
while the  thick red line is the revised best-fit model ($f_{GR} =
2.47$; $f_{MB} = 0.66$). The left-hand column of panels shows, from top
to bottom, the evolution of the orbital period, the donor mass and the
mass ratio. The right-hand column of panels shows, from
top to bottom, the evolution of the systemic angular-momentum-loss
rate, the mass-transfer rate and the orbital period derivative.}
\label{fig:evo2}
\end{figure*}

The evolution of the {\em binary} parameters is shown as a function of
orbital period in Figure~\ref{fig:evo1} and as a function of time in
Figure~\ref{fig:evo2}. Again, we show both the standard model 
and our best-fit donor-based evolution track. As expected, the
evolution is initially faster in the standard model (since $f_{MB} =
0.65 < 1$ above the gap in the best-fit model), but this situation is
quickly reversed below the gap, where the AML rate of the
best-fit model is substantially higher ($f_{GR} = 2.47$) than that of
the standard model ($f_{GR} = 1$). In the standard model, the upper
edge of the period gap is reached after about $1.7\times 10^8$ years,
while in the best-fit model, systems enter the gap after about
$2.4\times 10^8$. Here and throughout, $t = 0$ refers to a system
reaching contact at a donor mass of $M_2 = 0.6 M_{\odot}$. Evolution through the
gap takes about 1.3 Gyrs in the standard model, but only about 0.4
Gyrs in the best-fit model, due to the higher AML rate in the
latter. Period bounce is reached after about 5.7 Gyrs (standard model) or 2.6
Gyrs (best-fit model). 

These time scale differences are important, because they directly
affect the 
number of systems we may expect to find in the different phases of
CV evolution. For example, we have seen that CVs evolve faster below
the period gap in the best-fit model and reach period bounce 
sooner. Thus CVs evolving according to our best-fit model spend a
smaller fraction of their life as short-period, pre-bounce CVs, but a
larger fraction as period bouncers. Correspondingly, we may expect
(and Section~\ref{sec:pdist} indeed confirms) that the best-fit model
predicts a lower percentage of short-period,
pre-bounce CVs among the Galactic CV population, but a higher
percentage of period bouncers. This change helps to reconcile some of
the long-standing clashes between the standard model and the
statistics of observational CV samples. 

\begin{figure*}
\centering
\includegraphics[height=17cm,angle=0]{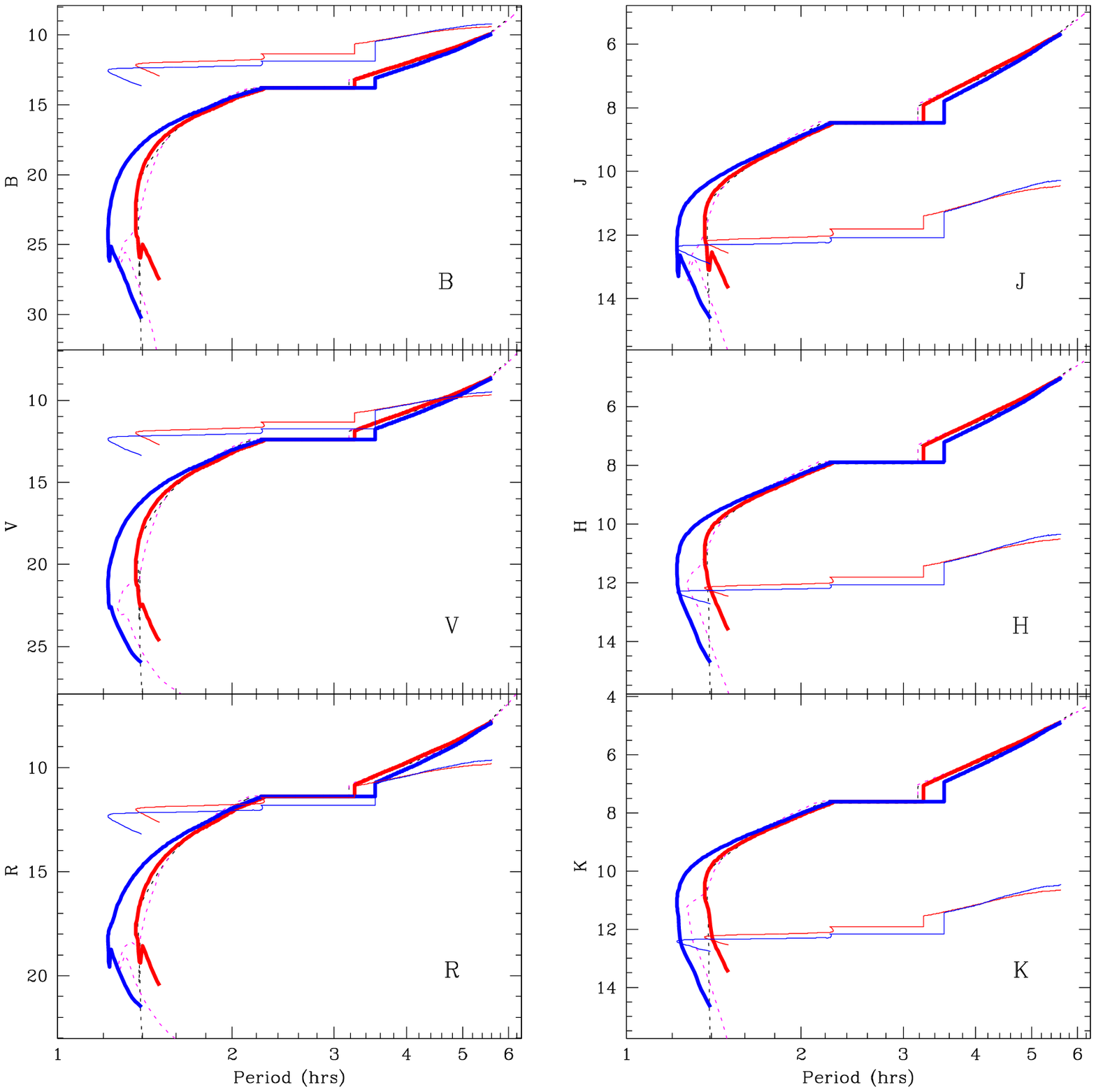}
\caption[The evolution of the absolute magnitudes of both binary
components as a function of orbital period along our self-consistent
evolution tracks.]
{The evolution of the absolute magnitudes of both binary
components as a function of orbital period along our self-consistent
evolution tracks. The left-hand column of panels shows three optical
bands (from top to bottom: $B$, $V$ and $R$), while the right-hand
column of panels shows three near-infrared bands (from top to bottom:
$J$, $H$ and $K$).  The thick solid blue and red lines represent the 
absolute magnitudes of the donor stars along the standard 
($f_{GR} = f_{MB} = 1$) and revised best-fit ($f_{GR} = 2.47$; $f_{MB}
= 0.66$) evolution sequence, respectively. The thin solid blue and red
lines shows the corresponding absolute magnitues of the accreting
WDs. In each panel, we also plot tracks corresponding to  
the original (thin magenta dashed line) and  updated (thin black
dashed line) broken-power-law donor sequences (see Section~\ref{sec:update}).}
\label{fig:phot}
\end{figure*}

Finally, in Figure~\ref{fig:phot}, we show the run of optical
($B$,$V$,$R$) and near-infrared ($J$,$H$,$K$) absolute magnitudes for
both the donor and the accreting WD along our self-consistent
evolution sequences. These plots confirm several pieces of
conventional wisdom. For example, only unusually bright
(e.g. nuclear-evolved) donors are expected to be easily detectable at
optical wavelengths, 
and particularly in the blue. Even the accreting WD competes with --
or even outshines -- the secondary at these wavelengths. In the
infrared, the situation is considerably more favourable, but a key
point to note is the steep drop-off in donor brightness near
$P_{min}$ in all bands. Around $P_{min}$, the WD begins to compete
with the donor even in the near-infrared. Given that the accretion
disk is likely to be considerably brighter than the WD in both the
optical and near-infrared, this highlights the observational challenge
facing us if we wish to directly detect and characterize the
sub-stellar donors in period bouncers.

\subsection{The Locations of Period Bounce and Period Gap}
\label{sec:pcrit}

\begin{figure*}
\centering
\includegraphics[height=21cm,angle=0]{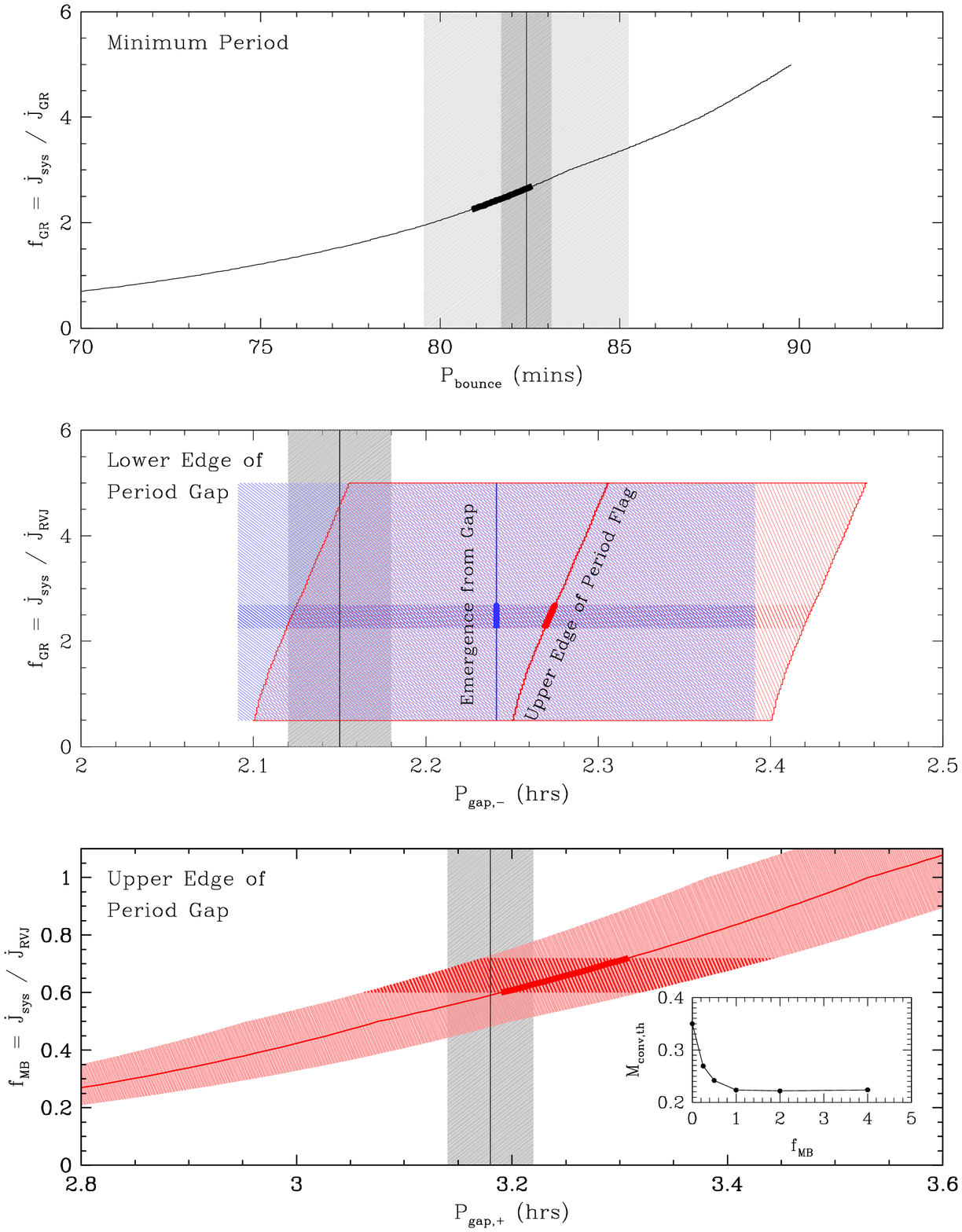}
\caption[Predicted and observed locations of the critical periods along a
CV evolution track.]
{Predicted and observed locations of the critical periods along a
CV evolution track. 
{\em Top panel:} The predicted location of the
minimum period as a function of $f_{GR}$, the angular-momentum-loss rate
relative to GR. The thick part of the curve marks the 1$\sigma$ confidence
interval around our best-fit value of $f_{GR} = 2.47 \pm 0.22$. The vertical
line is the observed location of the period spike $P_{min} = 82.4 \pm 0.7$~min
(G09). The vertical shaded regions mark the error on $P_{min}$ (dark
grey) and the intrinsic FWHM (5.7~min) of the spike (light grey). 
{\em Middle panel:} The predicted location of the lower edge of the
period gap as a function of $f_{GR}$. The blue line corresponds to the
period at which the system emerges from the period gap and
re-establishes contact; the red line is the upper edge of the period
flag, i.e. the longest period reached before evolution proceeds to
shorter periods again (see text for details). The thick portions on
both lines correspond to the 1$\sigma$ confidence interval around our
best-fit value of $f_{GR} = 2.47 \pm 0.22$. The blue- and red- shaded regions
extending left and right of these curves show the uncertainty on the
location of the lower gap edge arising from uncertainty in the donor
mass corresponding to the period gap ($M_{conv} = 0.20 \pm 0.02~
M_{\odot}$). The vertical black line and grey shaded region mark the
observed location of $P_{gap,-}$ and the error on this ($P_{gap,-} =
2.15 \pm 0.03$~hrs; K06). 
{\em Bottom panel:} The red line shows the predicted location of the
upper gap edge, $P_{gap,+}$, as a function of $f_{MB}$, the angular
momentum loss rate relative to the standard RVJ recipe with $\gamma =
3$. The thick part of the line marks the 1$\sigma$ confidence interval around our
best-fit value of $f_{MB} = 0.66 \pm 0.05$. The red-shaded region
extending left and right of this line again shows the uncertainty on the
location of the upper gap edge arising from uncertainty in the donor
mass corresponding to the period gap. The vertical black line and grey
shaded region mark the observed location of $P_{gap,+}$ and the error on this ($P_{gap,-} =
3.18 \pm 0.04$~hrs; K06). The inset in the bottom panel shows the mass
at which our stellar models formally become fully convective, as a
function of $f_{MB}$.}
\label{fig:extremep}
\end{figure*}

Figure~\ref{fig:extremep} presents a closer look at the locations of
the three critical orbital periods that delineate the life of a CV:
the minimum period, as well as the upper and lower edges of the period
gap. 

In the top panel, we show how the predicted location of $P_{min}$
depends on $f_{GR}$. The thick part of the curve marks the 1$\sigma$
confidence interval around our best-fit value of $f_{GR} = 2.47$. The
vertical line shows the observed location of the period spike, as
determined by G09. Their estimate of $P_{min}$
was determined by fitting a Gaussian to the spike, and the vertical
grey bands in the top panel mark their corresponding estimate of the
error on $P_{min}$ and the FWHM of the spike. 

The run of $P_{min}$ with $f_{GR}$ in the top panel confirms that, in
order to match the observed location 
of $P_{min}$, it is necessary to invoke AML rates in 
excess of GR below the period gap. More interestingly, it shows that
our best-fit model predicts a value for $P_{min}$ (81.8 $\pm$ 0.9
min) that is in excellent, quantitative agreement with the observed
location of the period spike (82.4 $\pm$ 0.7 min). Note that this
success was not guaranteed. Unlike for our
broken-power-law donor sequence, we do not impose $P_{min}$ as an
external constraint when fitting the data. Thus the predicted $P_{min}$
emerges simply from the fit to the mass-radius data shown in
Figure~\ref{fig:fits} -- a data set that is very different from the
SDSS sample used by G09.

Turning to the period gap, let us take the upper edge first (bottom
panel in Figure~\ref{fig:extremep}). In our models, we have assumed that 
systems enter the period gap when their donor mass reaches the 
empirically determined values of $M_{conv} = 0.20 \pm 0.02 M_{\odot}$
(K06). The notation ``conv'' here reflects the conventional notion
that the period gap is caused by a disruption of MB associated with
the transition of the donor star to a fully convective
structure. However, this empirically determined value for $M_{conv}$
does not have to correspond to the actual mass at which the models
predict a transition to a fully convective structure. In fact, the
inset in the bottom panel shows how this theoretically predicted
value, $M_{conv,th}$, varies with $f_{MB}$. Starting from the MS value
of $M_{conv,th} \simeq 0.35 M_{\odot}$ at $f_{MB} = 0$, $M_{conv,th}$
initially drops steeply with 
increasing $f_{MB}$, but then asymptotes around $M_{conv,th} \simeq
0.22 M_{conv}$ beyond $f_{MB} \simeq 1$. Based on a simple linear
interpolation, the value of $M_{conv,th}$ corresponding to our
best-fit estimate of $f_{MB} = 0.65 \pm 0.05$ is $M_{convm,th} = 0.236
\pm 0.002 M_{\odot}$. We conclude that the empirically determined
estimate of $M_{conv}$ is in reasonable, though by no means perfect
agreement with the idea that AML in CVs is suppressed exactly (and
instantaneously) when the donor star becomes fully convective. We also 
note that {\em if} this scenario is taken seriously, and {\em if} we
are willing to trust the stellar models in this respect, then
extremely low values of $f_{MB} \ltappeq 0.25$ would be ruled
out. However, these are both rather big ``ifs'', so we will continue
to adopt the empirically determined $M_{conv}$ in the construction of 
our model sequences.

Returning to the main plot in the bottom panel of 
Figure~\ref{fig:extremep}, the red line shows the predicted location 
of the upper gap edge as a function of $f_{MB}$. The thick part of
the curve marks the 1$\sigma$ range around our best-fit value of
$f_{MB} = 0.65 \pm 
0.05$. The red shaded region around this curve shows the uncertainty
associated with the $0.02 M_{\odot}$ error on $M_{conv}$. The 
best-fit model predicts $P_{gap,+} = 3.24 \pm 0.05$~hrs, which is an
excellent match to the observed location of the gap edge. By contrast,
the standard model -- which corresponds to the point on the curve
where $f_{MB} = 1$ --  predicts $P_{gap,+} = 3.52$~hrs, significantly
longer than observed.

We finally turn to the middle panel of Figure~\ref{fig:extremep},
which compares predicted and observed locations of the lower edge of
the gap. Here, we actually show two theoretically predicted
locations. The blue vertical line shows the period at which
the system is predicted to emerge from the period gap,
$P_{emerge}$. Since the 
donor is just a MS star in thermal equilibrium at this point, $P_{emerge}$
is independent of $f_{GR}$ or $f_{MB}$ -- it only depends on
$M_{conv}$ and the MS mass-radius relationship. The blue-shaded region
shows the uncertainty in the predicted location associated with the
error on $M_{conv}$. 

By contrast, the red curve and red shaded region show the maximum
extent of the so-called {\em period flag}.  
The period flag is a small ``loop'' in the $P_{orb}$ vs $\dot{M}_2$
plane that is executed by CVs as they emerge below the gap, immediately
after the turn-on of mass transfer and before settling on the standard
evolution track that will ultimately take them to $P_{min}$. The
period flag is clearly visible in the middle panel of
Figure~\ref{fig:evo1}, and the physics responsible for it have been
discussed in detail by \citet{1992A&A...259..159R} and
\citet{1996MNRAS.279..581S}. 

Briefly, the period flag arises because the mass-transfer
rate initially rises exponentially when contact between Roche lobe and
secondary is re-established
below the gap. Thus there is a brief window in which
significant mass loss has already begun, but the secondary has not yet
had 
time to relax thermally. During this window, the mass loss is
effectively adiabatic, and the response of the fully convective
donor to adiabatic mass loss is to swell up ($\zeta_{ad} \simeq
-1/3$). Following the discussion in Section~\ref{sec:bounce}, this means
that the orbital period must initially {\em increase}. However, the
secondary then relaxes thermally, causing $\zeta$ to increase and 
ultimately settle just below the MS value ($\zeta \simeq 0.6 -
0.7$). Along the way, $\zeta$ must pass through the critical value of
$1/3$. As discussed in Section~\ref{sec:bounce}, this value separates
evolution towards long periods from evolution 
towards short periods. The period at which $\zeta = 1/3$ is reached,
$P_{flag}$, therefore marks the upper edge of the period flag. 
All of these aspects of the donor response to the re-establishment of
mass loss are nicely visible in the bottom right panel of
Figure~\ref{fig:phys}.

In the middle panel of Figure~\ref{fig:extremep}, we show the
predicted $P_{flag}$ as a function of $f_{MB}$. Unlike $P_{emerge}$,
$P_{flag}$ {\em does} depend on the 
strength of the ongoing mass loss. More specifically,
Figure~\ref{fig:extremep} 
shows that $P_{flag}$ increases with $f_{MB}$. This is in line with
the findings of \citet{1996MNRAS.279..581S}. In any case, the main
conclusion to draw from this panel is simply that the bottom edge of
the period gap is in reasonable agreement with theoretical
predictions. However, since $P_{emerge}$, in particular, is completely
insensitive to mass loss, the level of agreement with $P_{gap,-}$ does
not provide a powerful way of distinguishing between different models.

\subsection{Donor Spectral Types}
\label{sec:spts}

As discussed in Section~\ref{sec:m_vs_t}, the effective temperature of
low-mass stars with a large convective envelope is virtually
unaffected by mass 
loss, so the $T_{eff,2}-M_2$ relationship of CV donors is the same as
for ordinary MS stars (at least to the point where the donors become
sub-stellar). Since the spectral type ($SpT$) of a star is essentially a direct
measure of its effective temperature, the $SpT_2-M_2$ relationship of CV
donors is also the same as for MS stars. However, mass loss does drive CV
donors slightly out of thermal equilibrium and causes them to be
bloated compared to MS stars of identical mass. As a consequence of
the period-density relationship (Equation~\ref{eq:dense}), a bloated,
mass-losing star will be found at longer $P_{orb}$ than an equivalent
Roche-lobe-filling MS star in thermal equilibrium. Thus, at fixed
period, CV donors should have lower mass, and hence later $T_{eff,2}$
and later $SpT$, than Roche-lobe-filling pure MS stars (see, for
example, \citealt{2000MNRAS.318..354B})

This theoretical expectation was initially verified for CVs above the
gap by \citet{1998A&A...339..518B} 
and later for both short- and
long-period CVs by K06. In K06, we also showed that our original
broken-power-law donor sequence was a good match to the empirical 
$P_{orb}-SpT$ relation for CV donors. Thus the degree of donor
inflation represented by that sequence was just right to produce the
observed later-than-MS $SpT$ at a given period.

\begin{figure*}
\centering
\includegraphics[height=15cm,angle=0]{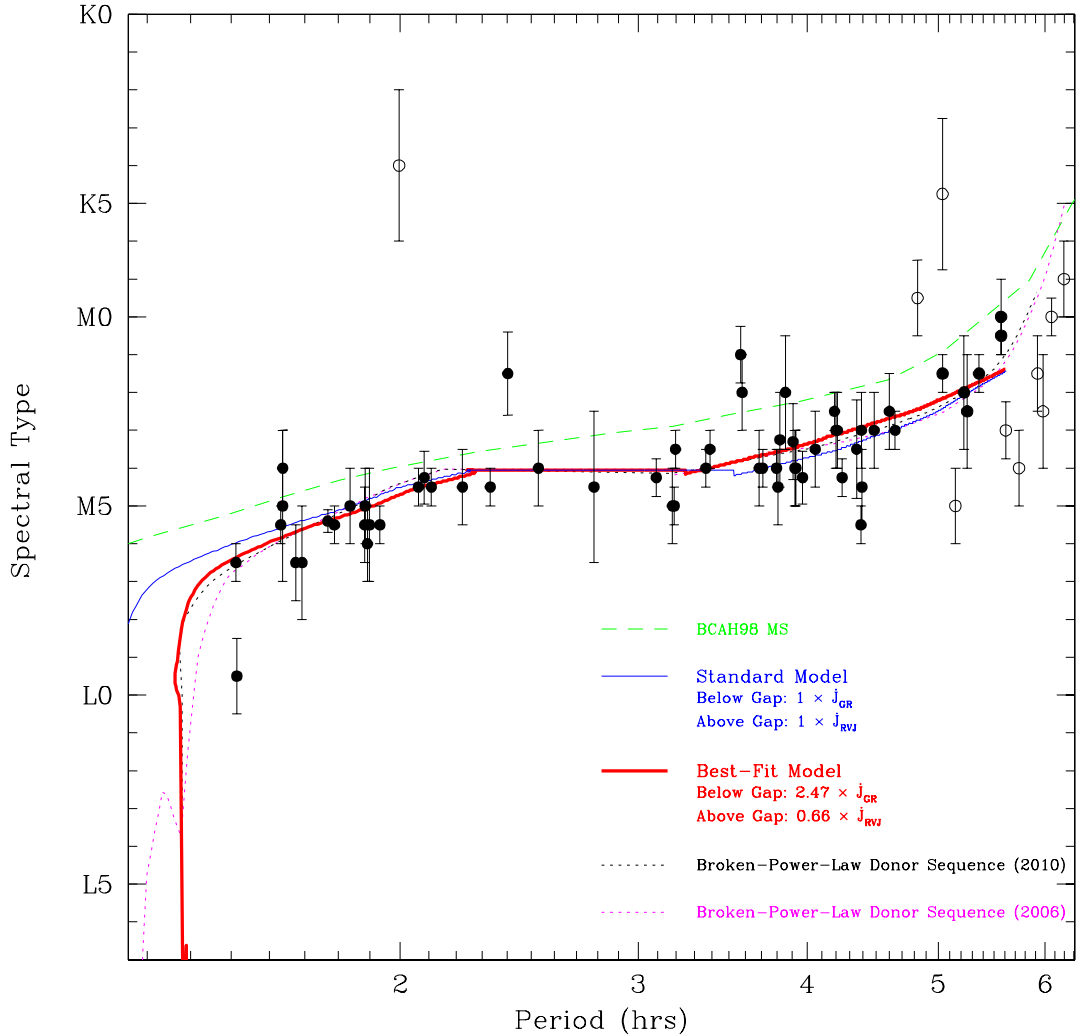}
\caption[Observed vs. predicted spectral types of CV donors as a
function of orbital period.]
{Observed vs. predicted spectral types of CV donors as a
function of orbital period. Points
correspond to empirically determined $SpT$s for CVs with \protect
$P_{orb} \protect \ltappeq  6$~hrs from K06 (with a few additions
noted in the text). Likely evolved donors are shown as open
symbols. The green dashed line shows the evolution of $SpT$ along a
standard MS BCAH98 5-Gyr track, assuming marginal contact, i.e. not
allowing for any thermal disequilibrium (see K06 for details). The
thin solid blue line shows the predicted $SpT-P_{orb}$ relation for
the standard model, the thick solid red line shows the same for the
revised (best-fit) model. For comparison, the thin dotted black
(magenta) line shows the $SpT-P_{orb}$ relation for our updated
(original; K06) broken-power-law donor sequence.}  
\label{fig:spt}
\end{figure*}

In Figure~\ref{fig:spt}, we now compare the standard model, the
best-fit model, as well as both original and updated broken-power-law
sequences to the observed $P_{orb}-SpT$ relation for CV donors. The
observational data here is the same as in K06, except that we have
include some additional long-period CVs from \citet{2010arXiv1009.1265T}. 
The calibration of $SpT$ based on $V-K$ photometric
colours is the same as in K06. The figure shows that all theoretical
tracks -- including 
the standard model -- do a reasonable job of matching these
observations. This is not unexpected, since all of these tracks
produce significant amounts of donor bloating. In fact, all but the
standard model are constrained by essentially the same donor
mass-radius data set, so they predictably produce almost identical
amounts of donor inflation. The donor bloating predicted by the
standard model is slightly different -- stronger above the gap, weaker
below; see Figure~\ref{fig:phys} -- but this difference is not large enough
to produce a significant difference in the predicted $SpT$s (at least
not in the period range for which we have data). 

Nevertheless, the observed $P_{orb}-SpT$ relation does confirm yet
again that CV donors are significantly bloated relative to MS
stars. The good agreement between predictions and observations in the 
$P_{orb}-SpT$ plane is therefore reassuring, even if the spectral type
data is not as powerful as the observed $M_2-R_2$ relation in
distinguishing between competing models.

\subsection{White Dwarf Temperatures}
\label{sec:WD}

\begin{figure*}
\centering
\includegraphics[height=17cm,angle=0]{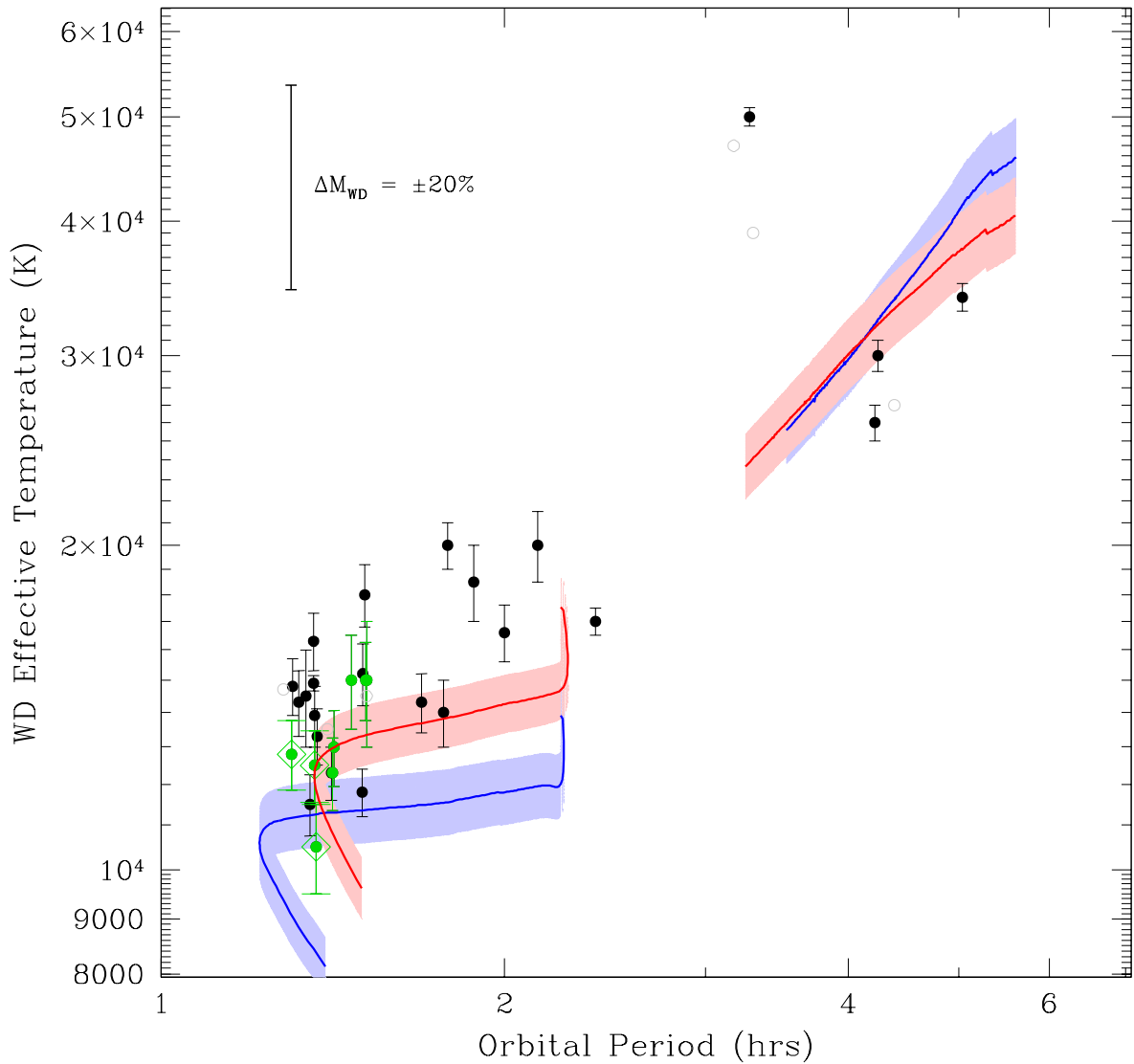}
\caption[Predicted and observed WD temperatures as a function
of orbital period.]
{Predicted and observed WD temperatures as a function
of orbital period. The data are mostly from the compilation by TG09,
but a few additional points have been added from L08. The latter 
are shown in green; candidate period bouncers with
sub-stellar secondaries among these are additionally marked by open
diamonds. Points without error bars are more uncertain and are plotted as
open circles. The thick solid blue line shows the long-term average
WD temperatures predicted by the standard model ($f_{GR} = f_{MB} =
1$). The thick solid red line shows the same for the revised
(best-fit) model ($f_{GR} = 2.47$; $f_{MB} = 0.66$). The shaded
regions around both curves show the uncertainty in $T_{eff,1}$
associated with the mass of the non-degenerate layer (or, equivalently,
with the time since the last nova eruption). The error bar in the top
left of the figure shows the effect of a 20\% change in $M_{1}$ on
the predicted temperatures.}
\label{fig:WD}
\end{figure*}

A final empirical check on our evolution track is presented in
Figure~\ref{fig:WD}. Here, we compare a set of observed WD temperatures to
the values predicted by both the standard model and our best-fit
alternative. The observational data are taken mainly from the
compilation provided by TG09, 
but supplemented by a few additional estimates from L08. 
Note that all of the observational estimates for dwarf
novae are obtained from observations in quiescence, when the WD
temperatures should best track the long-term-average accretion rate.

Taken at face value, Figure~\ref{fig:WD} suggests that neither the
standard model, nor our best-fit, donor-based one, provides a
particularly good match to the observed WD temperatures. The best-fit
model clearly lies closer to the data, but it still systematically
underpredicts the 
WD temperatures in short-period, pre-bounce CVs. The simplest 
interpretation of this result is that the WD temperatures point to 
an even stronger AML enhancement over GR than is allowed for in our
best-fit model. This apparent preference for faster-than-GR
mass transfer is in line with the findings of TG09.

However, as also noted by TG09, considerable care needs to be taken 
before accepting such a conclusion. First, and most fundamentally,
there is the issue already noted in Section~\ref{sec:intro}: even though
quiescent WD temperatures are tracers of $\dot{M}_{acc}$ averaged over
$10^3$~yrs - $10^4$~yrs, this time scale is short compared to the
time scale on which mass-transfer-rate fluctuations
associated with the donor star are expected to operate. Thus
$T_{eff,1}$ may simply not trace the secular mass-transfer rate
faithfully. This probably explains at least the huge scatter
in the (sparse) set of observations above the period gap. 
Second, the predicted WD
temperatures scale roughly as $T_{eff,1} \propto M_{1} \dot{M}_{acc}^{1/4}$
(TG09), so they are quite sensitive to the assumed WD
mass. No attempt has been made in Figure~\ref{fig:WD} to take this into
account. In particular, both models assume our usual $M_{1} =
0.75~M_{\odot}$, but the actual WD masses are not known for most of the sytems
plotted in Figure~\ref{fig:WD}. Third, the observed sample is likely to
be affected by selection effects. In particular, $T_{eff,1}$ is probably 
biased high in the sample, because hot WDs are easier to detect and
analyse. 

The best way to mitigate at least the last two concerns is to focus on
systems with well-determined $T_{eff,1}$ and $M_{1}$. For these, the 
accretion rate needed to heat the WD can be estimated and compared to
different models. A first attempt to do this was made by L08, based on
their small sample of eclipsing, short-period CVs. They concluded that
mass-transfer rates driven purely by GR -- as in the standard model --
provided a better match to the WD-based $\dot{M}_2$ estimates than an 
enhanced AML model. This is a rather surprising conclusion,
given the appearance of Figure~\ref{fig:WD}, 
%However, it appears to be
%due mainly to the fact that the typical WD mass in the L08 sample is
%considerably higher than the $0.75 M_{\odot}$ assumed in our models. 
and we will discuss the results of L08 in more detail below
(Section~\ref{sec:littlefair}). However, it would clearly be important to test
their conclusions by obtaining WD-based $\dot{M}_2$ estimates for a
larger sample of CVs with accurate $T_{eff,1}$ and $M_{1}$
measurements. At the moment, it may be premature to draw strong
conclusions about CV evolution scenarios on the basis of WD
temperatures.

\section{Implications and Applications}
\label{sec:implications}

In the previous section, we presented a semi-empirical CV evolution
track based on the observed mass-radius relationship of CV donor
stars. The inferred AML rates along our best-fit track are somewhat
lower (higher) than in the standard model of CV evolution above
(below) the period gap. We also compared its properties to
observed donor spectral types, the locations of the period gap edges,
the location of the period minimum and to observed WD
temperatures. Overall, our donor-based best-fit track appears to
match these observed properties of CV significantly better than the 
standard model. With the construction and testing of the donor-based
track concluded, it is time to consider some of its implications and
applications for observational and theoretical studies of CVs. This is
the goal of the present section. In Section~\ref{sec:dist}, we will show
how (lower limits on) distances towards CVs can be estimated via
simple donor-based photometric parallax in the near-infrared (NIR). In
Section~\ref{sec:stability}, we will compare the predicted mass-transfer rate
along the track to the critical rate for the occurence of dwarf nova
outbursts. Finally, in Section~\ref{sec:pdist}, we will calculate the
approximate orbital period distribution predicted by our donor-based
evolution scenario and compare this to both the standard model 
prediction and observations. 

\subsection{Donor-Based Photometric Parallax}
\label{sec:dist}

The physical and photometric properties of CV donors are expected to
change only on the secular evolution time scale. At fixed $P_{orb}$,
they are also quite insensitive to the detailed system parameters,
thanks largely to the period-density relationship
(Equation~\ref{eq:dense}). Thus all unevolved CV donors at the same orbital 
period should be quite similar. This makes them potentially quite
useful as distance indicators via photometric parallax, provided their
absolute magnitudes are reliably known as a function of $P_{orb}$. 

The absolute magnitude of a star is primarily a function of its radius
and effective temperature. Thus our semi-empirical donor sequences --
which were constructed by fitting the observed mass-radius
relationship of CV secondaries and tested against their observed $SpT$s --
should predict the absolute donor magnitudes at given $P_{orb}$ quite
well. As discussed in more detail in K06, we can therefore use 
these sequences to obtain lower limits on the distance towards any
CV with an unevolved donor star. For example, the lower limit on the
distance associated with a single epoch $K$-band measurement is 
\begin{equation}
\log{d} \geq \frac{K - M_{K,2}(P_{orb}) + 5}{5},
\label{eq:parallax}
\end{equation}
where $K$ is the apparent magnitude and $M_{K,2}$ is the absolute
$K$-band magnitude on the relevant donor sequence at the CV's orbital 
period. In principle, the apparent magnitude should be
extinction-corrected, but in practice this correction is usually
negligible for CVs in the infrared. 

This method only yields a lower limit on the distance, because the
fractional contribution of the donor to a single photometric
measurement is unknown. If an actual measurement of the {\em donor's}
apparent magnitude is available, $K$ in Equation~\ref{eq:parallax}
should be replaced with $K_2$, and the lower limit becomes an actual
estimate of the distance.

\begin{figure*}
\centering
\includegraphics[height=17cm,angle=0]{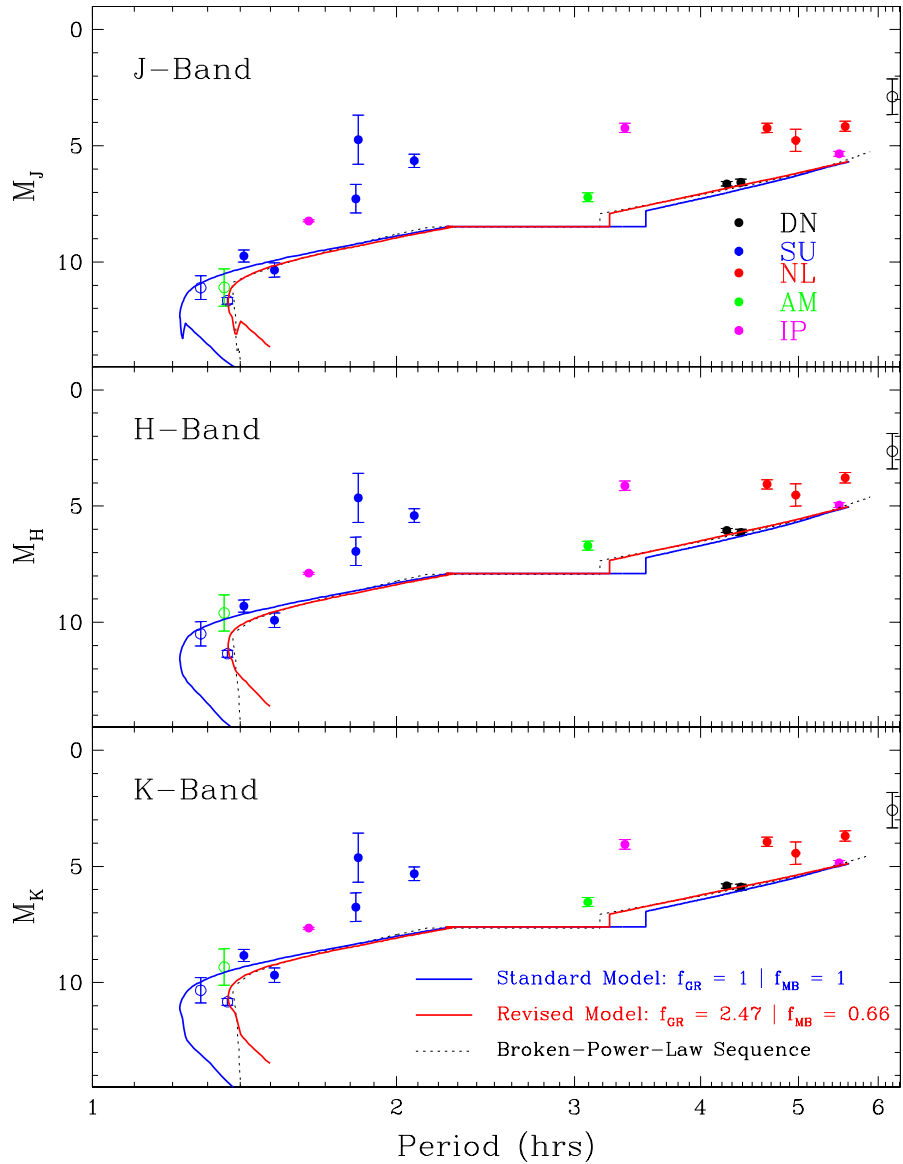}
\caption[Infrared absolute magnitudes of CVs as a function of
$P_{orb}$.]
{Infrared absolute magnitudes of CVs as a function of
$P_{orb}$. Points correspond to absolute magnitudes for CVs with
trigonometric parallaxes and reliable 2MASS observations, taken from
K06. Colours correspond to various CV sub-types, as indicated (DN = dwarf nova; SU
= SU UMa star; NL = non-magnetic nova-like; AM = AM Her star; IP =
intermediate polar). Systems with $P_{orb} < 82.4$~min or $P_{orb} >
6$~hrs are shown as open symbols and were excluded in calculating the 
quantities listed in Table~\ref{tab:IR}. 
The thick solid blue line shows the track predicted by the 
standard model ($f_{GR} = f_{MB} = 1$), the thick solid red line that
predicted by the revised (best-fit) model ($f_{GR} = 2.47$; $f_{MB} =
0.66$). For reference, the track predicted by our updated
broken-power-law donor sequences is also shown as a think dotted black
line.} 
\label{fig:IR}
\end{figure*}

In K06, we tested and calibrated this method for our original
broken-power-law donor sequence. Thus we compared the predicted NIR
donor absolute magnitudes along this track to the 
absolute magnitudes for CVs with distances determined by trigonometric
parallax. Figure~\ref{fig:IR} shows the same comparison for our updated
broken-power-law donor sequence (Section~\ref{sec:update}) and for the 
best-fit, self-consistent CV eolution track (Section~\ref{sec:newtrack}). For
reference, we also show the donor properties predicted by the standard 
model ($f_{MB} = f_{GR} = 1$). The observational data set is the same
as in K06. 

We already know from Figure~\ref{fig:phot} that all of these sequences
-- including the original broken-power-law one -- predict very similar
donor absolute magnitudes for pre-bounce CVs with $P_{orb} \gtappeq
90$~min. In line with this, we find that the scatter in the
observational data is considerably larger than any differences between
the models. None of the sequences are inconsistent with
these observations. The model tracks all trace a lower envelope
around the data points, as expected since the observed magnitudes may
include substantial contributions from system components other than
the donor star (e.g. the accretion disk). The predicted donor
magnitudes do differ quite significantly in the period bounce regime,
but here the stellar atmosphere models become unreliable anyway (see
discussion in Section~\ref{sec:final}), and even the WD starts to
compete with or outshine the donor in the NIR (Figure~\ref{fig:phot}). 

As in K06, we have estimated the average offsets ($\Delta J$, $\Delta
H$, $\Delta K$) between the donor tracks and the data, as well as the
scatter around the tracks if these offset are applied
($\sigma_J$,$\sigma_H$,$\sigma_K$).  
The results are listed in Table~\ref{tab:IR} and
are very similar to those given in K06 for the original
broken-power-law donor sequence. More specifically, for the small sample of
CVs shown in Figure~\ref{fig:IR}, the donor contribution to the total NIR 
light is about 23\% in J, 29\% in H, and 31\% in K. Thus, to the
extent that this sample is representative, lower
limits on CV distances (e.g. Equation~\ref{eq:parallax}) will typically
underestimate the true distance by about a factor of about 2.1 (J), 1.9 (H) 
and 1.8 (K). If such lower limits are converted into actual
distance estimates by appying these factors, the resulting values
will be uncertain by {\em at least} a factor of about 1.9 (J), 1.8 (H) and 
1.7 (K). The real uncertainties could be even larger, both because our 
small and heterogenous sample of CVs with trigonometric parallaxes may
not be representative, and also because there is no reason to think
that the typical donor contributions are actually constant along the
entire CV evolution track. The specific numbers quoted above are for
the best-fit self-consistent evolution track, but the other sequence
give similar estimates.  

Given that we have also calculated predicted WD temperatures and
absolute magnitudes along our evolution tracks, it is reasonable to
ask if improved distance estimates could be obtained by, for example,
replacing $M_{K,2}$ in Equation~\ref{eq:parallax} with the predicted absolute
magnitude of primary and secondary {\em combined}. This may also seem
to be a sensible way of extending the usefulness of the method to
shorter wavelengths, where the WD contribution can be dominant at all
orbital periods. However, we think the drawbacks of such an approach
would outweigh its benefits. In particular, as discussed in
Section~\ref{sec:fluc}, mass-transfer-rate fluctuations may cause WD
temperatures (and hence brightnesses) to vary on time scales longer
than $\sim 10^5$~yrs, but still shorter than the evolutionary time scale. 
%Moreover, both the WD radius and the
%predicted temperatures are sensitive to the assumed WD mass (albeit in
%opposite senses).  
As a result, and unlike for donor stars, it is not clear that the
absolute magnitude of a WD in a CV can be reliably predicted solely on
the basis of the system's orbital period. If aditional information is
available -- e.g. if the WD is detected spectroscopically 
\citep[e.g.][]{2000ApJ...539L..49K,
2002ApJ...575..419H,
2003ApJ...583..437A,
2005ApJ...622..589A,
2009ApJ...697.1512L}
or via eclipses in multiple photometric bands (e.g. L08) -- excellent
WD-based distance estimates can, of course, nevertheless be
obtained. However, as a simple tool for setting reliable distance
limits based solely on $P_{orb}$ and a single-shot NIR magnitude, the 
pure donor sequence is hard to beat. 

\subsection{Disk (In)stability: The Parameter Space of Dwarf Novae}
\label{sec:stability}

\subsubsection{Motivation and Observational Background}

As noted in Section~\ref{sec:fluc}, one of the reasons for suspecting CV
mass-transfer rates must vary on unobservably long time scales is the
fact that different {\em types} of CVs can apparently co-exist at the
same orbital period. It is interesting to explore this argument in
more detail. In particular, does it still hold for our new best-fit
evolution track?

According to the widely accepted disk instability model
\citep[DIM; e.g.][]{1996PASP..108...39O,2001NewAR..45..449L} 
the key difference between nova-likes 
and dwarf novae is the rate at which mass is supplied to the disk,
$\dot{M}_{disk}$. If this is above a certain critical rate,
$\dot{M}_{crit,hi}$, the disk is bright and stable, and the system is
a nova-like CV. Otherwise the system is unstable and cycles as a dwarf
nova between high- and low-states. (Actually, there is also a lower
critical rate, $\dot{M}_{crit,lo}$, below which the system is stable
again. However, unless the inner disk is truncated well before
reaching the WD surface, this rate is far below even the purely
GR-driven $\dot{M}_2$; see Equation~\ref{eq:mdot_crit_lo} below.)

Now the critical rate of mass supply to the accretion disk, 
$\dot{M}_{crit,hi}$, does depend on the system parameters, but only
weakly so at fixed $P_{orb}$ (see Equation~\ref{eq:mdot_crit_hi} below). 
Similarly, we know that unevolved CVs are expected to follow a 
unique evolutionary track (see Section~\ref{sec:adjust}), with only minor
sensitivity of $\dot{M}_2$ to other system parameters at given 
$P_{orb}$. It therefore follows that if $\dot{M}_{disk} = \dot{M}_2$, 
the unevolved CV population at any given $P_{orb}$ should be strongly
dominated by either dwarf novae or nova-likes. A population mix in
which both  types contribute significantly should only be possible
near orbital periods where $\dot{M}_{2}(P_{orb})$ crosses
$\dot{M}_{crit}(P_{orb})$, or is at least very close to it. 

\begin{figure*}
\centering
\includegraphics[height=17cm,angle=0]{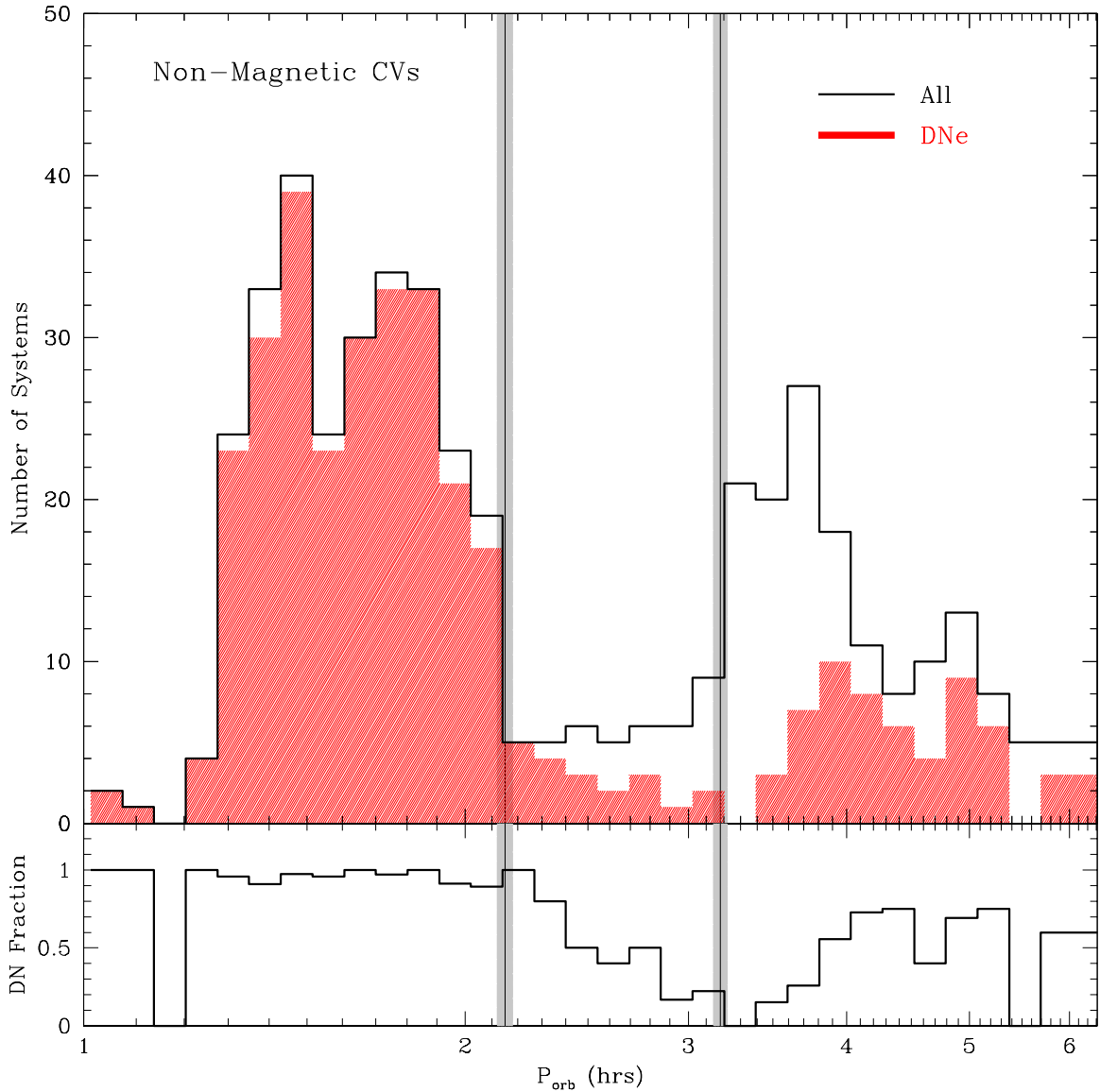}
\caption[The orbital period distribution of non-magnetic CVs and the
associated dwarf nova fraction.]
{The orbital period distribution of non-magnetic CVs and the
associated dwarf nova fraction. {\em Top panel:} The solid black
histogram shows the orbital period distribution of all non-magnetic CVs
in the Ritter \& Kolb catalogue (v7.11). The red-shaded histogram
shows the subset of dwarf novae within this sample. {\em Bottom
panel:} The dwarf nova fraction -- defined as the ratio of dwarf novae
to all non-magnetic CVs within a given period bin -- is shown as a
function of orbital period. In both panels, the vertical lines and
grey shaded regions mark the location of the upper and lower period
gap edges, along with their errors.}
\label{fig:dn_histo}
\end{figure*}

%\subsubsection{The Observed Dwarf Nova Fraction as a function of $P_{orb}$}

What is the observational situation? The top panel of
Figure~\ref{fig:dn_histo} shows the orbital period distribution of all
non-magnetic CVs in the \citet{2003A&A...404..301R} catalogue (version
7.11) compared to that of just the dwarf novae.
\footnote{It should be acknowledged that the type assignments in 
the Ritter \& Kolb catalogue are not
necessarily all that secure, especially for little-studied
systems. However, they are probably reliable enough for the basic
statistical analyses carried out in this section.}
The bottom panel shows the corresponding dwarf nova
fraction, i.e. the fractional contribution of dwarf novae to the
non-magnetic CV population as a function of $P_{orb}$. The 
short-period CV populations is essentially 100\% dominated by dwarf 
novae. However, the dwarf nova fraction then drops steadily through 
the period gap, reaches a minimum value of about 0\% just above the
gap, and then rises again to a value of about 75\% beyond periods of
about 4~hrs. Thus, in apparent violation of naive theoretical
expectations, significant populations of dwarf novae and nova-likes
happily co-exist above the period gap. As noted above, this has been
interpreted as evidence for mass-transfer-rate fluctuations on
unobservably long time scales.

\begin{figure*}
\centering
\includegraphics[height=17cm,angle=0]{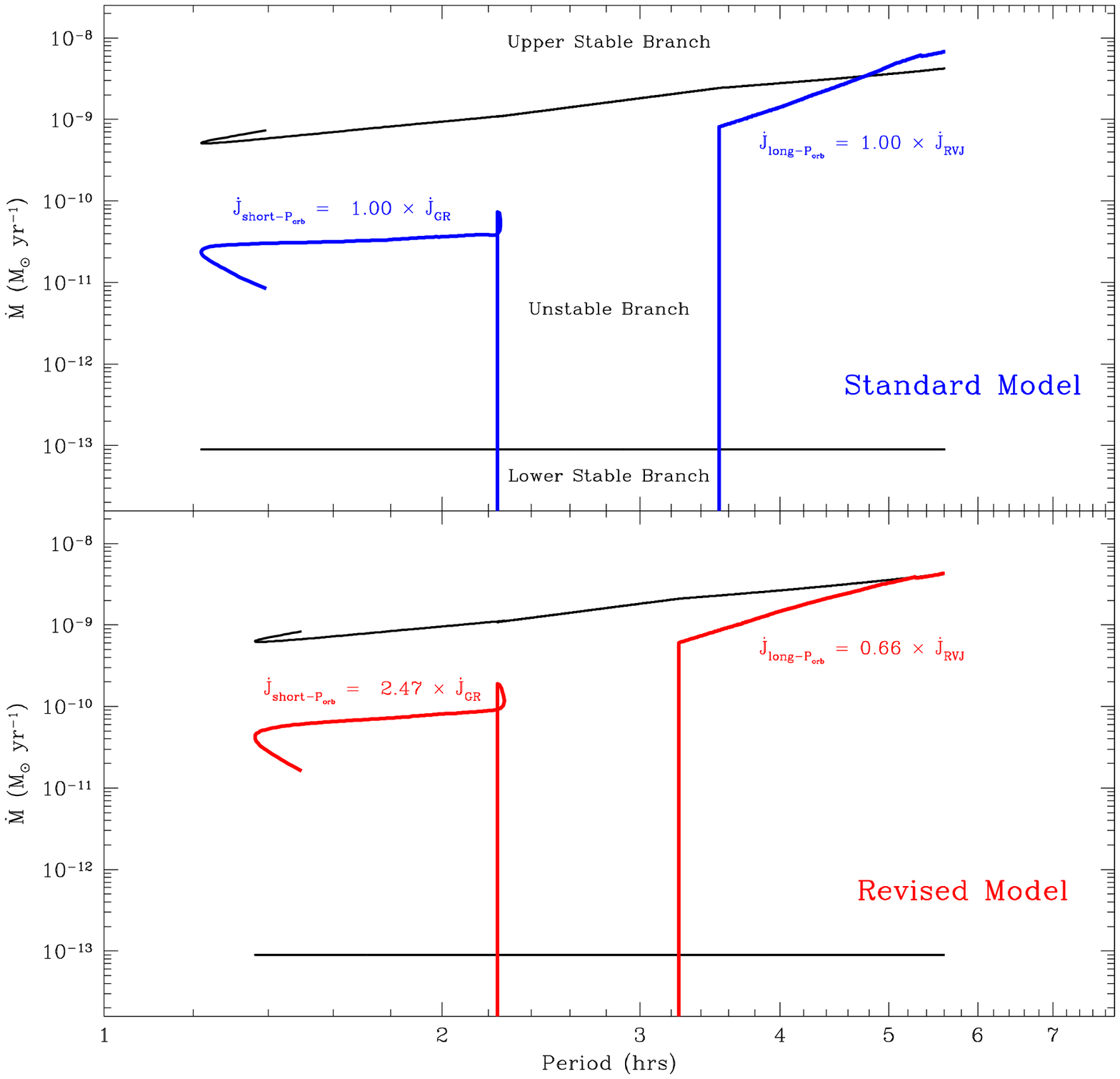}
\caption[A comparison of the predicted mass-transfer rates along our
self-consistent evolution sequences and the critical rates for the
occurence of dwarf nova eruptions.]
{A comparison of the predicted mass-transfer rates along our
self-consistent evolution sequences and the critical rates for the
occurence of dwarf nova eruptions. {\em Top panel:} The thick
solid blue line shows the run of $\dot{M}_2$ as a function of
$P_{orb}$ according to the standard model of CV evolution ($f_{GR} =
f_{MB} = 1$). The thin solid black lines show the corresponding upper
and lower boundaries of the unstable region in the
$P_{orb}$-$\dot{M}_2$ plane, as predicted by the thermal disk
instability model
(Equations~\ref{eq:mdot_crit_hi}~and~\ref{eq:mdot_crit_lo}). Only CVs
between these two 
regions are expected to be unstable, i.e. to undergo dwarf nova
eruptions. {\em Bottom panel:} Same as top panel, but here the thick
solid red line shows the revised model track ($f_{GR} = 2.47$, $f_{MB}
= 0.66$).}
\label{fig:dn_theo}
\end{figure*}

\subsubsection{Critical vs. Secular Mass-Transfer Rates}

Let us now consider the theoretical picture. After all, the simplest
explanation for the co-existence of dwarf novae and nova-likes would
be that the upper edge of the period gap is a crossing point for
$\dot{M}_{2}(P_{orb})$ and $\dot{M}_{crit,hi}(P_{orb})$. In this case,
differences in WD mass, or perhaps a modest amount of intrinsic
dispersion in $\dot{J}_{sys}$ (perhaps reflecting the initial donor
mass or system age) could be sufficient to explain the observations. 

With this in mind, Figure~\ref{fig:dn_theo} shows direct comparisons
of $\dot{M}_{2}(P_{orb})$ and $\dot{M}_{crit}(P_{orb})$ for both the
standard model of CV evolution (top panel) and our new best-fit model
(bottom panel). A very similar comparison was first
shown in \citet{1992ApJ...394..268S}.
The critical accretion rates adopted here are based on the
``grey atmosphere'' approximation and taken from
\citet{1998MNRAS.298.1048H}, i.e. 
\begin{equation}
\dot{M}_{crit,hi} = 8.0\times10^{15} \alpha_{hot}^{0.03}
\left(\frac{M_1}{M_{\odot}}\right)^{-0.89} 
\left(\frac{R_{disk}}{10^{10} {\rm cm}}\right)^{2.67} 
\label{eq:mdot_crit_hi}
\end{equation}
\begin{equation}
\dot{M}_{crit,lo} = 4.0\times10^{15} \alpha_{cold}^{-0.04}
\left(\frac{M_1}{M_{\odot}}\right)^{-0.89} 
\left(\frac{R_{1}}{10^{10} {\rm cm}}\right)^{2.67},
\label{eq:mdot_crit_lo}
\end{equation}
where $\alpha$ is the usual viscosity parameter. We adopt
$\alpha_{hot} = 0.3$, $\alpha_{cold} = 0.03$ and, as usual, $M_1 =
0.75~M_{\odot}$. We further assume that the disk radius is given by
$R_{disk} = 0.7 R_{L,1}$, based on the empirical study of
\citet{1996MNRAS.279..219H}, and take $R_{1} = 0.01R_{\odot}$, as
appropriate for our adopted WD mass.

In one sense, the theoretical curves are quite promising. For both
models, systems below the gap are predicted to be firmly in the
unstable regime, as is indeed observed. And above the gap, the
differences between $\dot{M}_{2}(P_{orb})$ and
$\dot{M}_{crit,hi}(P_{orb})$ are less than a factor of 3 everywhere.  
However, there are also two glaring inconsistencies. First, both models
predict that most unevolved systems above the gap should be unstable, at
least out to $P_{orb} \simeq 5$~hrs.
\footnote{And this is the period range that matters, since beyond this
evolved CVs may be 
expected to dominate the observed population in any case
\citep{2003MNRAS.340.1214P}.}
Yet, observationally, nova-likes dominate this distribution out to
about 3.8~hrs. Second, the {\em 
sense} in which the DN fraction changes above the period gap is
exactly opposite  to what one might expect theoretically. Both
models predict that CVs should become increasingly stable as we
move towards larger $P_{orb}$ above the period gap. Thus the dwarf nova
fraction is expected to {\em decrease}. However, the observed dwarf
nova fraction actually {\em increases} (from 0\% to 75\%)  between
$P_{orb} \simeq$3~hrs and $\simeq$4~hrs. 

\subsubsection{Reconciling Theory and Observation}
\label{sec:reconcile}

How can these discrepancies be resolved? The often-invoked
mass-transfer rate fluctuations -- be they driven by irradiation
(Section~\ref{sec:irrad_cycle}) or nova eruptions 
(Section~\ref{sec:hibernation} -- offer one possibility. Most nova-likes
below $P_{orb} \simeq 4-5$~hrs would then have to be systems caught in
the high state of such a cycle. In addition, the dearth of dwarf novae
in the period range just above the gap would imply that low-state systems
there must be essentially unobservable. 
%This follows from the fact
%that the predicted $\dot{M}_{2}$ would still be a good approximation
%to the secular mean transfer rate. So, by definition, the low-state
%transfer rate must be smaller than this, and hence would be in the
%unstable regime, unless it is actually below 
%$\dot{M}_{crit,lo}$. Since we do not see (m)any such unstable or
%ultra-faint CVs just above the gap, the low-state systems must 
%be undetectable as CVs in this phase. The most likely possibility is that
%they would actually detach during the low  
%state (as in irradiation driven cycles). 
Similarly, the
absence of nova-likes {\em below} the period gap would imply that, if
such cycles exist at all in this period range, the 
mass-transfer rate in the high state must be less than 
$\dot{M}_{crit,hi}(P_{orb})$.  Based on Figure~\ref{fig:dn_theo}, the 
high-state mass-transfer rate could thus be at most $\simeq 10$-times
higher than the secular $\dot{M}_{2}(P_{orb})$.

However, mass-transfer-rate fluctuations are not the only possible way
to account for the discrepancy between observed and predicted dwarf
nova fractions above the period gap. Errors or uncertainties in
$\dot{M}_2(P_{orb})$ and/or $\dot{M}_{crit,hi}(P_{orb})$ could, in
principle, also be to blame. 

Let us first assume that $\dot{M}_{crit,hi}(P_{orb})$ is correct. Is
it possible that we could have achieved an acceptable fit to the
observed $M_2-R_2$ data with a functional for $\dot{M}_2(P_{orb})$
that would better match the observed dwarf nova fraction? Given that
we would like systems to be most stable just above the gap,
Figure~\ref{fig:dn_theo} implies that this would require $\dot{M}$ to be a
flat or decreasing function of $P_{orb}$. Based on the preliminary
tests we carried out before settling on a functional form for
$\dot{M}_2(P_{orb})$ (Section~\ref{sec:form}), such models would produce a
significantly worse fit to the data. However, the mass-radius
data set above the gap remains fairly sparse, and additional
measurements are highly desirable to tie down the exact form of AML
above the gap.

What about $\dot{M}_{crit,hi}(P_{orb})$? The disk instability model on
which Equation~\ref{eq:mdot_crit_hi} is based has been around for
several decades and, overall, has been very successful
\citep[e.g.][]{1996PASP..108...39O,2001NewAR..45..449L}.
However, Equation~\ref{eq:mdot_crit_hi} is only an
approximation to 
numerical results that themselves depend on various
assumptions. Perhaps most importantly, it does not include the heating
of the outer disk that may be expected due to tidal dissipation and
irradiation \citep[e.g.][]{2001MNRAS.323..584S,2001A&A...369..925B}.
Depending on the system parameters, this can lower the
$\dot{M}_{crit,hi}(P_{orb})$ by a factor of several
\citep{2001A&A...369..925B}, which might be enough to push the
predicted $\dot{M}_2(P_{orb})$ into the stable regime above the period
gap.
\footnote{However, it is also worth noting that there is at least one
well-known, long-period dwarf nova -- the famous SS Cygni --
that manages to be unstable despite apparently satisfying 
$\dot{M}_2(P_{orb}) > \dot{M}_{crit,hi}(P_{orb})$ by a wide margin
(regardless of whether outer disk heating is accounted for; 
\citealt{2002A&A...382..124S,2007A&A...473..897S}).}
It is not clear if such an adjustment would tend to stabilize {\em
all} unevolved, long-period CVs. Even if it did, the rise in the dwarf
nova fraction with increasing $P_{orb}$ could perhaps still be
understood. It might simply reflect the increasing contribution of
evolved systems to the observed population, since these are indeed
expected to have  lower $\dot{M}_2$ at given $P_{orb}$ than unevolved
systems \citep{2003MNRAS.340.1214P}

Finally, Stuart Littlefair has suggested another possible explanation
to us: according to some recent studies of MB in non-interacting
stars \citep[][also see discussion in Section~\ref{sec:discuss_plausibility}]
{2003ApJ...586..464B,2007ApJ...669.1167B,2007MNRAS.377..741I,2010ApJ...722..222B}, 
even partly radiative stars begin their life with spin-down rates
comparable to fully convective stars. They only switch to a
faster, Skumanich-like spin-down sequence on a mass-dependent
time-scale that can reach $\simeq 1$~Gyr. This raises 
the possibility that there could be genuine $\dot{J}_{sys}$ and
$\dot{M}_2$ differences above the period gap, with the ``hidden
parameters'' being the initial donor mass and/or the age of the
system. It is not clear to us if 
this idea can be reconciled with the absence of significant
scatter in the $M_2-R_2$ relation above the period gap, but the data
is still somewhat sparse in this regime. In any case, this is an
interesting possibility that deserves to be explored more carefully.

Where does all of this leave us? In our view, the comparison of
Figures~\ref{fig:dn_histo} and~\ref{fig:dn_theo} does reveal a clear
and interesting discrepancy between theory and observation. This
discrepancy affects both the standard model of CV evolution and our
new best-fit donor-based one, so it cannot be used to distinguish between them. 
Long-term mass-transfer-rate fluctuations provide one possible
explanation for the observations, but so do uncertainties in the 
$\dot{M}_{crit,hi}$ and -- perhaps -- also  uncertainties in the form
of $\dot{M}_2$.
Thus, on its own, the co-existence of dwarf novae and
nova-likes above the period gap provides at most circumstantial
evidence for long-term mass-transfer-rate fluctuations.

\subsubsection{Coda: The Surprising Decline of Dwarf Novae in the Period Gap}

%Before we leave our discussson of different CV sub-types, 
There is one final feature of Figure~\ref{fig:dn_histo} that deserves comment:
the smooth and monotonic drop in the dwarf nova fraction from $\simeq
100\%$ to $\simeq 0\%$ inside the period gap is actually rather
surprising. In the disrupted MB framework, the existence of CVs 
in the gap is explained by noting that some systems will necessarily
first come into contact within the gap
\citep[e.g.][]{2001ApJ...550..897H}. 
Such systems will typically have fully convective secondaries with
masses $0.2 M_{\odot} < M_2 < 0.35 M_{\odot}$, and, in the standard
model, will evolve through the gap as purely GR-driven systems. More
generally, they are essentially short-period CVs that find themselves 
within the gap by an accident of birth. However, this immediately
implies that, just like other short-period systems, CVs within the period
gap would be expected to be located firmly within the unstable domain
in Figure~\ref{fig:dn_theo}. The dwarf nova fraction would therefore be
expected to be close to $100\%$ throughout the period gap, followed by
a sharp cut-off at the upper gap edge. This is not what is observed.

One possible explanation is that the CV population within the 
gap includes a significant proportion of MB-driven systems with
slightly lower metallicity secondaries. Such donors are expected to
become 
fully convective at shorter $P_{orb}$ than their solar metallicity
counterparts. As a result, stable, long-period, low-metallicity CVs
may `invade'' the period gap from above \citep{1997A&A...320..136S,
2002MNRAS.335....1W}.

\subsection{Orbital Period Distributions}
\label{sec:pdist}

\subsubsection{Motivation and Construction}

We have already seen that the best-fit donor-based evolution track 
successfully predicts the observed location of the period minimum,
thus resolving one of the long-standing conflicts between the standard
model of CV evolution and observations (Section~\ref{sec:bounce}). A similar
conflict exists concerning the ratio of long-period CVs to
short-period, pre-bounce CVs, with the standard model predicting a far
lower ratio than is observed, even after allowing for selection
effects
\citep{2007MNRAS.374.1495P,2008MNRAS.385.1471P,2008MNRAS.385.1485P}. 
Does the revised donor-based model resolve this conflict as well? 

Intuitively, it is reasonable to hope that it might. Relative to the
standard model, the best-fit donor-based evolution track is
characterized by a slower AML rate above the gap and a faster 
one below. CVs evolving along the best-fit track may therefore be
expected to spend a relatively larger fraction of their lives above
the period gap. Other things being equal, this should lead to a higher
fraction of long-period systems in the predicted CV population. 

A full population synthesis calculation (as carried in out various 
CV contexts by \citealt{1993A&A...271..149K, 1997MNRAS.287..929H,
2001ApJ...550..897H, 2004ApJ...604..817P, 2005ApJ...635.1263W})
would be required to properly
verify and quantify this prediction, and any quantitative comparison
with observed period distributions would also have to carefully deal
with selection effects \citep{2007MNRAS.374.1495P}. This is well 
beyond the scope of our present, already overlong study. 

Here, we instead limit ourselves to the simplest possible population
models that can be constructed from our evolution tracks. Thus we
assume that all CVs in our model populations have $M_{1} =
0.75~M_{\odot}$ and start their life as CVs with $M_2 = 0.6~M_{\odot}$
at 
$P_{orb} \simeq 5.6$~hrs. We also adopt a constant CV formation rate
over the age of the Galactic disk, for which we take $t_{gal} =
10$~Gyr; we assume that the pre-CV evolution phase is negligibly
short compared to $t_{gal}$. With these assumptions, the number of CVs
expected in a given orbital period bin is simply proportional to the
total amount of time a single CV evolving from $t = 0$ to $t =
t_{gal}$ spends inside this bin. The distribution is normalized by
taking the theoretical CV birth rate density to be $2 \times
10^{-15}$~yr$^{-1}$~pc$^{-3}$ \citep{1996ApJ...465..338P}. This is
$5\times$ lower than the estimate given by
\citet{1992A&A...261..188D}, but yields a CV space
density ($\simeq 2 \times 10^{-5}$~pc$^{-3}$) that is in better
agreement with observations
\citep[e.g.][]{1998PASP..110.1132P,2007MNRAS.382.1279P}.
We also take the space density to be constant throughout whatever
volume we consider.

\begin{figure*}
\centering
\includegraphics[height=23cm,angle=0]{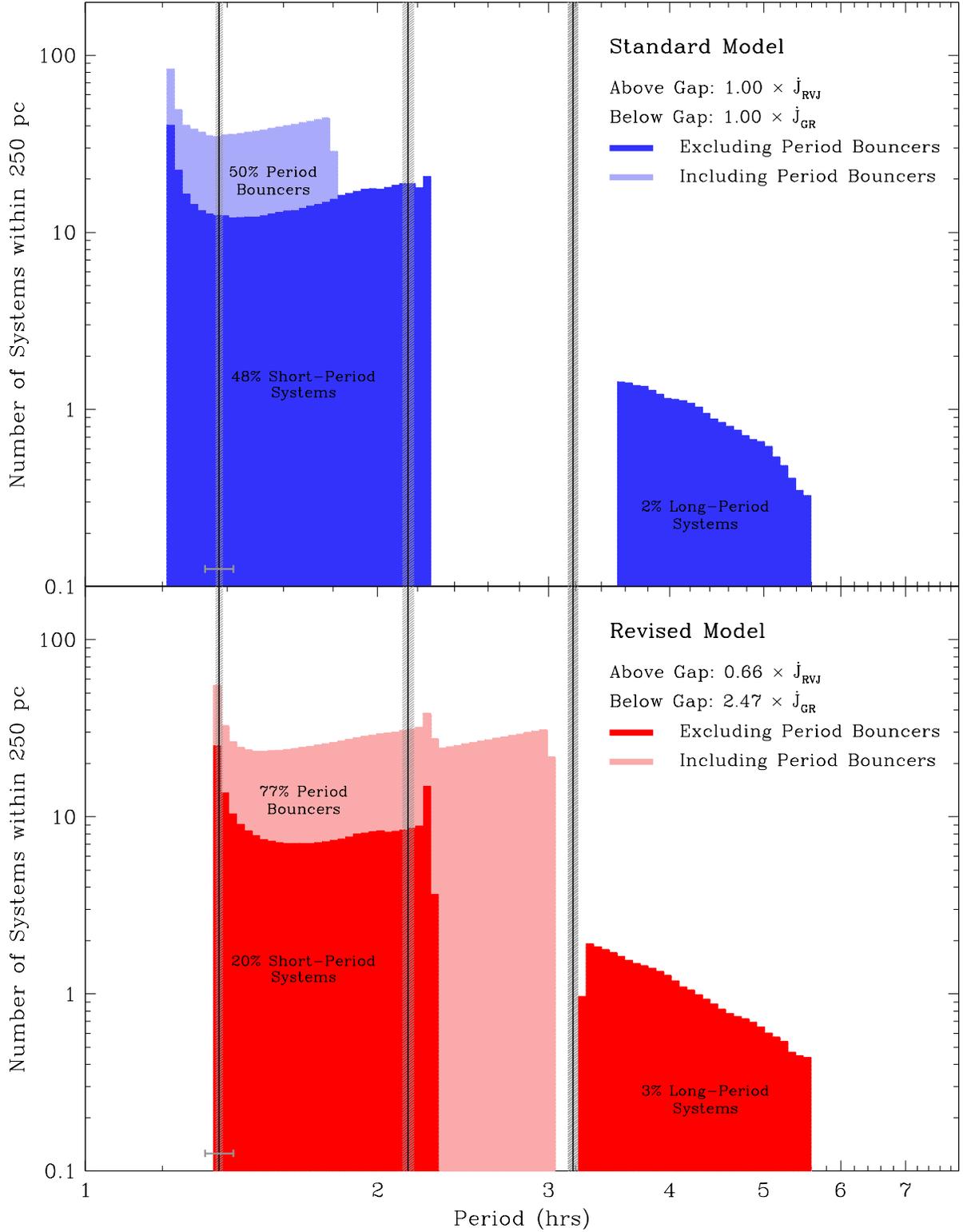}
\caption[Approximate period distributions calculated for the standard 
model and the revised best-fit model.]
{Approximate period distributions calculated for the standard 
model ($f_{GR} = f_{MB} = 1$; the blue-shaded histogram in the top
panel) and for the revised best-fit model ($f_{GR} = 2.47$, $f_{MB} =
0.66$; the red-shaded histogram in the bottom panel). The contributions
of period bouncers to the overall distributions are shown separately by
light shaded regions. The vertical lines show the observed locations
of the minimum period (as estimated from the recently discovered
period spike [G\"{a}nsicke et al. 2009]), and of the lower and upper
edges of the period gap (K06). The grey shaded regions around these
lines are the errors on these quantities. The error bars shown near 
the bottom of each panel on top of the $P_{min}$ estimate show the
intrinsic FWHM of the period spike. Note that these predicted period
distributions are highly approximate and purely illustrative. In
particular, they assume that all CVs are born at $P_{orb} \simeq
6$~hrs above the gap (see text for details).}
\label{fig:porb_dist}
\end{figure*}

%% \clearpage

\subsubsection{Basic Results}

The resulting theoretical orbital period distribution for both the
standard and the best-fit model are shown in 
Figure~\ref{fig:porb_dist}; some associated statistics are listed in
Table~\ref{tab:pdist}. It should be obvious that these distributions are highly
approximate and should not be compared directly to
observations. For example, the age-dependent scale height of the
Galactic disk should really be taken into account in such
comparisons \citep{2007MNRAS.374.1495P}. 
Perhaps more importantly, our distributions ignore 
the fact that CVs are expected to be borne with a wide range of ages
and orbital periods \citep{1992A&A...261..188D,1996ApJ...465..338P}.
In fact, the
CV birthrate distribution itself will be different for 
different evolution tracks, since the rate at which detached systems
are driven into contact depends on the adopted AML rates. However, our 
approximate theoretical distributions are nevertheless useful.
First, their very simplicity makes them easy to interpret
and thus helps to build intuition. Second, {\em differential}
comparisons between competing models are likely to be less affected by
our simplifications than direct comparisons to observational
data. Third, the qualitative impact of relaxing our most restrictive
assumptions is generally easy to predict.

With these points in mind, what do the predicted period distributions
tell us? Perhaps most importantly, the predicted ratio of long-period
to short-period, pre-bounce CVs ($f_{l:s}$) is indeed higher in the
best-fit model ($f_{l:s} = 0.125$) than in the standard model
($f_{l:s} = 0.042$), by a factor of 3.0. This is in reasonable
agreement with the naive expectation that this ratio should reflect
the change in the relative AML normalizations between the two models
(2.47/0.65 = 3.8). The agreement is not perfect, because rescaling 
$\dot{J}_{MB}$ and $\dot{J}_{GR}$ is not exactly equivalent to
rescaling $\dot{P}_{orb}$. In any case, this predicted factor of 3
increase in the relative abundance of long-period CVs should be fairly 
robust. 
%For example, suppose that a fraction $f_{bs}$ of CVs
%are actually born below the period gap. Let us also assume for simplicity
%that these systems are borne exactly at the lower gap edge and with
%the same age as the CVs emerging from the gap. The net result would
%then simply be to lower the predicted ratio of long-period CVs to all
%other systems by a factor of $(1-f_{bs})$. But this will be true of
%{\em any} model, so a comparison of such ratios {\em between}
%different models would remain unaffected. Unless the birth rate
%distributions for the models are very different, the factor of 3
%increase in the long-to-short period CV ratio produced by our best-fit
%model should therefore be fairly reliabe. 
This is promising, as just such a factor may be required to reconcile
theory with observations \citep{2008MNRAS.385.1485P}.

\subsubsection{Normalization and Space Densities}

The normalization we have adopted for our period distributions yields
quite reasonable numbers. For example,
Table~\ref{tab:pdist} shows that the best-fit (standard) model predicts
$\simeq 3$ ($\simeq 2$) long-period and $\simeq 16$ ($\simeq 35$)
short-period, pre-bounce CVs within  
100~pc. Observationally, there are two known long-period, non-magnetic 
nova-like systems with $P_{orb} \ltappeq 6$~hrs that {\em might} fall
within this volume: IX Vel \citep{1997A&A...323L..49P,2007ApJ...662.1204L}
and V3885 Sgr 
\citep{1997A&A...323L..49P,2009ApJ...703.1839L}. Both have $d
\simeq 100$~pc as measured by Hipparcos. Only two other nova-likes are
comparably bright, and neither is likely to be as close. For
reference, those objects are RW Sex 
\citep{1992A&A...256..433B,1997A&A...323L..49P,2010ApJ...719..271L}
and TT Ari \citep{1999A&A...347..178G}. 
Amongst the long-period dwarf novae with $P_{orb} <
6$~hrs, only the famous U~Gem is likely to lie within $d < 100$~pc
\citep{2000AJ....120.2649H,2006A&A...460..783B}.
All three of these objects are 
9$^{th}$-10$^{th}$ magnitude objects (in the case of U~Gem, this is 
the magnitude at the peak of an outburst). So one might hope that this
small sample of nearby long-period CVs is reasonably complete. 
On the short-period side, the recent compilation of
\citet{2009arXiv0903.1006P}
includes 6 short-period dwarf novae with estimated distances of $d \leq
100$~pc. However, these systems can be much fainter and may erupt
infrequently, so incompleteness may be a serious issue here. Given all
the uncertainties that affect the predicted and observed numbers, we
will not attempt to draw conclusions about the relative merits of the
two models on this basis. However, a more general point {\em can} be
made with some confidence: it would clearly be
hard to reconcile either model with space densities at the high
end of theoretical predictions, i.e. $\rho_{cv} \gtappeq
10^{-4}$~pc$^{-3}$ \citep{1992A&A...261..188D}. This qualitative
conclusion agrees quite well with the quantitative estimate 
$\rho_{cv} = 1.1^{+2.3}_{-0.7} \times 10^{-5}$~pc$^{-3}$ derived by
\citet{2007MNRAS.382.1279P} from a small, but purely flux-limited,
X-ray-selected sample.

\subsubsection{The Shapes of the Period Distribution and the Spike at the Lower Gap Edge} 
 
Turning to the {\em shapes} of the orbital period distributions, we
note that these are quite similar for both models. This similarity
reflects the identical functional forms we adopted for the AML laws in
the models. The two most obvious differences are the wider long-period
CV distribution in the  best-fit model, and the much longer maximum
$P_{orb}$ it predicts for the period bouncer distribution. The first 
difference arises simply 
because the upper gap edge is located at a shorter period in the
best-fit model, the second because faster-than-GR evolution below
the gap allows donors to evolve much further within $t_{gal}$. 

A less obvious, but interesting difference is the period spike seen at
the lower edge of the period gap in the best-fit 
model. This arises for essentially same reason as the period spike at
$P_{min}$. As explained in Section~\ref{sec:pcrit}, a system emerging
from the period gap is expected to briefly evolve to longer periods, as
the donor adjusts itself the mass loss it suddenly experiences,
before settling down onto the standard evolution track that takes it
to shorter periods and, ultimately, $P_{min}$. As a result,
$\dot{P}_{orb} = 0$ at the upper limit of the period flag. Both the
period flag itself and the associated local minimum in
$|\dot{P}_{orb}|$ are nicely visible in Figure~\ref{fig:evo1}. Just as for
the period spike at $P_{min}$, we might therefore expect the upper
limit of the period flag to be marked by a local maximum in the CV
period distribution. 

The real question, however, is why this second period spike is so
obvious in the best-fit model, but effectively absent in the standard
model. The answer turns out to be linked to fundamental stellar
physics. As discussed in Section~\ref{sec:adjust},
\citet{1996MNRAS.279..581S} 
showed 
that the time it takes a donor to adjust its radius following the
(re)-establishment of mass loss is $\tau_{per} \simeq 0.05
\tau_{{kh},eq}$. However, they also estimated the time it takes the
system to then reach the upper limit of the period flag, which turns 
out to be $\tau_{flag} = \frac{2}{3\zeta_{eq}} \tau_{per} \simeq
\tau_{per}$. This result implies that the amount of time a CV spends on
the increasing-period branch of the period flag is, to a first 
approximation, fixed solely by the stellar equilibrium
properties. As noted by 
\citet{1996MNRAS.279..581S}, this is indeed what is
seen in numerical calculations, and indeed it is what we see in our
models. Once the system has passed the upper limit of the period flag,
normal CV evolution on the AML time scale gradually takes over.

What does all of this mean for the predicted strength of the period 
spike at the lower gap edge? If $\tau_{flag} << \tau_{ev}$, the
turn-around happens too quickly to leave its imprint on the period
distribution. Conversely, if $\tau_{flag} \geq \tau_{ev}$, the time it
takes to execute the period flag is at least comparable to the
evolutionary time scale, so a clearly visible spike should be
expected. {\em Thus the period spike at $P_{gap,-}$ 
can, in principle, be used as a direct observational tracer of the AML
rate below the gap (since this is what determines $\tau_{ev}$).} 

A key point here is that, in predicted CV populations normalized to
the same total number of systems, the number of CVs contributing to
the spike is purely a property of stellar properties on the MS. In
particular, it is 
invariant under changes in the AML rate. The visibility of
the spike is therefore actually driven by the numbers of CVs
found in adjoining period bins, since these do depend on the local AML
rates. More specifically, as the AML rate below the gap is increased,
CV numbers in orbital period bins just below $P_{gap,-}$ drop and
eventually reveal the spike. This is exactly what is observed in
Figure~\ref{fig:porb_dist}. 

In reality, the situation is considerably more complicated, of
course. One issue is that the predicted strength of the spike will
depend on the CV birthrate distribution. More specifically, the
population of CVs born with $P_{orb} < P_{gap,-}$ will add to the
``continuum'' of short-period systems below the spike and reduce the
spike's visibility. Another complication is that the spike is
only a measure of the {\em instantaneous} mass- and angular
momentum-loss rate at $P_{gap,-}$, immediately following the
re-establishment of contact. Given that the exact shape of the AML law
is not determined very precisely by the donor mass-radius data, it
may be fairly straightforward to modify the adopted shape to increase
or decrease the visibility of the spike. 
%In particular, we suspect that the
%AML model adopted here -- $f_{GR} \times \dot{J}_{GR}$ -- could be replaced
%by the equally simple $\dot{J}_{GR} + {\rm Constant}$, without
%upsetting the fit to the data very much. In such a model, the period 
%spike at $P_{gap,-}$ would be reduced, since the constant 
%term would be less important at longer periods. At shorter periods,
%however, evolution would be considerably faster than GR evolution. If
%the constant AML term acted even in systems with substellar donors, it
%would quickly push the period bouncers to extremely long periods and
%low donor masses. It could perhaps even kill them within $t_{gal}$,
%which might actually a desirable property. Investigating
%such a model in detail would be worthwhile, but is once again beyond
%the scope of our study here.

%Returning to the predicted spike at $P_{gap,-}$, 
Is there any observational evidence for a spike at $P_{gap,-}$ in the
CV period distribution? Realistically, the only CV sample currently worth  
checking for this is that provided by the SDSS \citep{
2002AJ....123..430S,
2003AJ....126.1499S,
2004AJ....128.1882S,
2005AJ....129.2386S,
2006AJ....131..973S,
2007AJ....134..185S,
2009AJ....137.4011S}.
At present, this does not appear to show evidence for such
a spike (see Figure~2 of G09). However, the
data is still quite sparse, and even the period gap itself remains
poorly defined in the SDSS sample. 

\subsubsection{The Spectre of Mass-Transfer-Rate Fluctuations}

One final can of worms we can't help but open concerns the potential
impact of long-term mass-transfer-rate fluctuations
(Section~\ref{sec:fluc}) on the CV period distribution. If such
fluctuations exist, they could conceivably alter 
the observable distribution dramatically. For example, it seems likely 
that the low state of any such cycles is essentially a detached phase,
during which CVs are unrecognizable as such (see
Section~\ref{sec:fluc}). Thus the 
observed period distribution of CVs undergoing such cycles would be
that of systems caught in the high state. This distribution would 
therefore depend on the duty cycle of the fluctuations, i.e. the
percentage of time actually spent in the high state. If this is a
strong function of $P_{orb}$, the period distribution will be
significantly modified. It is also possible that mass-transfer rate
fluctuations could operate only within a certain range of donor masses
(and hence orbital periods). This, too, would leave an imprint in the
observed period distribution, as systems not undergoing such cycles
would appear to be overabundant. 

It is interesting to note here that studies of irradiation-driven
mass-transfer cycles show that long-period CVs are much more
susceptible to the instability producing such cycles than   
short-period CVs (see Section~\ref{sec:irrad_cycle}). Similarly,
long-period CVs might be more likely to undergo nova-induced
hibernation, since FAML may be less efficient in these systems
\citep[][]{1991A&A...246...84L}. If
mass-transfer-rate fluctuations exist, but are limited to long-period
CVs, they would tend to {\em reduce} the observed ratio of long-period
to short-period pre-bounce systems. This would exacerbate the
discrepancy with the standard model, as the observed ratio is actually
too {\em high}. Thus mass-transfer-rate fluctuations are unlikely to
be a panacea for the ailments afflicting the standard model.

\section{Discussion}
\label{sec:discuss}

The main result of our study is that the observed CV donor mass-radius
relation seems to demand a modest, but significant revision
of the standard model for CV evolution. More specifically, we have found
that AML rates derived from donors below the 
period gap are $\simeq 2.5$ times higher than expected for pure GR
driving, and that donor-based AML rates above the gap are $\simeq 1.5$
times smaller than in the standard RVJ $\gamma=3$ model for
MB. Allowing for these modifications produces a much improved match to
the donor mass-radius data, predicts a minimum period in agreement with
observations and might explain why long-period CVs are much more
abundant in observed CV samples than expected in the standard
model. These are nice returns for a relatively small price. 

In this section, we will try to tie up a few remaining loose
ends. First, we will discuss our results in the light of two recent
studies that have come to rather different conclusions regarding CV
evolution. Next, we will explore the sensitivity of our results to 
the assumed offset in radius between ``standard'' stellar models and
observations of non-interacting stars, which may be due to magnetic
activity and rapid rotation (see Section~\ref{sec:bloat}). This is arguably
the main area of uncertainty in our method. We will then briefly
consider other limitations and biases of our approach, and finally
discuss whether enhanced AML below the gap is {\em physically} 
plausible.

\subsection{A Comparison to Littlefair et al. (2008)}
\label{sec:littlefair}

L08 carried out detailed eclipse
analyses for a set of eight short-period CVs. Interestingly, they 
found that the donors in their sample were inflated by $\simeq 10\%$
relative to equal-mass MS stars. This agrees with our results (and 
L08's measurements are, in fact, included in our
mass-radius data base). However, L08 also
estimated the effective temperatures of the accreting WDs in their CV
sample and calculated the corresponding long-term accretion rates
(see Section~\ref{sec:WD}; 
\citealt{2002ApJ...565L..35T,2003ApJ...596L.227T,2004ApJ...600..390T}). 
They argue
that these are broadly consistent with purely GR-driven AML losses,
but not with substantially higher AML loss rates.

A comparison of the WD temperatures predicted by our model sequences
to the observational data has already been presented in
Figure~\ref{fig:WD} and briefly discussed in Section~\ref{sec:WD}. Here, we
therefore focus only 
on the subset of data points included in L08. These points are shown
in green in Figure~\ref{fig:WD}, 
and those systems with sub-stellar 
donors are additionally marked by open diamonds. At first glance, our
revised model would seem to provide quite a good match to this subset
of the data; if anything, it slightly underpredicts the effective
temperatures of the three longest-period systems. By contrast, the
standard model does not seem to match any of the data points. 

However, this comparison is quite sensitive to the mis-match between
the WD mass assumed in our models ($M_{1} = 0.75~M_{\odot}$) and the
typical WD mass in the L08 sample (the unweighted mean
and dispersion is $<M_{1}> = 0.87 \pm 0.06~M_{\odot}$). Thus, for a
fairer comparison, the model curves would need to be shifted upwards by
slightly less than half the error bar shown in the top left corner of
the plot. This would clearly improve the match to the standard model
considerably, in line with the findings of L08.

Would such a shift also destroy the agreement between the
revised model and the data? Based on Figure~\ref{fig:WD}, we don't think 
so. In fact, an upward 
shift would bring the three longest-period objects in the sample into
{\em better} agreement with the model, yet would still likely provide
an adequate match to the shorter-period objects as well. 
Why then did
L08 find that enhanced angular momentum loss
models cannot work? Inspection of their Figure~5 suggests that the
culprit is the particular {\em type} of model they chose to
test. Their enhanced-AML model of choice invokes CAML via a
circumbinary disk \citep{2001ApJ...561..329T,
2002ApJ...569..395D, 2003ApJ...592.1124T,2005ApJ...635.1263W}. 
This happens to produce an almost constant, and rather high, 
$\dot{M}_2$ for systems that have evolved beyond $P_{min}$. This is
indeed inconsistent with the data. However, in our scaled-GR model,
$\dot{M}_2$ drops sharply in systems evolving through the minimum
period (see Figure~\ref{fig:evo1}), 
producing a wide range of low mass-transfer rates in CVs with periods
close to $P_{min}$. The systems argued by L08 to 
be inconsistent with enhanced-AML are all located in this $P_{orb}$
regime. Even allowing for the expected offset due to the assume WD mass,
our revised model is likely to cover the observed accretion rates
fairly well. We therefore do not think that there is a conflict
between L08's data and our revised model, even
without invoking the possibility that WD-based temperatures may not
faithfully track the secular mass-transfer rates (Sections~\ref{sec:fluc}
and~\ref{sec:WD}). 

\subsection{A Comparison to Sirotkin \& Kim (2010)}

A second recent study of the evolution of short-period CVs has been
carried out by \citet[][hereafter SK10]{2010ApJ...721.1356S}. Like us,
they attempt to infer 
$\dot{M}_2(P_{orb})$ from the observed donor mass-radius
relation. More specifically, they adopt the broken-power-law
approximation to the $M_2-R_2$ relationship from K06 and use a
simplified analytical description of the donor response to mass loss
to infer the corresponding mass-loss rate. Their results contrast with
ours in two ways. First, they find considerably higher mass-loss rates
than we do at essentially all orbital periods (compare their Figure~5 to our
Figure~\ref{fig:evo1}). Second, despite this, SK10 argue that their
inferred mass-loss rates are consistent with purely GR-driven AML. 

There are quite a few important differences between our donor-based
reconstruction of the $\dot{M}_2(P_{orb})$ relation and 
theirs. For example, since the period spike had not yet been
discovered at the 
time, the value for $P_{min}$ adopted in the K06 $M_2-R_2$
relationship is probably incorrect. Moreover, the broken-power-law
approximation itself is, of course, only that -- an approximation. In
reality, $\zeta$ presumably varies smoothly along the CV evolution
track (see e.g. Figure~\ref{fig:phys}). Also, the impact of mass-loss
history on the donor radii (see 
Section~\ref{sec:adjust}) is ignored in SK10's 
treatment. Last, but not least, where we use state-of-the-art stellar
models to describe the donor response to mass loss, SK10
use a simplified analytical description of the donor. The latter
is definitely useful qualitatively, and helps to develop intuition,
but it is not obvious to us whether (or to what degree) it is
quantitatively reliable. It should be acknowledged, however, that
their method did succeeed in reproducing the $\dot{M}_{2}(P_{orb})$
relationship along a self-consistent donor sequence calculated by
\citet{1999MNRAS.309.1034K} 
to within $\simeq 30\%-40\%$, by using the numerical
$M_2-R_2$ relationship along this sequence as input.

\begin{figure}
\centering
\includegraphics[height=8cm]{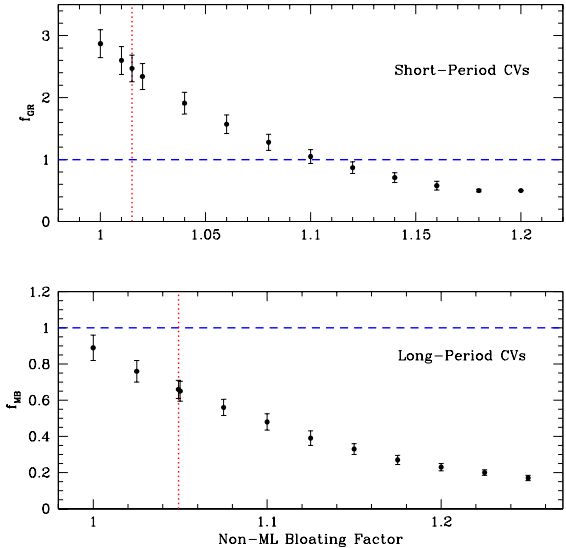}
\caption[The sensivitity of the inferred angular-momentum-loss rates
to the assumed amount of non-mass-loss-related donor bloating.]
{The sensivitity of the inferred angular-momentum-loss rates
to the assumed amount of non-mass-loss-related donor bloating (in
addition to that produced by tidal and rotational deformation; see
Section~\ref{sec:deform}). {\em Top panel:} The best-fit angular-momentum
loss-rate amongst short-period CVs, $f_{GR}$, expressed as a multiple
of the purely GR-driven rate, is shown as a function of the assumed
bloating factor. The blue horizontal dashed 
line marks the standard model assumption -- $f_{GR} = 1$ -- and
requires $\simeq 10\%$ bloating. Our actual best estimate of this bloating is 1.5\%;
this is marked by the red vertical dotted line and corresponds to our
previously quoted estimate of $f_{GR} = 2.47 \pm 0.22$. {\em Bottom
panel:} Same as top panel, but for long-period CVs, and with the
relative angular-momentum-loss rate, $f_{MB}$, defined with reference
to the standard RVJ recipe with $\gamma = 3$. Here, the standard model
-- $f_{MB} = 1$ -- would require no bloating at all. Our best
estimate of the bloating factor in this regime is $4.9\%$, which
corresponds to our previously adopted $f_{MB} = 0.66 \pm 0.05$.} 
\label{fig:sens}
\end{figure}

However, in spite of all these technical differences, we suspect
that the main reason for the higher mass-transfer rates inferred by
SK10 is that they did not correct for either 
tidal/rotational deformation (Section~\ref{sec:deform}) or for other
non-ML-related donor bloating (Section~\ref{sec:bloat}). As we have seen,
the former amounts to $\simeq 4.5\%$ for short-period CVs, the latter
to $\simeq 1.5\%$. The combined $\simeq 6\%$ non-ML-related bloating is
a non-negligible fraction of the total observed donor inflation below the
period gap. Figure~\ref{fig:sens} -- which is discussed in more detail
below -- suggests that this may be expected to
produce a factor of $\simeq 2$ difference in the inferred
mass-transfer rates. This probably accounts for much of the difference 
between our inferred $\dot{M}_2$ estimates and theirs. 

What is more puzzling to us is SK10's finding that their
very high inferred mass-transfer rates are consistent with purely
GR-driven AML. 
As far as we know, no other recent study of CV evolution has predicted
such high rates for pure GR-driven evolution 
\citep[e.g.][]{1993A&A...271..149K,1997MNRAS.287..929H,
1999MNRAS.309.1034K,2001ApJ...550..897H,2003MNRAS.340.1214P}.  
For example, at $P_{orb} \simeq 2$~hrs, the mass-transfer rates
inferred by SK10 are about 4-5 times higher than those on a
corresponding purely GR-driven evolution track calculated with our 
donor models (here we have made none of the corrections discussed in
Section~\ref{sec:skeletons}, to ensure the comparison with SK10 is fair). 

We can also turn this argument around. If we adopt the
broken-power-law mass-radius relationship of K06 -- which is also the
basis of SK10's method -- and then use Equation~\ref{eq:mdot} with
$\dot{J}_{sys} = \dot{J}_{GR}$, as given in
Equation~\ref{eq:jdot_gr}, we obtain an estimate of $\dot{M}_{GR}
\simeq 4-5 
\times 10^{-11} M_{\odot} {\rm yr}^{-1}$ near $P_{orb} \simeq
2$~hrs. This is again 4-5 times smaller than the mass-transfer rates 
estimated by SK10 in this regime. We thus do not understand how the
mass-transfer rates estimated by SK10 can be consistent with pure GR
driving.

\subsection{Sensitivity to Non-Mass-Loss-Related Donor Bloating}
\label{sec:discuss_bloat}

How sensitive are our results to the particular non-ML-related
bloating we have assumed? This question is explored quantitatively in
Figure~\ref{fig:sens}, where we plot the 
best-fit estimates of $f_{GR}$ and $f_{MB}$ as a function of the 
assumed non-ML-related donor bloating factor. Note that this factor is
{\em in addition} to the bloating factor associated with tidal and
rotational deformation, since the latter is relatively well understood
and should always be applied.
\footnote{However, see the discussion of the discrepant results
by \citet{2009A&A...494..209L} in Section~\ref{sec:deform}.}

Figure~\ref{fig:sens} shows that consistency with the standard model would
require non-ML-related bloating at a level of $\simeq 10\%$ in fully
convective donors below the gap and $\simeq 0\%$ in partially
radiative donors above. The former number is higher than suggested by
our analysis of the data in Figure~\ref{fig:mr_iso}, while the latter
is lower. Thus {\em some} movement in the standard model parameters
towards our revised model seems hard to avoid.  

In our default models, we correct for a $1.5\%$ ($4.9\%$) radius
offset between stellar models and actual stars in the fully convective
(partially radiative) regime. As discussed in Section~\ref{sec:bloat}, the
actual offset seen in Figure~\ref{fig:mr_iso} is $4.5\%$ ($7.9\%$),
but we follow \citet{2010ApJ...718..502M} in allowing for a $\simeq
3\%$ star-spot-induced bias in the observationally inferred radii. 
If we correct for the {\em full} offsets, we obtain $f_{GR} = 1.82
\pm 0.17$ and $f_{MB} = 0.55 \pm  0.05$. Figure~\ref{fig:sens} also
confirms that the potential impact 
of irradiation-driven bloating ($\ltappeq 1-2\%$ below the gap and
$\ltappeq 3\%$ above; see Section~\ref{sec:irrad}) is modest.

Overall then, we feel that the $M_2-R_2$ data provides fairly strong
evidence for a departure from the standard model, even allowing for
the systematic biases and uncertainties associated with non-ML-induced
donor inflation. In particular, it seems difficult to make purely
GR-driven AML below the gap consistent with the data without pushing
the systematics discussed above (and perhaps also those in
Section~\ref{sec:discuss_bias} below) to their limits and in the same
direction.

\subsection{Other Biases and Limitations}
\label{sec:discuss_bias}

We have already explored most of the assumptions and limitations of
our method in exhaustive (exhausting?) detail, so here we will only
consider a few additional issues that have not yet been covered
elsewhere.

First, it is important to understand that the exact functional
forms we have adopted for AML in our fully self-consistent models are
necessarily arbitrary to some degree. By this we do not just mean that
the detailed form is not perfectly constrained by the data, or that a
wider exploration of models predicting different shapes for
$\dot{M}_2(P_{orb})$ might have yielded an equally good or better fit
to the data. (Both of these statements are true, however.) Rather, we
mean that donor radii are sensitive only to the form and normalization
of $\dot{M}(M_2)$, whereas physical AML recipes can depend on all binary
parameters ($M_1$, $M_2$, $R_2$, $P_{orb}$, $a$). In a semi-detached
setting, these parameters are related by the period-density relation and
Kepler's third law. Thus we cannot expect to cleanly isolate the
dependence of $\dot{J}_{sys}$ on each individual parameter in this
setting. The important point to remember is that {\em all AML recipes
that produce similar runs of $\dot{M}_2(M_2)$ will produce virtually
identical donor mass-radius relationships.}

Second, we have adopted a unique WD mass of $M_{1} = 0.75~M_{\odot}$ in
all of our modelling. If this turns out not to be representative, our
results will necessarily be biased. Moreover, the sample of CVs for
which we have donor masses and radii does, of course, include systems
with a range of WD masses. Thus fitting a single model to all these
data is not really quite self-consistent. 

Fortunately, neither of these issues is too serious. At fixed $M_2$,
the mass-transfer rate depends on $M_{1}$ roughly as $\dot{M}_{2,GR}
\propto M_{1}^{0.6-0.7}$ and $\dot{M}_{2,RVJ} \propto M_{1}^{-(0.9 -
  1.7)}$ in our scaled-GR and scaled-RVJ models, respectively. As
discussed in K06, the dispersion in observed WD masses in CVs is
roughly 20\%. At fixed $M_2$, this therefore corresponds to an
expected dispersion in the mass-loss rates of about 10\% - 15\% below
the gap and 15\% - 35\% above. The effect of WD mass on the minimum
period is also weak. \citet{1983ApJ...268..825P} show that this
$P_{min}$ scales approximately as $P_{min} \propto (M_1 +
M_2)^{0.1}$. Thus an upward shift in the mean WD mass from $0.75
M_{\odot}$ to $0.80 M_{\odot}$, for example
(c.f. last paragraph of Section~\ref{sec:update}), would only move
$P_{min}$ by $\simeq 0.5$~min. Nevertheless, the $\simeq 20\%$
{\em dispersion} in observed WD masses (P05, K06) is large enough to
smear the period spike by $\simeq 3.2$~min (FWHM). For comparison, the
{\em actual} FWHM of the period spike in the SDSS sample has been
estimated to be
$5.7 \pm 1.7$~min (G09). Thus WD mass dispersion alone is likely to
account for at least a significant fraction of the observed spike width.
\footnote{The quoted dispersion in WD masses corresponds to
the {\em standard deviation} ($\sigma$) of the observed mass
distribution. In order to estimate the FWHM of the corresponding shift
in $P_{min}$ we have used the conversion FWHM = 2.3548$\sigma$, which
is appropriate for a Gaussian distribution.}

Third, we have assumed throughout that CV donors start their life as
semi-detached stars on the zero-age MS (ZAMS). This may seem like a
poor approximation, but \citet{2003MNRAS.340.1214P} have shown that
significantly evolved donors (with central Hydrogen fractions of $X_c
< 0.4$) only begin to dominate the CV population above periods of
$\simeq 5$~hrs. This is in line with the CV birthrate distributions
calculated by \citet{1992A&A...261..188D} and
\citet{1996ApJ...465..338P}, which show that long-period CVs typically
have ages $\ltappeq 1$~Gyr at the onset of mass-transfer. Even for a
solar-mass secondary, this amounts to only $\simeq 10\%$ of its MS
life time. CVs born in or below the gap can be considerably older, but
at the relevant masses ($\ltappeq 0.35~M_{\odot}$), even the age of
the Galactic disk is only a fraction of the MS life time. Once
mass transfer starts, the secondary's state of nuclear evolution is
essentially frozen. Thus the assumption that the CV population below
5~hrs is dominated by unevolved systems with initially ZAMS donors
should be quite good.

However, there might be one interesting exception to this
statement. CV born within the last $\simeq 5$~Gyr with donors in the
mass range $0.25~M_{\odot} \ltappeq M_2 \ltappeq 0.35~M_{\odot}$ --
{\em i.e. most of the CVs born inside the period gap}. The BCAH98
models suggest that the secondary stars in such systems will be
slightly, but significantly, inflated relative to their ZAMS radius
(by $\simeq 1\% - 4\%$). The calculations by 
\citet{1996ApJ...465..338P} and \citet{1992A&A...261..188D} suggest 
that such systems might contribute $\simeq 20\% - 30\%$ to the
present-day CV population.

\subsection{Is Enhanced AML below the Gap Plausible?}
\label{sec:discuss_plausibility}

As noted in Section~\ref{sec:mb}, the assumption that MB stops
when the donor loses its radiative core was originally rooted in the
idea that the  magnetic fields in low-mass stars are generated by
dynamo action in the tachocline (the boundary between the radiative
core and the overlying convection zone;
e.g. \citealt{1997ApJ...486..484M, 1997ApJ...486..502C}). If so,
then stars without a radiative core -- which do not possess a
tachocline -- would be incapable of generating significant magnetic
fields. Consequently, there would be no magnetically channelled stellar
winds in fully convective stars, and no magnetic braking.

At some level, we have known for a long time that these arguments
cannot be entirely correct. After all, fully convective stars do
display all aspects of stellar activity, from variability to emission
line, ultraviolet, X-ray and radio emission 
\citep[e.g.][]{1986ApJ...305..784G, 1994ApJ...427..808S, 1995ApJ...455..670L,
1995MNRAS.274..869H, 1995ApJ...450..401F, 1998A&A...331..581D}.
Thus there is little question that fully convective stars are capable
of generating significant magnetic fields. Given only this
information, one might begin to wonder if there is any change in the
magnetic field across the fully convective boundary at all. 

Thankfully, there is at least some observational support for
this key idea. Broadly speaking, there is evidence for systematic
changes in both magnetic field strength and topology with spectral type 
\citep[e.g.][]{2008ApJ...684.1390R,2009A&A...496..787R,2010ApJ...710..924R,
2008MNRAS.390..545D,2010MNRAS.407.2269M}. It is
tempting to interpret these changes as signalling a switch from an
interface dynamo anchored in the tachocline of partially radiative
stars to a turbulent dynamo in fully convective ones. It should be
noted, however, that the existing data already point to considerably
more complex behaviour than a simple switch. 

From the point of view of CV evolution, the big question
is, of course, whether and how {\em magnetic braking} changes across
the fully convective boundary. There is certainly {\em some} evidence
for this as well \citep[see, for example, Figs. 9 and 10
in][]{2008ApJ...684.1390R}. A particularly nice series of studies in this
context is that by \citet{2003ApJ...586..464B,2007ApJ...669.1167B}.
He showed that data on stellar rotation in open clusters showed
evidence for two distinct spin-down sequences: the so-called
I-sequence (``I'' for interface) and the so-called C-sequence (``C''
for convective). Partially radiative stars are born on the C-sequence,
but then evolve onto the I-sequence, where they spin down according to
a standard Skumanich-type law, with $\dot{J} \propto \Omega^3$. Fully
convective stars remain on the C-sequence throughout their
lives. Spin-down on the C-sequence is slower than on the I-sequence,
but still significant. 

This semi-empirical picture suggests an obvious application to
CVs. Above the period gap, donors have (mostly? -- see
discussion in Section~\ref{sec:stability}) joined the 
I-sequences and are being braked by a standard Skumanich-type MB
law. This is consistent with our finding that this type of braking law
provides a good fit to the mass-radius data for long-period CVs. As
CVs approach the gap, the donors undergo a switch from an
interface dynamo to a turbulent one, before ultimately joining the
C-sequence. Since MB is weaker on the C-sequence, this amounts to a
disruption of AML and produces the period gap. The transition from the
I-sequence to the C-sequences may not be instantaneous -- the opposite
transition in isolated stars apparently takes a few $\times 10^8$~yrs
\citep{2003ApJ...586..464B}. The transition may therefore not be entirely
uneventful, and this might have something to do with the huge spread in
WD-based accretion rates in the 3~hr - 4~hr period range (which is
also the location where the peculiar SW Sex stars are preferentially
located, e.g. \citealt{2007MNRAS.377.1747R}). The scenario sketched above
is similar, though not identical, to the double-dynamo model proposed
over a decade ago by \citet{1997MNRAS.289...59Z}. 

In any case, it remains to be seen whether the braking rates predicted
for the I- and C-sequences by 
\citet{2003ApJ...586..464B,2007ApJ...669.1167B} and \citet{2010ApJ...721..675B}
are quantatitvely consistent with the AML rates we have inferred for long- and
short-period CVs, respectively. So far, we have only carried out some
very preliminary investigations along these lines, which have not been
too promising. However, we plan to carry out more detailed tests in 
the near future.

\section{Summary}
\label{sec:summ}

We have used state-of-the-art stellar models to reconstruct the
evolutionary track followed by cataclysmic variables from the
observed properties of their donor stars. Along the way, we have
updated the semi-empirical donor sequence of K06 and also reviewed the
close connection between the evolution of CVs and the properties of
their secondaries. Our main results and conclusions are as follows.

\begin{enumerate}
\item Mass transfer in CVs is thought to be driven by systemic
angular momentum losses, on a time scale, $\tau_{ev}$, that is
comparable to the thermal time scale of the secondary star. As a
result, the donor is driven slightly out of thermal equilibrium and
has a larger radius than an isolated star of equal mass. The degree of
donor inflation increases with increasing $\dot{M}_2$ and can
therefore be used as a tracer of the secular mass-transfer rate.
\item We provide an update of the semi-empirical donor sequence
presented in K06, in which the donor mass-radius
relationship is approximated as a broken power law
(Figure~\ref{fig:brokenpower} and Table~\ref{tab:seq_bpl}. The
update is necessary mainly to include some new observational data and to take
into account the recent discovery of the period spike (G09). 
The original and updated sequences both confirm that 
CV donors are indeed larger than equal-mass MS stars. This is true
both above and below the period gap. The gap itself corresponds to a
discontinuity in the mass-radius relation, which is in line with the
idea that the period gap is produced by a sudden reduction in the
prevailing angular-momentum-loss rate.
\item The angular-momentum-loss rates predicted by a variety of 
commonly used magnetic braking recipes differ enormously, both in
magnitude and the form of their dependence 
on binary parameters (Figure~\ref{fig:mb}) These differences are not
only found between 
``classic'' and ``saturated'' recipes, but even between conceptually  
similar ones. A more reliable description of magnetic braking is
therefore desperately needed to to improve our understanding of binary
evolution, in general, and CV evolution, specifically.
\item The time scale on which the donor adjusts its radius 
to the prevailing mass-loss rate, $\tau_{adj}$, is $\simeq 5\%$ of its
equilibrium thermal time scale. So long as $\tau_{adj}$ is very short
compared to $\tau_{ev}$, the donor radius traces the
``instantaneous'' mass-loss rate (averaged over the last few
$\tau_{adj}$). However, $\tau_{ev}$ becomes comparable to
$\tau_{adj}$ at critical times in a CV's evolution (e.g. near the
upper edge of the period gap and on passage through the
minimum period). Thus, in practice, donor radii do retain some
sensitivity to their previous secular mass loss history
(Figure~\ref{fig:convergence}).
\item If CVs undergo long-term mass-transfer-rate fluctuations,
these are likely to take place on time scales of $\tau_{var} \sim
10^4 - 10^9~{\rm yrs}$ (Figure~\ref{fig:tau}). Such fluctuations would
confound 
most attempts to estimate the secular mass-transfer rates in CVs. In
particular, even the thermal time scale of the accreted layer on the
WD primary is far shorter than $\tau_{var}$. Thus the quiescent
effective temperatures of WDs in CVs would track these variations,
albeit with reduced amplitude.
\item Mass-transfer-rate estimates derived from the
mass-loss-induced bloating of the donor stars in CVs are reasonably 
likely to be valid even in the presence of such fluctuations. This is
partly because the radius adjustment time scale is long, and partly
because the donor swelling due to irradiation -- one of the possible
drivers of mass-transfer cycles -- is only expected to be $\sim 1\%$.
\item Based on a new compilation of observational data for 
non-interacting low-mass stars, stellar models appear to underestimate
the radii of fully 
convective stars by $4.5\% \pm 0.5\%$ and those of partially radiative
stars by $7.9\% \pm 1.2\%$ (Figure~\ref{fig:mr_iso}). 
Moreover, there is no compelling evidence for
significant intrinsic scatter in the observed radii for fully
convective stars, whereas a significant 5\% dispersion is seen among
partially radiative stars. Approximately 3\% of the discrepancy
between models and observations is likely to be an artefact of the
eclipse modelling from which most observational radii are
derived. The remaining offsets are probably due to star spots and/or a
reduction in convective efficiency associated with fast rotation or
magnetic fields.
\item Tidal and rotational distortions cause Roche-lobe-filling
stars to be larger than isolated, spherical stars, even in the absence
of mass loss (Figure~\ref{fig:siro2}). The size of this effect depends
on the effective 
polytropic index of the star (and hence on the stellar mass) and also
weakly on mass ratio. Below the period gap, the bloating associated
with tidal and rotational distortions is $\simeq 4.5\%$; above the
gap, it reaches $\simeq 6.8\%$ at $P_{orb} \simeq 5.6$~hrs ($M_2 =
0.6M_{\odot}$).
\item Irradiation can also lead to non-ML-related swelling of the
donor star, but the size of this effect is expected to be relative
modest in CVs: $<1-2\%$ below the period gap and $<3\%$ above
(Figure~\ref{fig:irrad}; see also discussion in Section~\ref{sec:irrad}).
\item Taking the corrections described in points (7) and (8) above into
account, we model the observed donor mass-radius relationship (as
presented in P05 and K06) using
self-consistent, state-of-the-art models of mass-losing stars. For
simplicity, we assume that the mass transfer in our models is driven
by two simple AML recipes: a scaled version of the standard
gravitational radiation loss rate below the gap, and a scaled version
of the \citet{1983ApJ...275..713R} MB law with $\gamma = 3$
above the gap. With suitable normalization parameters, $f_{GR}$ and
$f_{MB}$, these recipes provide acceptable matches to the observed
data and make it easy to assess the evidence
for deviations from the 
standard model of CV evolution, in which $f_{GR} = f_{MB} = 1$. 
The best fit to the mass-radius data is obtained for
$f_{GR} = 2.47 \pm 0.22$ (below that gap) and $f_{MB} = 0.66 \pm 0.05$
(above the gap), where the quoted uncertainties are purely
statistical. This revised model matches the observations 
significantly better than the standard model (Figure~\ref{fig:fits}).
\item Our revised model predicts a location for the minimum
period that is in excellent agreement with the observed location of
the period spike, $P_{min} \simeq 82$~min
(Figure~\ref{fig:extremep}). Even allowing for donor
bloating unrelated to mass loss -- see points (7) and (8) above -- the standard
model predicts a significantly shorter $P_{min}$. The revised model
also predicts the correct location for the upper edge of the period
gap, $P_{gap,+}$, while the standard model predicts a significantly
longer value. Both standard and revised models are consistent with the
observed location of the lower gap edge, $P_{gap,-}$, since this is
primarily determined by the equilibrium properties of main sequence
stars. 
\item Both standard and revised models predict donor spectral
types that are in good agreement with observations
(Figure~\ref{fig:spt}). The same is true
for both the original and updated broken-power-law donor
sequences. Ordinary main sequence stars would predict significantly
earlier spectral types at fixed orbital period.
\item We have used the method of
\citet{2003ApJ...596L.227T,2004ApJ...600..390T} to 
predict the effective temperatures of the accreting WD along both
standard and revised evolution tracks. A naive comparison to the
observed data compiled by TG09 (Figure~\ref{fig:WD}) suggests
that, if anything, an even higher value of $f_{GR}$ than adopted in
our revised model would provide a better fit to the data. However,
there are several possible biases and problems that may make such a
comparison unreliable. In particular, the predicted effective
temperatures are quite sensitive to the assumed WD mass. The issue
noted in point (5) above may also be a concern, particularly above the
period gap.
\item The predicted near-infrared absolute magnitudes of CV
donors generally trace the lower envelope of observed absolute
magnitudes for systems with well-determined parallax distances
(Figure~\ref{fig:IR}). This  
is true for all the sequences we consider (standard, revised and
broken power law). Any of these sequences can therefore be used to set
a firm lower limit on the distance to a CV, given only a single-epoch
measurement of its apparent magnitude in the near-IR. However, the
offset between the predicted and observed magnitudes indicates that
the donors only contribute $\simeq 20\% - 30\%$ to the total near-IR
flux of CVs. Distance {\em estimates} (as opposed to lower limits)
obtained by correcting for this offset are uncertain by at least a
factor of $\simeq 2$. 
\item We have carried out a comparison of the mass-transfer
rates along the standard and revised model sequences to the critical
rate for the thermal disk instability that is thought to be
responsible for dwarf nova eruptions (Figure~\ref{fig:dn_theo}). 
Below the gap, both models 
predict mass-transfer rates well below the critical rate and thus
firmly in the unstable regime. This is consistent with the
observations that nearly all CVs below the period gap are dwarf
novae. However, above the period gap, the mass-transfer rates
predicted by both sequences decline faster with orbital period than
the critical rate. The curves cross near $\simeq 5$~hrs in both cases,
thus naively predicting that stable nova-like CVs should only become
the dominant CV population at periods longer than this. By contrast,
observations show that the dwarf-nova fraction reaches a minimum at
the upper edge of the period gap and then gradually {\em rises}
towards longer periods (Figure~\ref{fig:dn_histo}). 
The origin and severity of this discrepancy
are not clear at this point. Long-term mass-transfer cycles offer one
possible explanation. 

\item Based on {\em highly approximate} predicted orbital period
distributions for the standard and revised sequences, we show that the
revised model predicts roughly $3\times$ more long-period CVs,
relative to short-period, pre-bounce CVs (Figure~\ref{fig:porb_dist}
and Table~\ref{tab:pdist}). This may be enough to
resolve a long-standing discrepancy between the standard model and
observations
\citep{2007MNRAS.374.1495P,2008MNRAS.385.1471P,2008MNRAS.385.1485P}.
The faster evolution below the gap in the revised model also
predicts that a higher fraction of CVs should be ``period bouncers''
,i.e. systems that have already evolved beyond the minimum period. For
an assumed space density of $\rho \simeq 2\times 10^{-5}$~pc$^{-3}$, the
predicted numbers of (especially long-period) CVs within 100~pc are in
good agreement with observations. Much higher space densities would
predict too many very bright and nearby CVs.
\item We draw attention to a peculiar feature in the period
distribution at the lower edge of the period gap
(Figure~\ref{fig:porb_dist}). The 
re-establishment of mass transfer causes CVs to execute a
loop (the so-called ``period flag'') in the $P_{orb}-\dot{M}_2$ plane
at this point. Since $\dot{P}_{orb} = 0$ during the evolution across
the flag, one might expect a local maximum in the predicted period
distribution at this point. It can be shown that the visibility of this
local maximum increases with increasing $\dot{M}_2$ below the gap. In
the standard model, this local maximum is not visible in the predicted
period distribution, but in the revised model, it might be. As yet,
there is no observational evidence for an accumulation of systems near
the lower edge of the gap, although none of the existing CV samples
are really suitable for testing this prediction.
%The absence of a spike in
%$P_{min}$ is found in better/larger samples would imply that,
%immediately below the period gap, the angular-motmentum-loss and
%mass-loss rates adopted in the revised model are too high. Since
%angular-momentum-loss rates stronger than GR are necessary to fit the
%mass-radius data, this would most likely point to a different {\em
%form} for the non-GR contribution to the angular momentum losses in
%short-period CVs.
%
\item Our result that $f_{GR} > 1$ contrasts with the study by L08,
who have argued that enhanced angular momentum loss below the 
gap is inconsistent with the WD temperatures in their sample
of eclipsing CVs. However, their findings appear to be applicable only
to the particular {\em type} of enhanced angular momentum loss they
assumed (CAML associated with a circumbinary disk). This predicts
relatively constant and high $\dot{M}_2$ for period bouncers, whereas
$\dot{M}_2$ drops off steeply in this regime in our simple scaled-GR
model. Based on a direct comparison (Figure~\ref{fig:WD}), our revised
model does not seem to be in conflict with the data presented in L08. 
\item An independent attempt to infer $\dot{M}_2(P_{orb})$ for
short-period CVs from the observed donor mass-radius relation has
recently been carried out by SK10. They estimate higher
mass-transfer rates than we do (compare their Figure~5 to our
Figure~\ref{fig:evo1}), but argue that these rates are
nevertheless compatible with purely GR-driven angular momentum
losses. There are numerous technical differences between their
approach and ours, but we believe the main reason for the differences
between the inferred mass-transfer rates is that SK10 
did not correct for donor bloating unrelated to mass loss
(points (7) and (8) above). However, we do not understand how the very
high rates they infer could be driven by GR alone. 
\item We consider the physical plausibility that fully 
convective donors below the period gap should experience residual
magnetic braking. Observations of non-interacting stars certainly
prove that they are capable of generating substantial magnetic
fields. Whether they can also produce MB strong enough to be
consistent with our revised model for CV 
evolution remains to be seen. Recent developments in the study of
isolated low-mass stars may make it possible to test this hypothesis
in the near future. 
\item In order to facilitate the use of our results, we make all
of the donor and evolution sequences we have constructed available in 
electronically readable form. More 
specifically, our broken-power-law donor sequence provides all of the
stellar and photometric properties describing CV secondary stars as a
function of orbital period(Table~\ref{tab:seq_bpl}. The full
self-consistent CV evolution sequences -- both standard and revised --
provide the same, plus a complete set of all other binary parameters
that are relevant to CV evolution (\ref{tab:seq_phys_stan}-\ref{tab:seq_wd_rev}).
\end{enumerate}

\acknowledgments

We are extremely grateful to Jean-Marie Hameury for providing us with
an extended version of Table~2 in \citet{1997A&AS..123..273H}, to
Fedir Sirotkin for supplying the fit coefficients listed in 
Table~\ref{tab:siro}, and to Boris G\"{a}nsicke for sharing with us
some of the results described in G09 prior to their
publication. We would also like to thank Juhan Frank, Boris
G\"{a}nsicke, Natasha Ivanova, Woong-Tae Kim, Ulrich Kolb, Natalia
Landin, Stuart Littlefair, Retha Pretorius, Hans Ritter, Alison Sills,
Fedir Sirotkin, Dean Townsley and Brian Warner for helpful discussions
and much-needed advice. We also gratefully acknowledge financial
support via grants to JP from the National Science Foundation
(AST-0908363) and the Mount Cuba Astronomical Foundation.

%\begin{thebibliography}{}
%\end{thebibliography}
\bibliography{apj-jour,kbp09_v18}

\clearpage
%\begin{landscape}
\begin{turnpage}
%\centering
%\flushleft

%\tabletypesize{\scriptsize}
\begin{deluxetable*}{ccccccccccc}
%\tablewidth{10cm}
%\tablecaption[Binary and evolution parameters along the standard model track.]
\tablecaption{Binary and evolution parameters along the standard model track.}
\tablehead{
\colhead{$M_1 (M_{\odot})$} &
\colhead{$M_2 (M_{\odot})$} &
\colhead{$P_{orb} (hr)$} &
\colhead{$a (R_{\odot})$} &
\colhead{$\log{J}$} &
\colhead{$\log{\dot{J}_{sys}}$} &
\colhead{$\log{\dot{J}_{GR}}$} &
\colhead{$\log{\dot{J}_{CAML}}$} &
\colhead{$\log{\dot{M}_{2}}$} &
\colhead{$\log{\dot{P}_{orb}}$} & 
\colhead{$\log{t [{\rm yrs}]}$}}{Binary and evolution parameters 
along the standard model track.} 
\startdata
% M1    &     M2  &              P  &      A    &     LJ       LJDOT_SYS     LJDOT_GR    LJDOT_CAML  & LMDOT  &       PDOT  &    LTIME
0.75  &    0.040  &      1.3939157  &    0.583  &   50.193  &   32.923  &    32.923  &   30.448  &   -11.077  &     1.648e-14  &  9.85600  \\  
0.75  &    0.045  &      1.3192333  &    0.563  &   50.235  &   33.080  &    33.080  &   30.690  &   -10.930  &     1.525e-14  &  9.82339  \\  
0.75  &    0.050  &      1.2695493  &    0.550  &   50.274  &   33.208  &    33.208  &   30.896  &   -10.806  &     1.447e-14  &  9.79791  \\  
0.75  &    0.055  &      1.2358468  &    0.542  &   50.311  &   33.316  &    33.316  &   31.067  &   -10.712  &     9.841e-15  &  9.77726  \\  
0.75  &    0.060  &      1.2211928  &    0.538  &   50.346  &   33.402  &    33.402  &   31.202  &   -10.646  &     2.183e-15  &  9.75937  \\  
0.75  &    0.062  &      1.2200589  &    0.539  &   50.360  &   33.431  &    33.431  &   31.250  &   -10.625  &    -6.433e-16  &  9.75271  \\  
0.75  &    0.065  &      1.2225936  &    0.540  &   50.380  &   33.469  &    33.469  &   31.314  &   -10.600  &    -5.749e-15  &  9.74308  \\  
0.75  &    0.070  &      1.2367626  &    0.545  &   50.413  &   33.520  &    33.520  &   31.409  &   -10.567  &    -1.279e-14  &  9.72770  \\  
0.75  &    0.075  &      1.2617622  &    0.554  &   50.445  &   33.558  &    33.558  &   31.486  &   -10.549  &    -2.019e-14  &  9.71273  \\  
0.75  &    0.080  &      1.2962487  &    0.565  &   50.476  &   33.585  &    33.585  &   31.553  &   -10.537  &    -2.545e-14  &  9.69775  \\  
0.75  &    0.085  &      1.3366576  &    0.578  &   50.506  &   33.605  &    33.605  &   31.614  &   -10.529  &    -2.889e-14  &  9.68260  \\  
0.75  &    0.090  &      1.3804350  &    0.591  &   50.535  &   33.620  &    33.620  &   31.673  &   -10.520  &    -3.043e-14  &  9.66719  \\  
0.75  &    0.095  &      1.4259075  &    0.606  &   50.562  &   33.632  &    33.632  &   31.724  &   -10.517  &    -3.259e-14  &  9.65147  \\  
0.75  &    0.100  &      1.4727480  &    0.620  &   50.588  &   33.642  &    33.642  &   31.775  &   -10.511  &    -3.313e-14  &  9.63535  \\  
0.75  &    0.110  &      1.5676297  &    0.649  &   50.637  &   33.658  &    33.658  &   31.869  &   -10.502  &    -3.359e-14  &  9.60172  \\   
0.75  &    0.120  &      1.6605382  &    0.677  &   50.681  &   33.672  &    33.672  &   31.954  &   -10.493  &    -3.373e-14  &  9.56611  \\  
0.75  &    0.130  &      1.7504148  &    0.704  &   50.722  &   33.685  &    33.685  &   32.035  &   -10.482  &    -3.293e-14  &  9.52822  \\  
0.75  &    0.140  &      1.8350365  &    0.729  &   50.759  &   33.698  &    33.698  &   32.116  &   -10.466  &    -3.117e-14  &  9.48793  \\  
0.75  &    0.150  &      1.9129707  &    0.752  &   50.794  &   33.713  &    33.713  &   32.190  &   -10.451  &    -3.072e-14  &  9.44525  \\  
0.75  &    0.160  &      1.9873895  &    0.774  &   50.826  &   33.727  &    33.727  &   32.258  &   -10.441  &    -3.189e-14  &  9.39957  \\  
0.75  &    0.170  &      2.0626921  &    0.797  &   50.856  &   33.739  &    33.739  &   32.319  &   -10.429  &    -3.135e-14  &  9.34963  \\  
0.75  &    0.180  &      2.1348046  &    0.818  &   50.884  &   33.751  &    33.751  &   32.380  &   -10.417  &    -3.123e-14  &  9.29493  \\  
0.75  &    0.190  &      2.2070274  &    0.840  &   50.911  &   33.761  &    33.761  &   32.426  &   -10.416  &    -3.427e-14  &  9.23373  \\  
0.75  &    0.200  &      2.2406366  &    0.851  &   50.934  &   33.787  &    33.787  &   32.745  &   -10.135  &     7.110e-14  &  9.17150  \\  
0.75  &    0.200  &      3.5314866  &    1.153  &   51.000  &   35.114  &    33.326  &   33.852  &    -9.091  &    -6.723e-13  &  8.23972  \\  
0.75  &    0.250  &      3.8602676  &    1.244  &   51.102  &   35.272  &    33.415  &   34.203  &    -8.915  &    -7.986e-13  &  8.08974  \\  
0.75  &    0.300  &      4.1508174  &    1.327  &   51.185  &   35.403  &    33.485  &   34.493  &    -8.768  &    -1.010e-12  &  7.94344  \\  
0.75  &    0.350  &      4.4040975  &    1.402  &   51.253  &   35.516  &    33.546  &   34.738  &    -8.643  &    -1.388e-12  &  7.79551  \\  
0.75  &    0.400  &      4.6501833  &    1.476  &   51.313  &   35.613  &    33.594  &   34.952  &    -8.522  &    -1.642e-12  &  7.63661  \\  
0.75  &    0.450  &      4.8819269  &    1.546  &   51.365  &   35.701  &    33.634  &   35.144  &    -8.412  &    -2.005e-12  &  7.45650  \\  
0.75  &    0.500  &      5.0929786  &    1.612  &   51.411  &   35.780  &    33.671  &   35.326  &    -8.306  &    -2.467e-12  &  7.23612  \\  
0.75  &    0.550  &      5.3192987  &    1.681  &   51.453  &   35.854  &    33.699  &   35.470  &    -8.219  &    -3.283e-12  &  6.90486  \\  
0.75  &    0.600  &      5.6150965  &    1.765  &   51.493  &   35.914  &    33.708  &   35.586  &    -8.167  &    -4.884e-12  &  5.00000  
\enddata
\tablecomments{Unless otherwise stated, all physical quantities are
given in cgs units. The sequence provided here is abbreviated. A more 
complete sequence sampled at higher resolution and providing
additional information is available in electronic form.}
\label{tab:seq_phys_stan}
\end{deluxetable*}

%\tabletypesize{\scriptsize}
\begin{deluxetable*}{ccccccccccc}
%\tablewidth{10cm}
%\tablecaption[Binary and evolution parameters along the revised (optimal) model track.]
\tablecaption{Binary and evolution parameters along the revised (optimal) model track.}
\tablehead{
\colhead{$M_1 (M_{\odot})$} &
\colhead{$M_2 (M_{\odot})$} &
\colhead{$P_{orb} (hr)$} &
\colhead{$a (R_{\odot})$} &
\colhead{$\log{J}$} &
\colhead{$\log{\dot{J}_{sys}}$} &
\colhead{$\log{\dot{J}_{GR}}$} &
\colhead{$\log{\dot{J}_{CAML}}$} &
\colhead{$\log{\dot{M}_{2}}$} &
\colhead{$\log{\dot{P}_{orb}}$} &
\colhead{$\log{t [{\rm yrs}]}$}
}{Binary and evolution parameters along the revised (optimal) model track.}
\startdata
%#M1  &         M2 &           P      &         A         LJ      LJDOT_SYS    LJDOT_GR   LJDOT_CAML     LMDOT &       PDOT &     LTIME    
0.750 &     0.040 &        1.4987458 &     0.612 &    50.203 &    33.243 &     32.850 &    30.747 &    -10.789 &      2.635e-14 &     9.52767    \\  
0.750 &     0.045 &        1.4370365 &     0.596 &    50.248 &    33.386 &     32.993 &    30.981 &    -10.650 &      2.508e-14 &     9.49134    \\  
0.750 &     0.050 &        1.3951191 &     0.586 &    50.288 &    33.505 &     33.113 &    31.177 &    -10.539 &      2.039e-14 &     9.46222    \\  
0.750 &     0.055 &        1.3713966 &     0.581 &    50.326 &    33.604 &     33.211 &    31.339 &    -10.453 &      1.127e-14 &     9.43766    \\  
0.750 &     0.060 &        1.3627148 &     0.579 &    50.362 &    33.684 &     33.291 &    31.477 &    -10.386 &      4.699e-16 &     9.41603    \\  
0.750 &     0.061 &        1.3626145 &     0.580 &    50.369 &    33.698 &     33.305 &    31.501 &    -10.375 &     -1.791e-15 &     9.41196    \\  
0.750 &     0.065 &        1.3665367 &     0.582 &    50.396 &    33.749 &     33.356 &    31.594 &    -10.335 &     -1.084e-14 &     9.39629    \\  
0.750 &     0.070 &        1.3809809 &     0.587 &    50.429 &    33.801 &     33.408 &    31.693 &    -10.298 &     -2.326e-14 &     9.37770    \\  
0.750 &     0.075 &        1.4046595 &     0.595 &    50.461 &    33.842 &     33.449 &    31.779 &    -10.270 &     -3.400e-14 &     9.35972    \\  
0.750 &     0.080 &        1.4345706 &     0.604 &    50.491 &    33.875 &     33.482 &    31.857 &    -10.247 &     -4.201e-14 &     9.34208    \\  
0.750 &     0.085 &        1.4687294 &     0.615 &    50.520 &    33.902 &     33.509 &    31.927 &    -10.229 &     -4.845e-14 &     9.32457    \\  
0.750 &     0.090 &        1.5058189 &     0.627 &    50.547 &    33.924 &     33.532 &    31.992 &    -10.214 &     -5.410e-14 &     9.30707    \\  
0.750 &     0.095 &        1.5453577 &     0.639 &    50.574 &    33.943 &     33.551 &    32.052 &    -10.201 &     -5.801e-14 &     9.28941    \\  
0.750 &     0.100 &        1.5861744 &     0.651 &    50.599 &    33.960 &     33.567 &    32.108 &    -10.190 &     -6.135e-14 &     9.27154    \\  
0.750 &     0.110 &        1.6698073 &     0.677 &    50.646 &    33.987 &     33.594 &    32.211 &    -10.169 &     -6.470e-14 &     9.23484    \\   
0.750 &     0.120 &        1.7537710 &     0.702 &    50.689 &    34.010 &     33.617 &    32.304 &    -10.151 &     -6.673e-14 &     9.19654    \\  
0.750 &     0.130 &        1.8356292 &     0.726 &    50.729 &    34.030 &     33.637 &    32.391 &    -10.134 &     -6.739e-14 &     9.15630    \\  
0.750 &     0.140 &        1.9129674 &     0.749 &    50.766 &    34.049 &     33.656 &    32.475 &    -10.113 &     -6.533e-14 &     9.11396    \\  
0.750 &     0.150 &        1.9860910 &     0.771 &    50.799 &    34.068 &     33.675 &    32.552 &    -10.095 &     -6.588e-14 &     9.06922    \\  
0.750 &     0.160 &        2.0578237 &     0.793 &    50.831 &    34.085 &     33.692 &    32.618 &    -10.083 &     -6.962e-14 &     9.02130    \\  
0.750 &     0.170 &        2.1300629 &     0.814 &    50.861 &    34.099 &     33.706 &    32.684 &    -10.068 &     -6.916e-14 &     8.96904    \\  
0.750 &     0.180 &        2.1996207 &     0.835 &    50.889 &    34.113 &     33.720 &    32.747 &    -10.052 &     -6.840e-14 &     8.91176    \\  
0.750 &     0.190 &        2.2614846 &     0.853 &    50.914 &    34.129 &     33.736 &    32.828 &    -10.008 &     -4.677e-14 &     8.84910    \\  
0.750 &     0.200 &        2.2406366 &     0.851 &    50.934 &    34.180 &     33.787 &    33.160 &     -9.720 &      2.087e-13 &     8.79656    \\  
0.750 &     0.200 &        3.2511822 &     1.091 &    50.988 &    34.974 &     33.410 &    33.716 &     -9.218 &     -4.536e-13 &     8.38885    \\  
0.750 &     0.250 &        3.5602183 &     1.179 &    51.090 &    35.131 &     33.497 &    34.060 &     -9.048 &     -5.799e-13 &     8.24679    \\  
0.750 &     0.300 &        3.8453095 &     1.261 &    51.174 &    35.259 &     33.563 &    34.343 &     -8.907 &     -7.467e-13 &     8.10919    \\  
0.750 &     0.350 &        4.1104217 &     1.339 &    51.244 &    35.368 &     33.616 &    34.581 &     -8.789 &     -1.048e-12 &     7.97036    \\  
0.750 &     0.400 &        4.3833222 &     1.419 &    51.304 &    35.461 &     33.654 &    34.784 &     -8.681 &     -1.311e-12 &     7.82103    \\  
0.750 &     0.450 &        4.6675438 &     1.501 &    51.358 &    35.542 &     33.680 &    34.961 &     -8.589 &     -1.711e-12 &     7.64906    \\  
0.750 &     0.500 &        4.9503757 &     1.582 &    51.407 &    35.614 &     33.700 &    35.129 &     -8.498 &     -2.117e-12 &     7.43343    \\  
0.750 &     0.550 &        5.2504943 &     1.667 &    51.451 &    35.681 &     33.712 &    35.269 &     -8.419 &     -2.667e-12 &     7.10266    \\  
0.750 &     0.600 &        5.6066878 &     1.764 &    51.493 &    35.736 &     33.710 &    35.386 &     -8.366 &     -3.598e-12 &     5.00000  
\enddata                                                                                                       \tablecomments{Unless otherwise stated, all physical quantities are
given in cgs units. The sequence provided here is abbreviated. A more 
complete sequence sampled at higher resolution and providing
additional information is available in electronic form.}
\label{tab:seq_phys_rev}
\end{deluxetable*}

%\tabletypesize{\scriptsize}
\begin{deluxetable*}{cccccccccccccccc}
%\tablewidth{10cm}
%\tablecaption[Physical and photometric donor properties along the standard model track.]
\tablecaption{Physical and photometric donor properties along the standard model track.}
\tablehead{
\colhead{$P_{orb} (hr)$} &
\colhead{$M_2 (M_{\odot})$} &
\colhead{$R_2 (M_{\odot})$} &
\colhead{$T_{eff,2} (K)$} &
\colhead{$\log{g_2}$} &
\colhead{$\log{L_2}$} &
\colhead{$\zeta_2$} &
\colhead{$M_U$} &
\colhead{$M_B$} &
\colhead{$M_V$} &
\colhead{$M_R$} &
\colhead{$M_I$} &
\colhead{$M_J$} &
\colhead{$M_H$} &
\colhead{$M_K$} &
\colhead{$SpT$} 
}{Physical and photometric donor properties along the standard model track.}
\startdata
%   p2   &     m2  &        r2  &    t2    &     g2   &    l2   &     XI    &     all_ux  &    all_b  &   all_v  &   all_r  &   all_i  &   all_j  &   all_h  &   all_k  &   spt
 1.3939  &  0.0400  &   0.098  &    1091  &   5.054  &   28.68  &   -0.015  &    36.29  &    30.29  &   26.00  &   21.51  &   18.56  &   14.62  &   14.73  &   14.68  &    T    \\
 1.3192  &  0.0450  &   0.098  &    1306  &   5.104  &   28.99  &    0.044  &    33.24  &    27.87  &   24.91  &   20.67  &   17.73  &   13.81  &   13.80  &   13.66  &    T    \\
 1.2695  &  0.0500  &   0.099  &    1507  &   5.143  &   29.25  &    0.094  &    31.50  &    26.26  &   23.75  &   19.65  &   16.71  &   13.11  &   13.02  &   12.96  &    T    \\
 1.2358  &  0.0550  &   0.101  &    1702  &   5.173  &   29.47  &    0.168  &    30.23  &    25.14  &   22.61  &   18.73  &   15.96  &   12.63  &   12.42  &   12.32  &    T    \\
 1.2212  &  0.0600  &   0.103  &    1910  &   5.194  &   29.68  &    0.275  &    28.95  &    24.61  &   21.42  &   18.36  &   15.95  &   12.43  &   11.69  &   11.20  &   L0.3  \\
 1.2201  &  0.0620  &   0.104  &    1986  &   5.199  &   29.76  &    0.313  &    28.74  &    24.51  &   21.33  &   18.24  &   15.82  &   12.32  &   11.58  &   11.11  &   L0.1  \\
 1.2226  &  0.0650  &   0.105  &    2094  &   5.205  &   29.87  &    0.375  &    27.74  &    23.54  &   20.72  &   17.99  &   15.52  &   12.00  &   11.29  &   10.87  &   M9.7  \\
 1.2368  &  0.0700  &   0.109  &    2254  &   5.210  &   30.02  &    0.461  &    26.07  &    21.87  &   19.68  &   17.53  &   14.98  &   11.44  &   10.76  &   10.45  &   M9.1  \\
 1.2618  &  0.0750  &   0.113  &    2404  &   5.209  &   30.16  &    0.554  &    24.26  &    20.52  &   18.59  &   16.79  &   14.34  &   11.06  &   10.40  &   10.11  &   M8.0  \\
 1.2962  &  0.0800  &   0.117  &    2529  &   5.203  &   30.29  &    0.629  &    22.75  &    19.50  &   17.68  &   16.05  &   13.75  &   10.80  &   10.16  &    9.85  &   M7.2  \\
 1.3367  &  0.0850  &   0.122  &    2630  &   5.195  &   30.39  &    0.680  &    21.58  &    18.69  &   16.95  &   15.43  &   13.27  &   10.57  &    9.94  &    9.63  &   M6.8  \\
 1.3804  &  0.0900  &   0.127  &    2716  &   5.185  &   30.48  &    0.708  &    20.62  &    18.02  &   16.33  &   14.91  &   12.86  &   10.37  &    9.75  &    9.44  &   M6.6  \\
 1.4259  &  0.0950  &   0.132  &    2786  &   5.174  &   30.56  &    0.744  &    19.89  &    17.49  &   15.85  &   14.49  &   12.53  &   10.20  &    9.59  &    9.28  &   M6.3  \\
 1.4727  &  0.1000  &   0.137  &    2844  &   5.163  &   30.63  &    0.752  &    19.31  &    17.05  &   15.45  &   14.14  &   12.25  &   10.06  &    9.45  &    9.14  &   M6.1  \\
 1.5676  &  0.1100  &   0.148  &    2938  &   5.141  &   30.75  &    0.768  &    18.39  &    16.36  &   14.81  &   13.56  &   11.80  &    9.78  &    9.19  &    8.88  &   M5.8  \\
 1.6605  &  0.1200  &   0.158  &    3006  &   5.120  &   30.85  &    0.775  &    17.74  &    15.85  &   14.34  &   13.14  &   11.46  &    9.57  &    8.97  &    8.67  &   M5.5  \\
 1.7504  &  0.1300  &   0.168  &    3062  &   5.102  &   30.93  &    0.769  &    17.24  &    15.45  &   13.96  &   12.80  &   11.18  &    9.37  &    8.78  &    8.48  &   M5.2  \\
 1.8350  &  0.1400  &   0.178  &    3105  &   5.085  &   31.01  &    0.741  &    16.84  &    15.12  &   13.65  &   12.52  &   10.95  &    9.20  &    8.62  &    8.32  &   M5.0  \\
 1.9130  &  0.1500  &   0.187  &    3155  &   5.071  &   31.08  &    0.733  &    16.42  &    14.78  &   13.33  &   12.23  &   10.72  &    9.04  &    8.46  &    8.17  &   M4.8  \\
 1.9874  &  0.1600  &   0.196  &    3199  &   5.058  &   31.14  &    0.737  &    16.05  &    14.48  &   13.05  &   11.98  &   10.52  &    8.90  &    8.32  &    8.03  &   M4.5  \\
 2.0627  &  0.1700  &   0.205  &    3221  &   5.045  &   31.20  &    0.742  &    15.84  &    14.30  &   12.87  &   11.81  &   10.37  &    8.78  &    8.20  &    7.91  &   M4.4  \\
 2.1348  &  0.1800  &   0.214  &    3251  &   5.033  &   31.24  &    0.736  &    15.60  &    14.09  &   12.68  &   11.63  &   10.22  &    8.65  &    8.08  &    7.79  &   M4.3  \\
 2.2070  &  0.1900  &   0.223  &    3273  &   5.021  &   31.29  &    0.783  &    15.40  &    13.91  &   12.51  &   11.48  &   10.08  &    8.54  &    7.97  &    7.69  &   M4.1  \\
 2.2406  &  0.2000  &   0.229  &    3289  &   5.019  &   31.33  &   -0.177  &    15.27  &    13.79  &   12.40  &   11.38  &    9.99  &    8.47  &    7.90  &    7.61  &   M4.0  \\
 3.5315  &  0.2000  &   0.310  &    3296  &   4.756  &   31.59  &    0.636  &    14.48  &    13.09  &   11.76  &   10.74  &    9.35  &    7.80  &    7.22  &    6.94  &   M4.2  \\
 3.8603  &  0.2500  &   0.356  &    3373  &   4.734  &   31.75  &    0.618  &    13.90  &    12.54  &   11.23  &   10.24  &    8.91  &    7.42  &    6.84  &    6.57  &   M3.8  \\
 4.1508  &  0.3000  &   0.398  &    3428  &   4.715  &   31.88  &    0.609  &    13.47  &    12.12  &   10.83  &    9.87  &    8.56  &    7.12  &    6.54  &    6.28  &   M3.6  \\
 4.4041  &  0.3500  &   0.438  &    3491  &   4.699  &   31.99  &    0.623  &    13.06  &    11.73  &   10.45  &    9.51  &    8.24  &    6.85  &    6.27  &    6.02  &   M3.2  \\
 4.6502  &  0.4000  &   0.477  &    3548  &   4.683  &   32.10  &    0.636  &    12.72  &    11.38  &   10.11  &    9.20  &    7.96  &    6.61  &    6.03  &    5.79  &   M3.0  \\
 4.8819  &  0.4500  &   0.515  &    3614  &   4.668  &   32.19  &    0.655  &    12.39  &    11.04  &    9.78  &    8.88  &    7.69  &    6.38  &    5.79  &    5.57  &   M2.7  \\
 5.0930  &  0.5000  &   0.551  &    3690  &   4.654  &   32.29  &    0.650  &    12.07  &    10.71  &    9.44  &    8.57  &    7.43  &    6.17  &    5.57  &    5.36  &   M2.3  \\
 5.3193  &  0.5500  &   0.590  &    3784  &   4.636  &   32.39  &    0.739  &    11.74  &    10.34  &    9.07  &    8.22  &    7.15  &    5.94  &    5.31  &    5.13  &   M1.9  \\
 5.6151  &  0.6000  &   0.634  &    3899  &   4.612  &   32.51  &    0.855  &    11.38  &     9.92  &    8.66  &    7.84  &    6.84  &    5.68  &    5.04  &    4.89  &   M1.4 
\enddata
\tablecomments{Unless otherwise stated, all physical quantities are
given in cgs units. UBVRI
magnitudes are given on the Johnson-Cousins system
\citep{1990PASP..102.1181B}, JHK are given on the CIT system
\citep{1982AJ.....87.1029E,1982AJ.....87.1893E}. The sequence provided here is abbreviated. A more 
complete sequence sampled at higher resolution and providing
additional information is available in electronic form.}
\label{tab:seq_donor_stan}
\end{deluxetable*}

%\tabletypesize{\scriptsize}
\begin{deluxetable*}{cccccccccccccccc}
%\tablewidth{10cm}
%\tablecaption[Physical and photometric donor properties along the revised (optimal) model track.]
\tablecaption{Physical and photometric donor properties along the revised (optimal) model track.}
\tablehead{
\colhead{$P_{orb} (hr)$} &
\colhead{$M_2 (M_{\odot})$} &
\colhead{$R_2 (M_{\odot})$} &
\colhead{$T_{eff,2} (K)$} &
\colhead{$\log{g_2}$} &
\colhead{$\log{L_2}$} &
\colhead{$\zeta_2$} &
\colhead{$M_U$} &
\colhead{$M_B$} &
\colhead{$M_V$} &
\colhead{$M_R$} &
\colhead{$M_I$} &
\colhead{$M_J$} &
\colhead{$M_H$} &
\colhead{$M_K$} &
\colhead{$SpT$} 
}{Physical and photometric donor properties along the revised (optimal) model track.}
\startdata
%   p2 &   & m2 &   r2   &   t2  &     g2 &   l2    &   XI &      all_ux &  all_b &  all_v &  all_r &  all_i &  all_j &  all_h &  all_k  &  spt
 1.4987 & 0.0400 &  0.103 &  1321 &  5.012 &  29.05 &  0.063 &    32.89 &  27.54 &  24.68 &  20.46 &  17.51 &  13.66 &  13.62 &  13.47 &  T      \\
 1.4370 & 0.0450 &  0.104 &  1538 &  5.055 &  29.32 &  0.104 &    31.17 &  25.92 &  23.42 &  19.35 &  16.45 &  12.96 &  12.82 &  12.75 &  T      \\
 1.3951 & 0.0500 &  0.106 &  1738 &  5.088 &  29.55 &  0.158 &    30.26 &  25.44 &  22.48 &  18.94 &  16.35 &  12.79 &  12.23 &  11.88 &  T      \\
 1.3714 & 0.0550 &  0.108 &  1932 &  5.113 &  29.75 &  0.232 &    28.81 &  24.47 &  21.29 &  18.19 &  15.78 &  12.28 &  11.56 &  11.08 &  L0.0   \\
 1.3627 & 0.0600 &  0.110 &  2109 &  5.130 &  29.92 &  0.304 &    27.48 &  23.28 &  20.54 &  17.84 &  15.36 &  11.85 &  11.15 &  10.73 &  M9.6   \\
 1.3626 & 0.0610 &  0.111 &  2143 &  5.133 &  29.95 &  0.320 &    27.10 &  22.92 &  20.31 &  17.75 &  15.25 &  11.73 &  11.03 &  10.64 &  M9.5   \\
 1.3665 & 0.0650 &  0.113 &  2265 &  5.141 &  30.07 &  0.374 &    25.78 &  21.66 &  19.53 &  17.44 &  14.88 &  11.33 &  10.66 &  10.34 &  M9.2   \\
 1.3810 & 0.0700 &  0.117 &  2404 &  5.146 &  30.20 &  0.447 &    24.05 &  20.41 &  18.50 &  16.72 &  14.27 &  10.98 &  10.33 &  10.02 &  M8.1   \\
 1.4047 & 0.0750 &  0.121 &  2518 &  5.147 &  30.31 &  0.512 &    22.68 &  19.48 &  17.67 &  16.04 &  13.73 &  10.74 &  10.10 &   9.79 &  M7.3   \\
 1.4346 & 0.0800 &  0.125 &  2618 &  5.145 &  30.41 &  0.559 &    21.55 &  18.69 &  16.96 &  15.44 &  13.27 &  10.53 &   9.91 &   9.59 &  M6.9   \\
 1.4687 & 0.0850 &  0.130 &  2698 &  5.140 &  30.49 &  0.599 &    20.65 &  18.06 &  16.38 &  14.95 &  12.88 &  10.35 &   9.73 &   9.41 &  M6.6   \\
 1.5058 & 0.0900 &  0.135 &  2767 &  5.134 &  30.56 &  0.630 &    19.94 &  17.54 &  15.90 &  14.53 &  12.55 &  10.18 &   9.57 &   9.25 &  M6.4   \\
 1.5454 & 0.0950 &  0.139 &  2825 &  5.127 &  30.63 &  0.655 &    19.37 &  17.12 &  15.52 &  14.19 &  12.29 &  10.05 &   9.44 &   9.13 &  M6.2   \\
 1.5862 & 0.1000 &  0.144 &  2877 &  5.120 &  30.69 &  0.674 &    18.86 &  16.73 &  15.16 &  13.88 &  12.04 &   9.91 &   9.31 &   9.00 &  M6.0   \\
 1.6698 & 0.1100 &  0.154 &  2958 &  5.104 &  30.80 &  0.696 &    18.11 &  16.15 &  14.62 &  13.39 &  11.66 &   9.67 &   9.08 &   8.77 &  M5.7   \\
 1.7538 & 0.1200 &  0.164 &  3020 &  5.089 &  30.89 &  0.710 &    17.54 &  15.69 &  14.19 &  13.01 &  11.35 &   9.47 &   8.88 &   8.57 &  M5.4   \\
 1.8356 & 0.1300 &  0.173 &  3069 &  5.074 &  30.96 &  0.713 &    17.10 &  15.34 &  13.86 &  12.71 &  11.10 &   9.30 &   8.71 &   8.41 &  M5.2   \\
 1.9130 & 0.1400 &  0.183 &  3112 &  5.061 &  31.04 &  0.698 &    16.71 &  15.02 &  13.56 &  12.43 &  10.87 &   9.13 &   8.55 &   8.25 &  M5.0   \\
 1.9861 & 0.1500 &  0.192 &  3162 &  5.049 &  31.10 &  0.698 &    16.30 &  14.69 &  13.24 &  12.15 &  10.65 &   8.98 &   8.40 &   8.10 &  M4.7   \\
 2.0578 & 0.1600 &  0.201 &  3199 &  5.038 &  31.16 &  0.722 &    15.98 &  14.42 &  12.99 &  11.92 &  10.46 &   8.84 &   8.27 &   7.97 &  M4.5   \\
 2.1301 & 0.1700 &  0.209 &  3228 &  5.026 &  31.21 &  0.717 &    15.75 &  14.22 &  12.81 &  11.75 &  10.32 &   8.73 &   8.15 &   7.86 &  M4.4   \\
 2.1996 & 0.1800 &  0.218 &  3251 &  5.015 &  31.26 &  0.715 &    15.55 &  14.04 &  12.64 &  11.59 &  10.18 &   8.61 &   8.04 &   7.75 &  M4.3   \\
 2.2615 & 0.1900 &  0.226 &  3273 &  5.007 &  31.31 &  0.615 &    15.36 &  13.88 &  12.48 &  11.45 &  10.06 &   8.51 &   7.94 &   7.66 &  M4.1   \\
 2.2406 & 0.2000 &  0.229 &  3289 &  5.019 &  31.33 & -0.240 &    15.27 &  13.79 &  12.40 &  11.38 &   9.99 &   8.47 &   7.90 &   7.61 &  M4.0   \\
 3.2512 & 0.2000 &  0.293 &  3296 &  4.804 &  31.55 &  0.620 &    14.62 &  13.22 &  11.87 &  10.85 &   9.47 &   7.92 &   7.35 &   7.06 &  M4.2   \\
 3.5602 & 0.2500 &  0.337 &  3373 &  4.781 &  31.70 &  0.629 &    14.03 &  12.66 &  11.34 &  10.35 &   9.02 &   7.54 &   6.96 &   6.69 &  M3.8   \\
 3.8453 & 0.3000 &  0.378 &  3428 &  4.759 &  31.83 &  0.636 &    13.58 &  12.23 &  10.93 &   9.97 &   8.67 &   7.23 &   6.66 &   6.40 &  M3.5   \\
 4.1104 & 0.3500 &  0.418 &  3491 &  4.739 &  31.95 &  0.660 &    13.16 &  11.83 &  10.54 &   9.60 &   8.34 &   6.95 &   6.38 &   6.13 &  M3.2   \\
 4.3833 & 0.4000 &  0.458 &  3548 &  4.718 &  32.06 &  0.700 &    12.81 &  11.47 &  10.19 &   9.28 &   8.05 &   6.70 &   6.12 &   5.88 &  M2.9   \\
 4.6675 & 0.4500 &  0.500 &  3614 &  4.694 &  32.17 &  0.749 &    12.45 &  11.10 &   9.83 &   8.94 &   7.75 &   6.45 &   5.86 &   5.64 &  M2.6   \\
 4.9504 & 0.5000 &  0.541 &  3690 &  4.670 &  32.27 &  0.762 &    12.11 &  10.75 &   9.48 &   8.61 &   7.47 &   6.21 &   5.61 &   5.40 &  M2.3   \\
 5.2505 & 0.5500 &  0.585 &  3793 &  4.643 &  32.39 &  0.839 &    11.74 &  10.33 &   9.07 &   8.22 &   7.15 &   5.95 &   5.32 &   5.15 &  M1.9   \\
 5.6067 & 0.6000 &  0.633 &  3908 &  4.612 &  32.51 &  0.930 &    11.36 &   9.90 &   8.64 &   7.82 &   6.83 &   5.68 &   5.03 &   4.88 &  M1.4 
\enddata
\tablecomments{Unless otherwise stated, all physical quantities are
given in cgs units. UBVRI magnitudes are given on the Johnson-Cousins system
\citep{1990PASP..102.1181B}, JHK are given on the CIT system
\citep{1982AJ.....87.1029E,1982AJ.....87.1893E}. The sequence provided here is abbreviated. A more  
complete sequence sampled at higher resolution and providing
additional information is available in electronic form.}
\label{tab:seq_donor_rev}
\end{deluxetable*}

%\tabletypesize{\scriptsize}
\begin{deluxetable*}{cccccccccccccccc}
%\tablewidth{10cm}
%\tablecaption[White dwarf properties along the standard model track.]
\tablecaption{White dwarf properties along the standard model track.}
\tablehead{
\colhead{$P_{orb} (hr)$} &
\colhead{$M_1 (M_{\odot})$} &
\colhead{$R_1 (10^{8}~cm)$} &
\colhead{$\log{g_1}$} &
\colhead{$<T_{eff,1} (K)>$} &
\colhead{$T_{eff,1,lo} (K)$} &
\colhead{$T_{eff,1,hi} (K)$} &
\colhead{$<\log{L_1}>$} &
\colhead{$M_U$} &
\colhead{$M_B$} &
\colhead{$M_V$} &
\colhead{$M_R$} &
\colhead{$M_I$} &
\colhead{$M_J$} &
\colhead{$M_H$} &
\colhead{$M_K$}
}{White dwarf properties along the standard model track.}
\startdata
%    p  &       m1  &     rad8  &     logg  &     teff  &  teff_lo  &  teff_hi  &     llum  &      u    &      b    &      v    &   r     &      i    &      j    &      h    &      k   
1.3939  &     0.75  &    7.289  &    8.273  &     8132  &     7647  &     8647  &    30.22  &    12.96  &    13.63  &    13.37  &  13.20  &    12.99  &    12.90  &    12.72  &    12.76  \\ 
1.3192  &     0.75  &    7.311  &    8.270  &     8852  &     8306  &     9434  &    30.37  &    12.57  &    13.26  &    13.03  &  12.91  &    12.75  &    12.73  &    12.59  &    12.64  \\ 
1.2695  &     0.75  &    7.324  &    8.268  &     9517  &     8912  &    10162  &    30.50  &    12.27  &    12.96  &    12.75  &  12.66  &    12.55  &    12.59  &    12.48  &    12.55  \\ 
1.2358  &     0.75  &    7.343  &    8.266  &    10064  &     9415  &    10758  &    30.60  &    12.06  &    12.74  &    12.55  &  12.49  &    12.40  &    12.49  &    12.40  &    12.48  \\ 
1.2212  &     0.75  &    7.353  &    8.265  &    10464  &     9779  &    11197  &    30.66  &    11.91  &    12.59  &    12.40  &  12.37  &    12.31  &    12.43  &    12.34  &    12.43  \\ 
1.2201  &     0.75  &    7.357  &    8.265  &    10594  &     9897  &    11341  &    30.69  &    11.87  &    12.55  &    12.36  &  12.34  &    12.28  &    12.40  &    12.33  &    12.42  \\ 
1.2226  &     0.75  &    7.362  &    8.264  &    10751  &    10040  &    11514  &    30.71  &    11.82  &    12.50  &    12.32  &  12.29  &    12.24  &    12.38  &    12.31  &    12.40  \\ 
1.2368  &     0.75  &    7.370  &    8.263  &    10966  &    10234  &    11751  &    30.75  &    11.76  &    12.44  &    12.26  &  12.24  &    12.20  &    12.35  &    12.29  &    12.37  \\ 
1.2618  &     0.75  &    7.373  &    8.263  &    11085  &    10342  &    11882  &    30.77  &    11.72  &    12.41  &    12.23  &  12.21  &    12.18  &    12.33  &    12.28  &    12.36  \\ 
1.2962  &     0.75  &    7.375  &    8.262  &    11165  &    10414  &    11971  &    30.78  &    11.70  &    12.38  &    12.21  &  12.20  &    12.16  &    12.32  &    12.27  &    12.36  \\ 
1.3367  &     0.75  &    7.377  &    8.262  &    11219  &    10463  &    12030  &    30.79  &    11.68  &    12.37  &    12.19  &  12.19  &    12.15  &    12.31  &    12.27  &    12.35  \\ 
1.3804  &     0.75  &    7.378  &    8.262  &    11280  &    10518  &    12097  &    30.80  &    11.67  &    12.35  &    12.18  &  12.17  &    12.14  &    12.31  &    12.26  &    12.35  \\ 
1.4259  &     0.75  &    7.379  &    8.262  &    11300  &    10536  &    12120  &    30.80  &    11.66  &    12.35  &    12.17  &  12.17  &    12.14  &    12.30  &    12.26  &    12.35  \\ 
1.4727  &     0.75  &    7.380  &    8.262  &    11341  &    10573  &    12165  &    30.81  &    11.65  &    12.33  &    12.16  &  12.16  &    12.14  &    12.30  &    12.26  &    12.34  \\ 
1.6605  &     0.75  &    7.383  &    8.262  &    11459  &    10682  &    12293  &    30.83  &    11.61  &    12.30  &    12.13  &  12.14  &    12.12  &    12.28  &    12.24  &    12.33  \\ 
1.7504  &     0.75  &    7.385  &    8.261  &    11528  &    10748  &    12364  &    30.84  &    11.59  &    12.28  &    12.11  &  12.12  &    12.11  &    12.28  &    12.24  &    12.33  \\ 
1.8350  &     0.75  &    7.388  &    8.261  &    11628  &    10844  &    12469  &    30.85  &    11.57  &    12.26  &    12.09  &  12.11  &    12.09  &    12.27  &    12.23  &    12.32  \\ 
1.9130  &     0.75  &    7.392  &    8.261  &    11723  &    10935  &    12567  &    30.87  &    11.55  &    12.24  &    12.07  &  12.09  &    12.08  &    12.26  &    12.22  &    12.31  \\ 
1.9874  &     0.75  &    7.394  &    8.260  &    11786  &    10996  &    12634  &    30.88  &    11.53  &    12.23  &    12.06  &  12.08  &    12.07  &    12.25  &    12.21  &    12.31  \\ 
2.0627  &     0.75  &    7.396  &    8.260  &    11863  &    11069  &    12714  &    30.89  &    11.51  &    12.21  &    12.05  &  12.06  &    12.06  &    12.24  &    12.20  &    12.30  \\ 
2.1348  &     0.75  &    7.399  &    8.260  &    11940  &    11143  &    12794  &    30.90  &    11.49  &    12.19  &    12.03  &  12.05  &    12.05  &    12.24  &    12.20  &    12.29  \\ 
2.2070  &     0.75  &    7.399  &    8.260  &    11947  &    11149  &    12801  &    30.90  &    11.49  &    12.19  &    12.03  &  12.05  &    12.05  &    12.24  &    12.20  &    12.29  \\ 
2.2406  &     0.75  &    7.445  &    8.254  &    13908  &    13062  &    14808  &    31.17  &    11.14  &    11.86  &    11.75  &  11.82  &    11.86  &    12.09  &    12.07  &    12.17  \\ 
3.5315  &     0.75  &    7.739  &    8.221  &    25581  &    23806  &    27487  &    32.26  &     9.37  &    10.46  &    10.62  &  10.74  &    10.89  &    11.28  &    11.32  &    11.43  \\ 
3.6024  &     0.75  &    7.757  &    8.219  &    26182  &    24345  &    28158  &    32.30  &     9.29  &    10.40  &    10.57  &  10.69  &    10.84  &    11.23  &    11.28  &    11.39  \\ 
3.8603  &     0.75  &    7.822  &    8.211  &    28579  &    26516  &    30802  &    32.46  &     9.01  &    10.18  &    10.37  &  10.50  &    10.66  &    11.07  &    11.13  &    11.25  \\ 
4.1508  &     0.75  &    7.917  &    8.201  &    31342  &    29018  &    33853  &    32.63  &     8.74  &     9.94  &    10.16  &  10.30  &    10.47  &    10.89  &    10.95  &    11.08  \\ 
4.4041  &     0.75  &    8.024  &    8.189  &    33935  &    31368  &    36712  &    32.78  &     8.51  &     9.74  &     9.99  &  10.13  &    10.30  &    10.73  &    10.80  &    10.92  \\ 
4.6502  &     0.75  &    8.119  &    8.179  &    36632  &    33803  &    39697  &    32.93  &     8.32  &     9.58  &     9.84  &   9.98  &    10.16  &    10.60  &    10.67  &    10.79  \\ 
4.8819  &     0.75  &    8.202  &    8.170  &    39294  &    36218  &    42631  &    33.06  &     8.18  &     9.45  &     9.72  &   9.87  &    10.05  &    10.50  &    10.56  &    10.69  \\ 
5.0930  &     0.75  &    8.290  &    8.161  &    42041  &    38714  &    45654  &    33.18  &     8.06  &     9.34  &     9.62  &   9.77  &     9.95  &    10.40  &    10.46  &    10.59  \\ 
5.3193  &     0.75  &    8.364  &    8.153  &    44370  &    40829  &    48218  &    33.29  &     7.96  &     9.25  &     9.54  &   9.69  &     9.87  &    10.32  &    10.39  &    10.52  \\ 
5.6151  &     0.75  &    8.415  &    8.148  &    45843  &    42161  &    49846  &    33.35  &     7.91  &     9.20  &     9.49  &   9.64  &     9.82  &    10.28  &    10.35  &    10.47   
\enddata
\tablecomments{Unless otherwise stated, all physical quantities are
given in cgs units. The three effective temperatures listed correspond to the secular mean temperature 
($<T_{eff,1}>$) and the temperatures 5\% ($T_{eff,1,lo}$) and 95\% ($T_{eff,1,hi}$) through a given nova cycle. 
UBVRI magnitudes are given on the Johnson-Cousins system
\citep{1990PASP..102.1181B}, JHK are given on the CIT system
\citep{1982AJ.....87.1029E,1982AJ.....87.1893E}. The sequence provided here is abbreviated. A more 
complete sequence sampled at higher resolution and providing
additional information is available in electronic form.}
\label{tab:seq_wd_stan}
\end{deluxetable*}

%\tabletypesize{\scriptsize}
\begin{deluxetable*}{cccccccccccccccc}
%\tablewidth{10cm}
%\tablecaption[White dwarf properties along the revised (optimal) model track.]
\tablecaption{White dwarf properties along the revised (optimal) model track.}
\tablehead{
\colhead{$P_{orb} (hr)$} &
\colhead{$M_1 (M_{\odot})$} &
\colhead{$R_1 (10^{8}~cm)$} &
\colhead{$\log{g_1}$} &
\colhead{$<T_{eff,1} (K)>$} &
\colhead{$T_{eff,1,lo} (K)$} &
\colhead{$T_{eff,1,hi} (K)$} &
\colhead{$<\log{L_1}>$} &
\colhead{$M_U$} &
\colhead{$M_B$} &
\colhead{$M_V$} &
\colhead{$M_R$} &
\colhead{$M_I$} &
\colhead{$M_J$} &
\colhead{$M_H$} &
\colhead{$M_K$} 
}{White dwarf properties along the revised (optimal) model track.}
\startdata
%    p      m1    rad8    logg     teff   teff_lo  teff_hi  llum    u        b        v       r          i     j      h        k    
1.4987 &  0.75 & 7.328 & 8.268 &   9613 &  9001 & 10267 &  30.51 & 12.23 & 12.92 & 12.72  &  12.63 & 12.52 & 12.57 & 12.46 & 12.54   \\
1.4370 &  0.75 & 7.352 & 8.265 &  10440 &  9757 & 11170 &  30.66 & 11.92 & 12.60 & 12.41  &  12.38 & 12.31 & 12.43 & 12.35 & 12.44   \\
1.3951 &  0.75 & 7.375 & 8.263 &  11152 & 10402 & 11956 &  30.78 & 11.70 & 12.39 & 12.21  &  12.20 & 12.17 & 12.32 & 12.27 & 12.36   \\
1.3714 &  0.75 & 7.391 & 8.261 &  11710 & 10922 & 12554 &  30.86 & 11.55 & 12.24 & 12.08  &  12.09 & 12.08 & 12.26 & 12.22 & 12.31   \\
1.3627 &  0.75 & 7.403 & 8.259 &  12140 & 11336 & 13001 &  30.93 & 11.45 & 12.15 & 12.00  &  12.02 & 12.03 & 12.22 & 12.18 & 12.28   \\
1.3626 &  0.75 & 7.405 & 8.259 &  12210 & 11404 & 13073 &  30.94 & 11.44 & 12.14 & 11.99  &  12.01 & 12.02 & 12.21 & 12.18 & 12.27   \\
1.3665 &  0.75 & 7.409 & 8.259 &  12470 & 11658 & 13340 &  30.98 & 11.39 & 12.10 & 11.94  &  11.98 & 11.99 & 12.19 & 12.16 & 12.25   \\
1.3810 &  0.75 & 7.415 & 8.258 &  12716 & 11897 & 13592 &  31.01 & 11.34 & 12.05 & 11.91  &  11.95 & 11.97 & 12.18 & 12.15 & 12.24   \\
1.4047 &  0.75 & 7.420 & 8.257 &  12912 & 12089 & 13791 &  31.04 & 11.31 & 12.02 & 11.88  &  11.93 & 11.95 & 12.17 & 12.14 & 12.23   \\
1.4346 &  0.75 & 7.424 & 8.257 &  13075 & 12249 & 13956 &  31.06 & 11.29 & 12.00 & 11.86  &  11.91 & 11.93 & 12.15 & 12.13 & 12.22   \\
1.4687 &  0.75 & 7.426 & 8.257 &  13204 & 12376 & 14086 &  31.08 & 11.26 & 11.98 & 11.85  &  11.90 & 11.92 & 12.14 & 12.12 & 12.21   \\
1.5058 &  0.75 & 7.428 & 8.256 &  13312 & 12483 & 14196 &  31.09 & 11.24 & 11.96 & 11.83  &  11.88 & 11.91 & 12.13 & 12.11 & 12.20   \\
1.5454 &  0.75 & 7.430 & 8.256 &  13407 & 12576 & 14292 &  31.10 & 11.23 & 11.94 & 11.82  &  11.87 & 11.91 & 12.13 & 12.11 & 12.20   \\
1.5862 &  0.75 & 7.431 & 8.256 &  13489 & 12656 & 14377 &  31.11 & 11.21 & 11.93 & 11.81  &  11.86 & 11.90 & 12.12 & 12.10 & 12.19   \\
1.6698 &  0.75 & 7.436 & 8.255 &  13647 & 12809 & 14540 &  31.14 & 11.18 & 11.91 & 11.79  &  11.85 & 11.89 & 12.11 & 12.09 & 12.18   \\
1.7538 &  0.75 & 7.441 & 8.255 &  13785 & 12943 & 14681 &  31.15 & 11.16 & 11.88 & 11.77  &  11.83 & 11.87 & 12.10 & 12.08 & 12.17   \\
1.8356 &  0.75 & 7.445 & 8.254 &  13915 & 13070 & 14816 &  31.17 & 11.13 & 11.86 & 11.75  &  11.82 & 11.86 & 12.09 & 12.07 & 12.16   \\
1.9130 &  0.75 & 7.450 & 8.254 &  14079 & 13228 & 14984 &  31.19 & 11.10 & 11.84 & 11.73  &  11.80 & 11.84 & 12.07 & 12.06 & 12.15   \\
1.9861 &  0.75 & 7.454 & 8.253 &  14220 & 13365 & 15130 &  31.21 & 11.08 & 11.81 & 11.71  &  11.78 & 11.83 & 12.06 & 12.05 & 12.14   \\
2.0578 &  0.75 & 7.457 & 8.253 &  14315 & 13455 & 15229 &  31.22 & 11.06 & 11.80 & 11.70  &  11.77 & 11.82 & 12.05 & 12.04 & 12.13   \\
2.1301 &  0.75 & 7.460 & 8.253 &  14433 & 13569 & 15353 &  31.24 & 11.03 & 11.78 & 11.68  &  11.76 & 11.81 & 12.05 & 12.03 & 12.13   \\
2.1996 &  0.75 & 7.463 & 8.252 &  14561 & 13691 & 15487 &  31.25 & 11.01 & 11.76 & 11.66  &  11.74 & 11.79 & 12.04 & 12.03 & 12.12   \\
2.2615 &  0.75 & 7.469 & 8.252 &  14918 & 14032 & 15861 &  31.29 & 10.94 & 11.70 & 11.62  &  11.70 & 11.76 & 12.01 & 12.00 & 12.09   \\
2.2406 &  0.75 & 7.533 & 8.244 &  17532 & 16488 & 18644 &  31.58 & 10.46 & 11.33 & 11.33  &  11.43 & 11.52 & 11.81 & 11.82 & 11.92   \\
3.2512 &  0.75 & 7.691 & 8.226 &  23669 & 22078 & 25375 &  32.12 &  9.60 & 10.65 & 10.77  &  10.89 & 11.03 & 11.40 & 11.43 & 11.54   \\
3.5602 &  0.75 & 7.759 & 8.218 &  26281 & 24434 & 28269 &  32.31 &  9.28 & 10.39 & 10.56  &  10.68 & 10.84 & 11.23 & 11.27 & 11.39   \\
3.8453 &  0.75 & 7.826 & 8.211 &  28723 & 26648 & 30961 &  32.47 &  9.00 & 10.16 & 10.36  &  10.49 & 10.65 & 11.06 & 11.12 & 11.24   \\
4.1104 &  0.75 & 7.899 & 8.203 &  30926 & 28640 & 33395 &  32.61 &  8.77 &  9.97 & 10.19  &  10.33 & 10.49 & 10.92 & 10.98 & 11.11   \\
4.3833 &  0.75 & 7.991 & 8.193 &  33126 & 30636 & 35818 &  32.74 &  8.58 &  9.80 & 10.04  &  10.18 & 10.35 & 10.78 & 10.85 & 10.97   \\
4.6675 &  0.75 & 8.070 & 8.184 &  35116 & 32435 & 38019 &  32.85 &  8.41 &  9.65 &  9.91  &  10.05 & 10.22 & 10.67 & 10.73 & 10.86   \\
4.9504 &  0.75 & 8.137 & 8.177 &  37191 & 34308 & 40316 &  32.96 &  8.29 &  9.55 &  9.82  &   9.96 & 10.13 & 10.58 & 10.64 & 10.77   \\
5.2505 &  0.75 & 8.197 & 8.171 &  39118 & 36059 & 42438 &  33.05 &  8.19 &  9.46 &  9.73  &   9.88 & 10.06 & 10.50 & 10.57 & 10.69   \\
5.6067 &  0.75 & 8.239 & 8.166 &  40464 & 37281 & 43918 &  33.11 &  8.12 &  9.40 &  9.67  &   9.82 & 10.00 & 10.45 & 10.51 & 10.64   
\enddata
\tablecomments{Unless otherwise stated, all physical quantities are
given in cgs units. The three effective temperatures listed correspond to the secular mean temperature 
($<T_{eff,1}>$) and the temperatures 5\% ($T_{eff,1,lo}$) and 95\% ($T_{eff,1,hi}$) through a given nova cycle. 
UBVRI magnitudes are given on the Johnson-Cousins system
\citep{1990PASP..102.1181B}, JHK are given on the CIT system
\citep{1982AJ.....87.1029E,1982AJ.....87.1893E}. The sequence provided here is abbreviated. A more 
complete sequence sampled at higher resolution and providing
additional information is available in electronic form.}
\label{tab:seq_wd_rev}
\end{deluxetable*}

\clearpage
%\end{landscape}
\end{turnpage}

\tabletypesize{\footnotesize}
\begin{deluxetable}{lcccccc}
\tablecolumns{7} 
%\tablewidth{9cm}
%\tablecaption[Mean offsets between observed absolute near-IR  
%magnitudes of CVs with trigonometric parallaxes and predicted
%absolute magnitudes for CV donors.]
\tablecaption{Mean offsets between observed absolute near-IR  
magnitudes of CVs with trigonometric parallaxes and predicted
absolute magnitudes for CV donors (see Figure~\ref{fig:IR}). The
dispersions around the predictions after applying these offsets are
also given.}
\tablehead{
\colhead{Prediction}   &
\colhead{~$\Delta J$}   & 
\colhead{~$\sigma_J$}   &
\colhead{~$\Delta H$}   &
\colhead{~$\sigma_H$}   &
\colhead{~$\Delta K$}   &
\colhead{~$\sigma_K$}
}{Mean offsets (and associated dispersions) between observed absolute near-IR  
magnitudes of CVs with trigonometric parallaxes and predicted
absolute magnitudes for CV donors.}
\startdata
Broken-power-law           & \multicolumn{6}{c}{}\\
~~~donor sequence               & 1.61 & 1.34 & 1.36 & 1.22 & 1.26   & 1.15\\
\\
Standard Model                  & 1.65 & 1.47 & 1.39 & 1.35 & 1.30  & 1.27\\
Revised Model                   & 1.63 & 1.38 & 1.37 & 1.27 & 1.27  & 1.19
\enddata
\label{tab:IR}
\end{deluxetable}

\tabletypesize{\footnotesize}
\begin{deluxetable}{lcc}
\tablecolumns{3} 
%\tablewidth{11cm}
%\tablecaption[Statistics of the approximate period distributions shown
%in Figure~\ref{fig:porb_dist}.]
\tablecaption{Statistics of the approximate period distributions shown
in Figure~\ref{fig:porb_dist}. The absolute numbers quoted are for a
uniform space density of $2 \times 10^{-5}$~pc$^{-3}$ within 100~pc.}
\tablehead{
\colhead{CV Type} &
\colhead{Standard} &
\colhead{Revised} \\ 
\colhead{} &
\colhead{Model} &
\colhead{Model} \\ \hline \\
\multicolumn{3}{c}{Numbers within 100~pc}
}{Statistics of the approximate period distributions shown in Figure~\ref{fig:porb_dist}.}
\startdata
Active CVs                       &  72.80  &  80.58     \\
Inactive CVs in the Period Gap   &  11.98  &  ~3.19     \\
Pre-Bounce Short-Period Systems  &  34.97  &  16.39     \\
Long-Period CVs                  &  ~1.46  &  ~2.05     \\
All Pre-Bounce CVs               &  36.43  &  18.44     \\
All Post-Bounce CVs              &  36.37  &  62.14     \\
\cutinhead{Percentages within the intrinsic active population}
Long-Period CVs                  &  ~2.00  &  ~2.55     \\  
Pre-Bounce Short-Period CVs      &  48.04  &  20.34     \\
Period Bouncers                  &  49.96  &  77.12     
\enddata
\label{tab:pdist}
\end{deluxetable}

\clearpage
\pagebreak

\appendix

\section{A Compendium of Magnetic Braking Recipes}
\label{app:mb}

Here, we present an overview of some of the most widely-used MB
prescriptions. Note that this is not not meant to be a  
comprehensive exploration of all AML mechanisms, or even all MB
formulations, that have been proposed. In particular, we will not
consider any mechanism unrelated to MB due to stellar winds, such as
circumbinary  disks \citep{2001ApJ...561..329T,2002ApJ...569..395D,
2003ApJ...592.1124T, 2005ApJ...635.1263W},
accretion disk winds \citep{1988ApJ...327..840C, 1994ApJ...427..956L},
or frictional drag associated
with nova eruptions \citep{1998MNRAS.297..633S}.
\footnote{Note that all of these alternatives are, in fact, examples 
of CAML \citep{1995ApJ...439..330K}.}
We have also not yet included some of the most recent MB recipes in
the literature, such as those developed in \citet{2007MNRAS.377..741I}
and \citet{2010ApJ...721..675B}. 
The particular MB prescriptions we do consider were selected 
mainly on the basis that they have proven themselves to be popular
(particularly in the CV literature) and/or are flexible (i.e. include
easily tuneable free parameters that permit a convenient exploration
of the viable parameter space).

\subsection{Verbunt \& Zwaan (1981)}
\label{app:vz}

In a hugely influential paper, \citet[][hereafter VZ81]{1981A&A...100L...7V}
pointed out that a simple extraploation of the observed spin-down
rates of solar-type stars due to MB implied an AML rate far in excess
of $\dot{J}_{GR}$ for CVs and low-mass X-ray binaries (LMXBs). The VZ81
formulation of MB has been widely used in theoretical studies of CV
evolution, and it is instructive to consider its derivation. 

VZ81 begin with the empirical observation due to
\citet{1972ApJ...171..565S} that the equatorial rotation speed of
slowly spinning ($v_{eq,2} \ltappeq 
100~{\rm km~s^{-1}}$) G-type MS stars declines with age, $t$, as 
\begin{equation}
v_{eq,2} = 10^{14} \, f \, t^{-1/2} \, {\rm {cm~s^{-1}}}, 
\label{eq:sku}
\end{equation}
where $f = 0.73$ according to \citet{1972ApJ...171..565S}, and $f = 1.78$
according to \citet{1979PASP...91..737S}. 
Here and below, we retain the subscript ``2'' 
to emphasize that, in the CV setting, the stars whose properties we
are considering are the secondaries. The spin angular momentum of the
star can be written as 
\begin{equation}
J_2 = I_2 \Omega = k_2^2 M_2 R_2^2 \Omega,
\end{equation}
where $I_2$ is the moment of inertia of the star, $k_2$ is its radius
of gyration ($k^2 \simeq 0.1$ for low-mass stars, although in detail
it varies slightly with the stellar properties), and $\Omega =
2\pi/P_{orb}$ is the angular velocity (where we have already assumed
perfect synchronization, i.e. $P_{spin,2} = P_{orb}$). Since we can
also write $\Omega = v_{eq,2}/R_2$, we have
\begin{equation}
J_2 = k_2^2 M_2 R_2 v_{eq,2} = 10^{14} \, f k_2^2 M_2 R_2  t^{-1/2}.
\end{equation}
Differentiating this with respect to $t$ (keeping all other variables
fixed) yields
\begin{equation}
\dot{J}_{2} = - 5\times 10^{13} \, f k_2^2 M_2 R_2  t^{-3/2}.
\end{equation}
The substitution 
\begin{equation}
t^{-3/2} = \left(\frac{v_{eq,2}}{10^{14} \, f}\right)^3 =
\left(\frac{\Omega R_2}{10^{14}\, f}\right)^3 
\end{equation}
then leads to the familiar expression for the VZ81 MB law,
\begin{equation}
\dot{J}_{VZ} = -5 \times 10^{-29} k_2^2f^{-2} M_2 R_2^4 \Omega^3,
\label{eq:jdot_vz}
\end{equation}
where we have changed the subscript to ``VZ'' to allow comparison with
other prescriptions later on. 

We have presented the derivation of this MB recipe in some detail here
in order to expose the weakness of its foundations. The key point 
-- which was already made by \citet{1984ApJS...54..443P} -- 
is the fact that Equation~\ref{eq:sku} was derived
exclusively from observations of 
{\em slowly rotating G-type stars}. Adopting it without change
for CV secondaries, which, in the period range we are
mainly interested in, are extremely fast rotating M-type stars, is
clearly fraught with uncertainties. In particular, even if the
time dependence of Equation~\ref{eq:sku} were universal, it seems
likely that its normalization would depend on the stellar properties 
(e.g. $M_2$ and $R_2$). In this case the braking rate would still 
scale as $\Omega^3$, but its dependence on the stellar properties
could be very different. In a CV setting, this means 
the entire prescription is highly uncertain. After all, 
for a Roche-lobe-filling CV secondary, $M_2$, $R_2$ and $\Omega$ are 
connected via the period-density relation
(Equation~\ref{eq:dense}). Thus any one of these three parameters can 
always be eliminated in favour of the other two. 

None of this is intended as a criticism of VZ81. For their purpose of
demonstrating the relevance of MB as an AML mechanism in CVs and
LMXBs, the order-of-magnitude approach they adopted is entirely
appropriate. Our point here is simply that neither the normalization
nor the functional form of Equation~\ref{eq:jdot_vz} can be regarded
as well-established.

\subsection{Rappaport, Verbunt \& Joss (1983)}
\label{app:rvj}

\citet[][hereafter RVJ83]{1983ApJ...275..713R} carried out one of
the first detailed investigations of MB-driven CV evolution. Realizing 
the uncertainties emphasized in the last section, they chose to
parameterize the AML rate due to MB as 
\begin{equation}
\dot{J}_{MB} = \dot{J}_{VZ} \left(\frac{R_2}{R_{\odot}}\right)^{\gamma-4}.
\label{eq:jdot_rvj}
\end{equation}
For $\gamma = 4$, this reduces to the VZ81 formulation. Other choices
provide a convenient way to assess how the shape and strength of the
MB law affects CV evolution. RVJ83 considered the range $0 \leq \gamma
\leq 4$ and noted that varying $\gamma$ within this range had rather 
dramatic effects on the resulting binary evolution. 

\subsection{Kawaler (1988)}
\label{app:k88}

A more physically motivated MB model was constructed by 
\citet[][hereafter K88]{1988ApJ...333..236K}, 
building on earlier theoretical work by \citet{1968MNRAS.138..359M}, 
\citet{1967ApJ...148..217W} and others. Kawaler's main
contribution was to 
derive an approximate expression for the braking torque exerted by a 
stellar wind in which the key uncertainties and parameter dependencies
are parameterized in a convenient way. 

The AML rate in all MB models can be calculated by assuming perfect
co-rotation of the wind out to the Alfv\'{en} radius, $R_A$
\citep{1987MNRAS.226...57M}.
In general, $R_A$ can be a function of polar angle, but
in the Kawaler model it is assumed to be spherical, so that
\begin{equation}
\dot{J}_{Kaw} = \frac{2}{3} \dot{M}_{w} R_2^2\Omega\left[\left(\frac{R_A}{R_2}\right)_{radial}\right]^{n_k}
\end{equation}
where $\dot{M}_{w}$ is the mass-loss rate in the stellar
wind, and $R_A$ is to be evaluated  for a purely radial field
geometry. The power-law index $n_k$ allows different field 
geometries to be considered, with $n_k=2$ being a purely radial field, 
and $n_k=3/7$ a purely dipolar one. 

In order to determine $(R_A/R_2)_{radial}$, K88 assumes that the
magnetic field of the star scales as
\begin{equation}
\left(\frac{B_2}{B_{\odot}}\right) = K_B \left(\frac{R_2}{R_{\odot}}\right)^{-2}\Omega^{a_k}.
\end{equation}

Most applications of this formalism have focused on the case $a_k =
1$ (based on an observational suggestion by
\citet{1987LNP...291...44L}), for which K88 adopts $K_B^2 \simeq 4.4
\times 10^{11} {\rm g \, cm^{-1}}$. 

The final expression for the AML loss rate in the K88 picture is
\begin{equation}
\small
\dot{J}_{Kaw} =
-K_{W}\Omega^{1+4a_kn_k/3}\dot{M}_{w,14}^{1-2n_k/3}\left(\frac{R_2}{R_{\odot}}\right)^{2-n_k}\left(\frac{M_2}{M_{\odot}}\right)^{-n_k/3},
\label{eq:jdot_k88}
\end{equation}
where $\dot{M}_{w,14} = \dot{M}_{w}/(10^{-14} {\rm M_{\odot} \,  yr^{-1}})$ and 
\begin{equation}
K_W = 2.035\times10^{33} \left(24.93K_V^{-1/2}\right)^{n_k}K_B^{4n_k/3},
\end{equation}
with $K_V$ defined as the ratio of the wind speed at the Alfv\'{e}n radius to
the escape speed there. In practice, K88 sets $K_v = 1$. In this
formulation of MB, the Skumanich relation is recovered for $n_k = 1.5$
(assuming $a_k = 1$), since this yields the required $\dot{J} \propto
\Omega^3$. 

\subsection{Mestel \& Spruit (1987)}
\label{app:ms87}

One of the main advantages of the K88 formulation is that it provides
a convenient parameterization for the dependence of the braking
rate on the magnetic field geometry inside $R_A$ (via the parameter
$n_k$). However, the way in which this dependence was calculated is
highly approximate. For example, a predominantly dipolar field
structure inside $R_A$ implies the existence of closed field
lines near the equatorial plane that cannot contribute to
MB. Moreover, the size of this ``dead zone'' will itself
depend on key parameters such as the magnetic field strength and the
rotation rate. These effects cannot be captured in single-zone outflow
descriptions like the K88 model.

In an attempt to overcome these limitations, \citet[][hereafter
  MS87]{1987MNRAS.226...57M} 
constructed a model in which the ``wind zone'' and ``dead zone'' are
both treated explicitly and simultaneously. The dead zone in this
picture is inside the Alfv\'{e}n surface and its equatorial extent, $R_d$, is
calculated self-consistently. The magnetic field structure is then
taken to be dipolar inside a radius $R_d$ and radial beyond this. 
%MB
%is weaker in this picture than for a purely radial wind, partly
%because gas trapped in 
%the dead zone cannot contribute to the AML, and partly because the
%geometry is different.
The AML rate produced by the stellar wind in this picture is again
found by assuming co-rotation out to the 
Alfv\'{e}n surface. Allowing for an angular dependence of $R_A$, this yields 
\begin{equation}
\dot{J}_{MB} = -4\pi\Omega\int_0^{\pi/2} (\rho_A v_A R_A^2) (R_A\sin{\theta})^2 \sin{\theta} d\theta,
\end{equation}
where $\rho_A$, $v_A$ and $R_A$ are the wind density, velocity and radius
at the Alfv\'{e}n surface, and $\theta$ is the polar angle ($\theta =
\pi/2$ corresponds to the equatorial plane). In the MS87 scenario, 
this can be rewritten as (their Equation 16)
\begin{equation}
\dot{J}_{MS} = -4 \pi \Omega \int_0^{1} \rho_d v_d R_d^2 R_A^2(1-\mu^2) d\mu,
\label{eq:jdot_ms87}
\end{equation}
where $\rho_d$, $v_d$, and $R_d$ now denote the density, velocity and
radius in the wind zone, but now evaluated at $R = R_d$. The actual
values of $R_A$ and $R_d$ can be calculated via two implicit equations
(Eqs 8 and 14 in MS87). 

We will not develop this derivation further here or repeat the implicit
equations for $R_A$ and $R_d$; for details, the reader is
referred to MS87 as well as \citet{1988MNRAS.231..535H,
1989ApJ...342.1019M, 1999MNRAS.302..203L, 2003ApJ...599..516I}.
\footnote{However, note that, as pointed out by
\citet{1989ApJ...342.1019M} and \citet{1999MNRAS.302..203L}, the
numerical coefficient given in MS87 
for the quantity $\chi l_d$ ($1.45\times10^{-4}$; their Equation~12)
is incorrect and 
should instead read $1.1 \times 10^{-4}$. As a result, the numerical
constant $7.25\times 10^{-5}$ appearing in Equation~4 of
\citet{1988MNRAS.231..535H} is also incorrect and should read $5.5 \times
10^{-5}$.} 
However, two key points regarding the way in which 
$\dot{J}_{MS}$ is actually calculated will be important to our discussion
below. First, in order to calculate $\dot{J}_{MS}$,  the density at
the base of the wind, $\rho_0$, and the magnetic field strength of the
star, $B_2$, are needed (this is partly because the implicit equations
for $R_A$ and $R_d$ depend on these quantities). We follow the
notation of \citet{1988MNRAS.231..535H} and parameterize $\rho_0$ as 
\begin{equation} 
\frac{\rho_0}{\rho_{0,\odot}} = \left(\frac{\Omega}{\Omega_{\odot}}\right)^{n_{ms}}
\end{equation}
and $B_2$ as 
\begin{equation} 
\frac{B_2}{B_{\odot}} = \left(\frac{\Omega}{\Omega_{\odot}}\right)^{p_{ms}}.
\end{equation}
Virtually all applications of the MS87 model take $p_{ms}=1$, but
different values for the power-law index $n_{ms}$ are sometimes 
considered (typically $0.5 \leq n_{ms} \leq 2$).\footnote{In order to avoid
confusion, we note that MS87 use the power law index $n$ differently in their
paper.} Second, in evaluating $\dot{J}_{MS}$, several approximations 
are commonly made. In particular, the angular dependence of $R_A$ is
usually ignored, the equatorial value being used for all angles. As
explained in MS87, this formally yields a lower limit on the AML rate,
but one that should be quite close to the correct answer. We will
return to this issue below in Section~\ref{app:it}.

\subsection{Andronov et al. (2003)}
\label{app:aps}

One potential problem with AML prescriptions that scale as $\Omega^3$
is that they are unable to explain the observed spin-down rates of
young stars in clusters. In particular, 
prescriptions in which $\dot{J}$ depends so steeply on $\Omega$
predict that  rapid rotators are braked extremely efficiently and
should therefore not be observed. By contrast, very rapidly rotating
stars stars have been found in many open
clusters \citep[e.g.][]{1987ApJ...318..337S}. In order to explain these
observations, \citet{2000ApJ...534..335S} modified the K88 prescription in
a way that makes the dependence of $\dot{J}$ on $\Omega$ much shallower
at high rotation rates. \citet[][hereafter
  APS03]{2003ApJ...582..358A} then
pointed out that adopting this modified prescription would have a
significant impact on CV evolution.

The \citet{2000ApJ...534..335S} MB recipe that was adopted by APS03 is given
by 
\begin{equation}
\dot{J}_{APS} =
-K_{w}\left(\frac{R_2}{R_{\odot}}\right)\left(\frac{M_{\odot}}{M_2}\right) 
\left\{
\begin{array}{lll}
\Omega^3              & {\rm for} & \Omega \leq \Omega_{crit} \\
& & \\
\Omega \Omega_{crit}^2 & {\rm for} & \Omega > \Omega_{crit} \\
\end{array}
\right. .
\label{eq:jdot_aps}
\end{equation}
Here, $K_W = 2.7 \times 10^{47} {\rm g~cm^2~s}$, and we have
intentionally used the same symbol for this normalization factor as for that
in the K88 prescription. This is because, in the
limit $\Omega < \Omega_{crit}$, the APS03 prescription reduces to the
default K88 MB law ($a_K = 1$, $n_K = 1.5$, $\dot{M}_{w,14} = 1$ and
$K_V = 1$). In principle, the critical 
angular velocity $\Omega_{crit}$ depends on the stellar properties,
but this is irrelevant for CV donors, since they all spin well above
$\Omega_{crit}$. 

As we shall see, the APS03 braking law produces extremely low braking
rates. These would not suffice to drive donors significantly out
of thermal equilibrium above the period gap and, if correct, would thus 
require a fundamental revision of CV evolution theory. It is therefore
worth noting that \citet{2003ApJ...586..464B,2007ApJ...669.1167B} 
has suggested a different
interpretation for the open cluster rotation data. In Barnes's 
picture, all stars above the fully convective boundary eventually
settle on a $\dot{J} \propto \Omega^3$ spin-down sequence (which Barnes
dubs the ``I-sequence''), consistent with the
\citet{1972ApJ...171..565S} relationship. However, stars do not reach
this sequence at a fixed 
time, so that the fastest rotators in a given young cluster will
always be those object that have not yet settled on the I-sequence. 

\subsection{Ivanova \& Taam (2003)}
\label{app:it}

The last MB prescription we will consider is that due to
\citet[][hereafter IT03]{2003ApJ...599..516I}. Taking the MS87 model
as their starting 
point, their goal was to identify an optimal value for $n_{ms}$
\footnote{Note that IT03
use the symbol $p$ for this (not to be confused with our definition of
$p_{ms}$ in Section~\ref{app:ms87}).}
by assuming that the X-ray luminosity of the star, $L_{x,2}$, is generated
by the dead zone and then requiring that the dependence of $L_{x,2}$ on
$\Omega$ should match the observations. The specific dependence they
aimed to reproduce was $L_{x,2} \propto \Omega^2$ at low $\Omega /
\Omega_{\odot} \ltappeq 2-12$ and $L_{x,2} \propto \Omega^0$ for higher rotation
speeds \citep{2003A&A...397..147P}. According to their calculations, this
behaviour could be matched within the MS87 framework by taking $n_{ms}
\simeq 0.6$. 

Crucially, IT03 found that the dependence of $\dot{J}_{MS}$ on
$\Omega$ turns over at fast rotation speeds, becoming substantially
flatter than the roughly $\dot{J} \propto \Omega^3$ Skumanich-like
relation they obtained at low rotation speeds. In fact, they
obtained this behaviour for essentially {\em all} values of $n_{ms}$
in the range $0 < n_{ms} < 1$. This is somewhat surprising at first
sight, since neither MS87, 
nor \citet{1988MNRAS.231..535H}, nor \citet{1989ApJ...342.1019M}, nor
we found such a flattening in their implementations of the MS87
picture. On the other hand, \citet{1995A&A...294..469K} also found
a flattening in $\dot{J}$ with $\Omega$ in their implementation of a
Weber-Davis magnetic wind model (from $\dot{J} \propto \Omega^3$ at
low rotation rates to $\Omega^3 \propto \Omega^2$ at high
rates). Moreover, Natasha Ivanova (private communication) has
suggested to us 
that the discrepancy between MS87 and IT03 may reflect different ways
of estimating conditions at the Alfv\'{e}n surface (specificially
whether one connects them to the physical parameters within the dead
zone or at the stellar surface). Pursuing this is beyond the scope of
the present paper. 

For our purposes here, it is possible to avoid this issue completely
by simply considering the analytical approximation for $\dot{J}$
suggested by IT03. This is  
\begin{equation}
\dot{J}_{IT} = -K_j \left(\frac{R_2}{R_{\odot}}\right)^4 \left\{
\begin{array}{lll}
\left(\frac{\Omega}{\Omega_{\odot}}\right)^3          & {\rm for} & \Omega \leq \Omega_{x} \\
& & \\
\frac{\Omega^{1.3} \Omega_{x}^{1.7}}{\Omega_{\odot}^3} & {\rm for} & \Omega > \Omega_{x} \\
\end{array}
\right. .
\label{eq:jdot_it}
\end{equation}
where $K_j = 6\times10^{30} {\rm ~ g ~ cm^2 ~ s^{-2}}$, and we have followed
\citet{2006ApJ...636..985I} in dropping the scaling with coronal
temperature in the formula given by IT03. This is a good
approximation, since the coronal temperature is otherwise approximated
by the virial temperature, which scales as $M_2 R_2^{-1}$ and is
therefore almost constant along the lower MS. As with the APS03
prescription, in principle $\Omega_{x}$ depends on the stellar
properties, but in practice all CV donors fall into the fast rotation
regime, $\Omega > \Omega_{x}$.

\section{A Compilation of Masses and Radii for Non-Interacting Low-Mass Stars}
\label{app:msmr}

The starting point for our compilation was Table~1 in
\citet{2007ApJ...660..732L} and Figure~13 in
\citet{2009A&A...504.1021V}. These
data sets were then further updated with several new and improved
measurements taken from the recent literature. Following
\citet{2007ApJ...660..732L}, we include both isolated stars and objects
found in detached binary systems. Our full data base of masses and
radii for non-interacting low-mass stars is listed in
Table~\ref{tab:msmr}. 

\LongTables
\begin{deluxetable}{lccccccc}
%\tablefontsize{\footnotesize}
\tablecolumns{8} 
\tablewidth{0pt}
%\tablecaption[Mass and radius measurements for non-interacting low-mass stars.]
\tablecaption{Mass and radius measurements for non-interacting low-mass stars.}
\tablehead{
\colhead{Name} &
\colhead{$M (M_{\odot})$} &
\colhead{$\sigma_M (M_{\odot})$} &
\colhead{$R (R_{\odot})$} &
\colhead{$\sigma_R (R_{\odot})$} &
\colhead{Type\tablenotemark{a}} &
\colhead{Reference\tablenotemark{b}} &
\colhead{Comments\tablenotemark{c}}
}{Mass and radius measurements for non-interacting low-mass stars.}
\startdata
GJ 15 A			       & 0.4040	& 0.0404	& 0.379	   & 0.006	& S 	& 1    	&   	      \\   
GJ 514			       & 0.5260	& 0.0526	& 0.611	   & 0.043	& S	& 1    	&   	      \\ 
GJ 526			       & 0.5020	& 0.0502	& 0.493	   & 0.033	& S	& 1    	&   	      \\ 
GJ 687			       & 0.4010	& 0.0401	& 0.492	   & 0.038	& S	& 1    	&   	      \\ 
GJ 752 A		       & 0.4840	& 0.0484	& 0.526	   & 0.032	& S	& 1    	&   	      \\ 
GJ 880			       & 0.5860	& 0.0586	& 0.689	   & 0.044	& S	& 1    	&   	      \\ 
GJ 205			       & 0.631	& 0.031		& 0.702	   & 0.063	& S	& 1    	&   	      \\ 
GJ 191			       & 0.281	& 0.014		& 0.291	   & 0.025	& S	& 1    	&   	      \\ 
GJ 699			       & 0.158	& 0.008		& 0.196	   & 0.008	& S	& 1    	&   	      \\ 
GJ 411			       & 0.403	& 0.020		& 0.393	   & 0.008	& S	& 1    	&   	      \\ 
GJ 380			       & 0.670	& 0.033		& 0.605	   & 0.020	& S	& 1    	&   	      \\ 
GJ 105 A		       & 0.790	& 0.039		& 0.708	   & 0.050	& S	& 1    	&   	      \\ 
YY Gem A		       & 0.5992	& 0.0047	& 0.6191   & 0.0057	& B	& 1    	&   	      \\ 
YY Gem B		       & 0.5992	& 0.0047	& 0.6191   & 0.0057	& B	& 1    	&   	      \\ 
CU Cnc A		       & 0.4333	& 0.0017	& 0.4317   & 0.0052	& B	& 1    	&   	      \\ 
CU Cnc B		       & 0.3980	& 0.0014	& 0.3908   & 0.0094	& B	& 1    	&   	      \\ 
GU Boo A		       & 0.610	& 0.007		& 0.623	   & 0.016	& B	& 1    	&   	      \\ 
GU Boo B		       & 0.599	& 0.006		& 0.620	   & 0.020	& B	& 1    	&   	      \\ 
BW3 V38 A		       & 0.44	& 0.07		& 0.51	   & 0.04	& B	& 1    	&   	      \\ 
BW3 V38 B		       & 0.41	& 0.09		& 0.44	   & 0.06	& B	& 1    	&   	      \\ 
TrES-Her0-07621 A	       & 0.493	& 0.003		& 0.453	   & 0.060	& B	& 1    	&   	      \\ 
TrES-Her0-07621 B	       & 0.489	& 0.003		& 0.452	   & 0.050	& B	& 1    	&   	      \\ 
2MASS J05162881+2607387 A      & 0.787	& 0.012		& 0.788	   & 0.015	& B	& 1    	&   	      \\ 
2MASS J05162881+2607387 B      & 0.770	& 0.009		& 0.817	   & 0.010	& B	& 1    	&   	      \\ 
2MASS J04463285+1901432 A      & 0.47	& 0.05		& 0.57	   & 0.02	& B	& 1    	&  (i) 	      \\ 
2MASS J04463285+1901432 B      & 0.19	& 0.02		& 0.21	   & 0.01	& B	& 1    	&  (i)	      \\ 
UNSW-TR-2 A		       & 0.529	& 0.035		& 0.641	   & 0.05	& B	& 1    	&   	      \\ 
UNSW-TR-2 B		       & 0.512	& 0.035		& 0.608	   & 0.06	& B	& 1    	&   	      \\ 
V818 Tau B		       & 0.7605	& 0.0062	& 0.768	   & 0.010	& B	& 1    	&   	      \\ 
FL Lyr B		       & 0.960	& 0.012		& 0.962	   & 0.028	& B	& 1    	&   	      \\ 
V1061 Cyg Ab		       & 0.9315	& 0.0068	& 0.974	   & 0.020	& B	& 1    	&   	      \\ 
V1061 Cyg B		       & 0.925	& 0.036		& 0.870	   & 0.087	& B	& 1    	&   	      \\ 
RW Lac B		       & 0.870	& 0.004		& 0.964	   & 0.004	& B	& 1    	&   	      \\ 
HS Aur B		       & 0.879	& 0.017		& 0.873	   & 0.024	& B	& 1    	& 	      \\ 
OGLE-TR-5 B		       & 0.271	& 0.035		& 0.263	   & 0.012	& B	& 1    	& 	      \\ 
OGLE-TR-6 B		       & 0.359	& 0.025		& 0.393	   & 0.018	& B	& 1    	& 	      \\ 
OGLE-TR-7 B		       & 0.281	& 0.029		& 0.282	   & 0.013	& B	& 1    	& 	      \\ 
OGLE-TR-18 B		       & 0.387	& 0.049		& 0.39	   & 0.04	& B	& 1    	& 	      \\ 
OGLE-TR-34 B		       & 0.509	& 0.038		& 0.435	   & 0.033	& B	& 1    	& 	      \\ 
OGLE-TR-78 B		       & 0.243	& 0.015		& 0.24	   & 0.013	& B	& 1    	& 	      \\ 
OGLE-TR-106 B		       & 0.116	& 0.021		& 0.181	   & 0.013	& B	& 1    	& 	      \\ 
OGLE-TR-120 B		       & 0.47	& 0.04		& 0.42	   & 0.02	& B	& 1    	& 	      \\ 
OGLE-TR-122 B		       & 0.092	& 0.009		& 0.120	   & 0.019	& B	& 1    	& 	      \\ 
OGLE-TR-125 B		       & 0.209	& 0.033		& 0.211	   & 0.027	& B	& 1    	& 	      \\ 
SDSS-MEB-1 A		       & 0.272	& 0.020		& 0.268	   & 0.010	& B	& 2    	& 	      \\ 
SDSS-MEB-1 B		       & 0.240	& 0.022		& 0.248	   & 0.0090	& B	& 2    	& 	      \\ 
NSVS01031772 A		       & 0.5428	& 0.0027	& 0.5260   & 0.0028	& B 	& 3    	& 	      \\ 
NSVS01031772 B		       & 0.4982	& 0.0025	& 0.5088   & 0.0030	& B	& 3    	& 	      \\ 
T-Lyr1-17236 A		       & 0.6795	& 0.0107	& 0.634	   & 0.043	& B 	& 4    	& 	      \\ 
T-Lyr1-17236 B		       & 0.5226	& 0.0061	& 0.525	   & 0.052	& B 	& 4    	& 	      \\ 
V405 And A		       & 0.49	& 0.05		& 0.78	   & 0.02	& B	& 5    	& (ii)	      \\ 
V405 And B		       & 0.21	& 0.04		& 0.24	   & 0.04	& B	& 5    	& (ii)	      \\ 
2MASS J0154 A		       & 0.66	& 0.03		& 0.64	   & 0.08	& B	& 6    	& 	      \\ 
2MASS J0154 B		       & 0.62	& 0.03		& 0.61	   & 0.09	& B	& 6    	& 	      \\ 
BD -22 5866 Aa		       & 0.5881	& 0.029		& 0.614	   & 0.045	& B	& 7    	& 	      \\ 
BD -22 5866 Ab		       & 0.5881	& 0.029		& 0.614	   & 0.045	& B	& 7    	& 	      \\ 
LP-133-373 A		       & 0.340	& 0.014		& 0.33	   & 0.02	& B	& 8    	& 	      \\ 
LP-133-373 B		       & 0.340	& 0.014		& 0.33	   & 0.02	& B	& 8    	& 	      \\ 
NSVS-02502726 A		       & 0.714	& 0.019		& 0.645	   & 0.006	& B	& 9    	& 	      \\ 
NSVS-02502726 B		       & 0.347	& 0.012		& 0.501	   & 0.005	& B	& 9    	& 	      \\ 
V471 Tau B		       & 0.93	& 0.07		& 0.96	   & 0.04	& W	& 10   	& 	      \\ 
RXJ2130 		       & 0.555	& 0.023		& 0.534	   & 0.017	& W	& 11   	& 	      \\ 
RR Cae B		       & 0.1825	& 0.0139	& 0.209	   & 0.0143	& W 	& 12   	& (iii)	      \\ 
EC 13471		       & 0.43	& 0.04		& 0.42	   & 0.02	& W	& 13   	& (iv) 	      \\ 
NN Ser B		       & 0.111	& 0.004		& 0.141	   & 0.002	& W	& 14   	& (v)	      \\ 
RXJ0239.1 A		       & 0.73	& 0.009		& 0.741	   & 0.004	& B	& 15   	& 	      \\ 
RXJ0239.1 B		       & 0.693	& 0.006		& 0.703	   & 0.002	& B	& 15   	& 	      \\ 
GJ166A			       & 0.877	& 0.044		& 0.770	   & 0.021	& S	& 16   	& 	      \\ 
GJ570A			       & 0.802	& 0.040		& 0.739	   & 0.019	& S	& 16   	& 	      \\ 
GJ845			       & 0.762	& 0.038		& 0.732	   & 0.006	& S	& 16   	& 	      \\ 
GJ879			       & 0.725	& 0.036		& 0.629	   & 0.051	& S	& 16   	& 	      \\ 
GJ887			       & 0.503	& 0.025		& 0.459	   & 0.011	& S	& 16   	& 	      \\ 
GJ551			       & 0.123	& 0.006		& 0.141	   & 0.007	& S	& 16   	& 	      \\ 
2MASSJ0746		       & 0.085	& 0.010		& 0.078	   & 0.010	& B	& 17   	& 	      \\ 
HIP96515 Aa		       & 0.59	& 0.03		& 0.64	   & 0.01	& B	& 18   	& 	      \\ 
HIP96515 Ab		       & 0.54	& 0.03		& 0.55	   & 0.03	& B	& 18   	& 	      \\ 
CMDra A			       & 0.2310	& 0.0009	& 0.2534   & 0.0019	& B	& 19   	& 	      \\ 
CM Dra B		       & 0.2141	& 0.0010	& 0.2396   & 0.0015	& B	& 19   	& (vi)	      \\ 
T-TR-205-013 B		       & 0.124	& 0.010		& 0.167	   & 0.006	& B	& 20   	& 	      \\ 
RXJ1547 A		       & 0.2576	& 0.0085	& 0.2895   & 0.0068	& B	& 21   	& (vii)	      \\ 
RXJ1547 B		       & 0.2585	& 0.0080	& 0.2895   & 0.0068	& B	& 21   	& (vii)	      \\ 
IM Vir A		       & 0.981	& 0.012		& 1.061	   & 0.016	& B	& 22   	& 	      \\ 
IM Vir B		       & 0.6644	& 0.0048	& 0.681	   & 0.013	& B	& 22   	& 	      
\enddata													      

\tablenotetext{a}{(S) = single star or star in very wide binary; (W) star in close binary with white dwarf companion; (B) star in other type of binary}

\tablenotetext{b}{(1) \citet{2007ApJ...660..732L};
(2) \citet{2008ApJ...684..635B};
(3) \citet{2006astro.ph.10225L};
(4) \citet{2008ApJ...687.1253D};
(5) \citet{2009A&A...504.1021V};
(6) \citet{2008MNRAS.386..416B};
(7) \citet{2008ApJ...682.1248S};
(8) \citet{2007ApJ...661.1112V};
(9) \citet{2009NewA...14..496C};
(10) \citet{2001ApJ...563..971O};
(11) \citet{2004MNRAS.355.1143M};
(12) \citet{2007MNRAS.376..919M};
(13) \citet{2003MNRAS.345..506O};
(14) \citet{2010MNRAS.402.2591P};
(15) \citet{2007ASPC..362...26L};
(16) \citet{2009A&A...505..205D};
(17) \citet{2009ApJ...695..310B};
(18) \citet{2009A&A...503..873H};
(19) \citet{2009ApJ...691.1400M};
(20) \citet{2007ApJ...663..573B};
(21) \citet{2009arXiv0907.2924H};
(22) \citet{2009ApJ...707..671M}}

\tablenotetext{c}{(i) This object is a member of a 150 Myr cluster;
(ii) Mass errors estimated from Fig. 13 in Vida et al. (2009); (iii)
The mass and radius given in the tabla correspond to the mid-points of
the model-dependent parameter ranges quoted in Maxted et al. (2007);
the half-range associated with the model uncertainties has been added
in quadrature to the quoted formal errors; (iv) This object may be the
secondary of a hibernating CV (O'Donoghue et al. 2003); we have not
attempted to correct the radius for tidal/rotational deformation; (v)
The radius listed in the table has been corrected for
irradiation-driven bloating, as described by Parsons et al. (2010;
also see Section~\ref{sec:irrad}); (vi) CM Dra may be a Population II
object; (vii) Since the mass ratio estimated from the radial velocity
curve of this binary was $q = 1.00 \pm 0.02$, Hartman et al. (2009)
assumed equal radii for both binary component in obtaining their final
parameter estimates.}

\label{tab:msmr}

\end{deluxetable}

\end{document}